\documentstyle[aps,eqsecnum,preprint,tighten]{revtex}
\begin{document}
\draft
\hyphenation{quasi-nucleon}
\def\vec#1{{\bf #1}}
\def\tr{{\rm tr}}
\def\Tr{{\rm Tr}}
\def\eqref#1{(\ref{#1})}
\def\Eq{E_q}
\def\Eqbar{\overline E_q}
\def\Psibra{\langle {\Psi_0}|}
\def\Psiket{|{\Psi_0} \rangle}
\def\ie{{\it i.e.},}
\def\dfour#1{\!{\rm d}^4 #1\,}
\def\dthree#1{\!{\rm d}^3 #1\,}
\def\qslash{\rlap{/}{q}}
\def\uslash{\rlap{/}{u}}
\def\Sigmas{\Sigma_{\rm s}}
\def\Sigmav{\Sigma_{\rm v}}
\def\xsq{x^2}
\def\qsq{q^2}
\def\Qsq{Q^2}
\def\musq{\mu^2}
\def\qdotu{q\cdot u}
\def\qzero{q_0}
\def\qzerobar{\overline\qzero}
\def\qvec{\vec q}
\def\qvecsq{\qvec^2}
\def\psibar{\overline{\psi}}
\def\qbar{\overline{q}}
\def\ubar{\overline{u}}
\def\dbar{\overline{d}}
\def\sbar{\overline{s}}
\def\chibar{\overline{\chi}}
\def\psibarpsi{\psibar \psi}
\def\qbarq{\qbar q}
\def\ubaru{\ubar u}
\def\dbard{\dbar d}
\def\sbars{\sbar s}
\def\smN{{\scriptscriptstyle\rm N}}
\def\lamN{\lambda_{\smN}^{\ast}}
\def\lamNp{\lambda_{\smN}}
\def\lamNsq{{\lambda_{\smN}^{\ast}}^2}
\def\lamsq{\lambda_{\smN}^2}
\def\etabar{\overline{\eta}}
\def\gammafive{\gamma^5}
\def\gammazero{\gamma^0}
\def\gammamul{\gamma_\mu}
\def\sigmamunu{\sigma^{\mu\nu}}
\def\Piu{\Pi_u}
\def\vector{\vec}
\def\MN{M_{\smN}}
\def\MNstar{\Mstar_{\smN}}
\def\Mstar{M^\ast}
\def\beq{\begin{equation}}
\def\eeq{\end{equation}}
\def\beqa{\begin{eqnarray}}
\def\eeqa{\end{eqnarray}}
\def\MeV{\nobreak\,\mbox{MeV}}
\def\GeV{\nobreak\,\mbox{GeV}}
\def\fm{\nobreak\,\mbox{fm}}
\def\ltsim{\hspace{3pt}\raisebox{-3.5pt}
           {$\stackrel{\displaystyle <}{\sim}$}\hspace{3pt}}
\def\partderiv#1#2{{\partial#1\over\partial#2}}
\def\gvec#1{\mbox{\boldmath $#1$}}
\def\rvec#1{{\bf #1}}
\def\transpose#1{#1^{\mbox{\tiny T}}}
\def\tr{{\rm tr}}
\def\Tr{{\rm Tr}}
\def\A{{\cal A}}
\def\G{{\cal G}}
\def\D{{D}}
\def\udotD{u\!\cdot\!\D}
\def\dfourslash#1{\!{{\rm d}^4 #1\over (2\pi)^4}\,}
\def\dthreeslash#1{\!{{\rm d}^3 #1\over (2\pi)^3}\,}
\def\doneslash#1{\!{{\rm d} #1\over 2\pi}\,}
\def\dfour#1{\!{\rm d}^4 #1\,}
\def\dthree#1{\!{\rm d}^3 #1\,}
\def\done#1{\!{\rm d} #1\,}
\def\qslash{\rlap{/}{q}}
\def\uslash{\rlap{/}{u}}
\def\xslash{\rlap{/}{x}}
\def\Dslash{\rlap{\,/}{\D}}
\def\overleftarrow#1{\stackrel{\leftarrow}{#1}}
\def\partialL{\raisebox{1.25pt}{$\overleftarrow{\partial}$}}
\def\DL{\raisebox{1.25pt}{$\overleftarrow{\D}$}}
\def\DslashL{\hspace{-3.25pt}\overleftarrow{\Dslash}\hspace{-3.25pt}}
\def\Pis{\Pi_{s}}
\def\PisE{\Pis^{\mbox{\tiny E}}}
\def\PisO{\Pis^{\mbox{\tiny O}}}
\def\Piq{\Pi_q}
\def\PiqE{\Piq^{\mbox{\tiny E}}}
\def\PiqO{\Piq^{\mbox{\tiny O}}}
\def\Piu{\Pi_u}
\def\PiuE{\Piu^{\mbox{\tiny E}}}
\def\PiuO{\Piu^{\mbox{\tiny O}}}
\def\Sigmas{\Sigma_{s}}
\def\Sigmav{\Sigma_{v}}
\def\xsq{x^2}
\def\qsq{q^2}
\def\Qsq{Q^2}
\def\musq{\mu^2}
\def\qdotu{q\!\cdot\!u}
\def\scap#1{{\scriptscriptstyle #1}}
\def\nuc{{\scriptscriptstyle \! N}}
\def\QCD{{\mbox{\tiny QCD}}}
\def\phen{{\mbox{\tiny phen}}}
\def\rhonuc{\rho_\nuc}
\def\Mnuc{M_\nuc}
\def\sigmanuc{\sigma_\nuc}
\def\Snuc{S_\nuc}
\def\mpi{m_\pi}
\def\fpi{f_\pi}
\def\psibar{\overline{\psi}}
\def\qbar{\overline{q}}
\def\ubar{\overline{u}}
\def\dbar{\overline{d}}
\def\sbar{\overline{s}}
\def\chibar{\overline{\chi}}
\def\psibarpsi{\psibar \psi}
\def\qbarq{\qbar q}
\def\ubaru{\ubar u}
\def\dbard{\dbar d}
\def\sbars{\sbar s}
\def\psidagger{\psi^\dagger}
\def\qdagger{q^\dagger}
\def\udagger{u^\dagger}
\def\ddagger{d^\dagger}
\def\sdagger{s^\dagger}
\def\psidaggerpsi{\psidagger \psi}
\def\qdaggerq{\qdagger q}
\def\udaggeru{\udagger u}
\def\ddaggerd{\ddagger d}
\def\sdaggers{\sdagger s}
\def\adagger{a^\dagger}
\def\bdagger{b^\dagger}
\def\mquark{m_q}
\def\mup{m_u}
\def\mdown{m_d}
\def\mstrange{m_s}
\def\lqsq{\lambda_q^2}
\def\gs{g_{\rm s}}
\def\alphas{\alpha_{\rm s}}
\def\gluonconA{{\displaystyle{\alphas\over\pi}}G^2}
\def\gluonconAtext{(\alphas/\pi)G^2}
\def\gluonconB{{\displaystyle{\alphas\over\pi}}
               \biggl[(u\!\cdot\!G)^2+(u\!\cdot\!\widetilde{G})^2\biggr]}
\def\gluonconBtext{(\alphas/\pi)
                   [(u\!\cdot\!G)^2+(u\!\cdot\!\widetilde{G})^2]}
\def\gluonconBrest{{\displaystyle{\alphas\over\pi}}
                   \biggl[(u^\prime\!\cdot\!G)^2
                       +(u^\prime\!\cdot\!\widetilde{G})^2\biggr]}
\def\gluonconBresttext{(\alphas/\pi)

[(u^\prime\!\cdot\!G)^2+(u^\prime\!\cdot\!\widetilde{G})^2]}
\def\gluonminus{{\displaystyle{\alphas\over\pi}}
                  \left(\rvec{E}^2-\rvec{B}^2\right)}
\def\gluonminustext{(\alphas/\pi)(\rvec{E}^2-\rvec{B}^2)}
\def\gluonplus{{\displaystyle{\alphas\over\pi}}
                  \left(\rvec{E}^2+\rvec{B}^2\right)}
\def\gluonplustext{(\alphas/\pi)(\rvec{E}^2+\rvec{B}^2)}
\def\threegluon{\gs^3 f G^3}
\def\mixbar{\gs\qbar\sigma\!\cdot\!\G q}
\def\mixuslash{\gs\qbar\uslash\sigma\!\cdot\!\G q}
\def\mixdagger{\gs\qdagger\sigma\!\cdot\!\G q}
\def\Nc{N_{\rm c}}
\def\Nf{N_{\rm f}}
\def\me#1{\langle{#1}\rangle}
\def\vacme#1{\me{#1}_{\lower2pt\hbox{\scriptsize $\!$vac}}}
\def\nucme#1{\me{#1}_{\lower2pt\hbox{$\scriptscriptstyle \! N$}}}
\def\rhome#1{\me{#1}_{\lower2pt\hbox{$\scriptstyle
            \!\rho_{\lower2pt\hbox{$\scriptscriptstyle \! N$}}$}}}
\def\meadjust#1{\left<{#1}\right>}
\def\vacmeadjust#1{\meadjust{#1}_{\lower2pt\hbox{\scriptsize $\!$vac}}}
\def\nucmeadjust#1{\meadjust{#1}_{\lower2pt\hbox{$\scriptscriptstyle \!
N$}}}
\def\rhomeadjust#1{\meadjust{#1}_{\lower2pt\hbox{$\scriptstyle
            \!\rho_{\lower2pt\hbox{$\scriptscriptstyle \! N$}}$}}}
\def\bra#1{\langle #1|}
\def\ket#1{| #1\rangle}
\def\braadjust#1{\left<{#1}\right|}
\def\ketadjust#1{\left|{#1}\right>}
\def\vacbra{\bra{{\rm vac}}}
\def\vacket{\ket{{\rm vac}}}
\def\nucbra{\bra{\widetilde N}}
\def\nucket{\ket{\widetilde N}}
\def\Psibra{\bra{\Psi_0}}
\def\Psiket{\ket{\Psi_0}}
\def\ie{{\it i.e.},}
\def\lambn{\lambda_\nuc^{*^{2}}}
\def\lamb0{\lambda_\nuc^{2}}
\def\masn{M^{*}_\nuc}
\def\msn2{M_\nuc^{*^{2}}}
\def\Eq{E_q}
\def\Eqb{\overline{E}_q}
\def\qv2{{\bf q}}
\def\borel{{\cal B}}
\def\PiEi{\Pi_i^{\mbox{\tiny E}}}
\def\PiOi{\Pi_i^{\mbox{\tiny O}}}
\def\e{e}

\preprint{\vbox{\hfill DOE/ER/40762--056\\
                \null\hfill UMPP \#95--108\\
                \null\hfill OSUPP \#94--152\\
                 \null\hfill TRI--95--7}}

\title{QCD Sum Rules and Applications to Nuclear Physics}

\author{T. D. Cohen$^{\rm a,b,}$%
\thanks{Supported in part by the U.S.\ Department of Energy and
              the U.S.\ National Science Foundation.},
R. J. Furnstahl$^{\rm c,}$%
\thanks{Supported in part by the U.S.\ National Science Foundation
                 and the Sloan Foundation.},
D. K. Griegel$^{\rm b,}$%
\thanks{Supported in part by the U.S.\ Department of Energy.},
Xuemin Jin$^{\rm d,}$%
\thanks{Supported in part by the National Sciences
             and Engineering Research Council of Canada.}}

\address{$^{\rm a}$Department of Physics, University of Maryland,
               College Park, MD\ \ 20742, USA}
\address{$^{\rm b}$Department of Physics and
     Institute for Nuclear Theory,\\ University of Washington,
               Seattle, WA\ \ 98195, USA}
\address{$^{\rm c}$Department of Physics, Ohio State University,
               Columbus, OH\ \ 43210, USA}
\address{$^{\rm d}$TRIUMF, 4004 Wesbrook Mall,
                  Vancouver, BC\ \ V6T 2A3, CANADA}

\date{December 1994}
\maketitle

\begin{abstract}
Applications of QCD sum-rule methods to the physics of nuclei are reviewed,
with an emphasis on calculations of baryon self-energies in infinite
nuclear matter.
The sum-rule approach relates spectral properties of hadrons propagating
in the finite-density medium, such as optical potentials for quasinucleons,
to matrix elements of QCD composite operators (condensates).
The vacuum formalism for QCD sum rules is generalized to finite density,
and the strategy and implementation of the approach is discussed.
Predictions for baryon self-energies are compared to those suggested by
relativistic nuclear physics phenomenology.
Sum rules for vector mesons in dense nuclear matter are also considered.

\end{abstract}
\vspace*{1cm}
\hspace{1.8cm}{\bf To appear in Progress in Particle and Nuclear Physics Vol.
35}

\noindent {\bf Contents}

\begin{itemize}
\item[\bf I] {\bf Introduction}
\begin{itemize}
\setlength{\itemindent}{12pt}
\item[A]Quantum Chromodynamics and Nuclear Physics
\item[B]QCD Sum Rules
\item[C]Dirac Phenomenology
\item[D]The Chiral Condensate
\item[E]Other Hadrons in Medium
\end{itemize}
\item[\bf II] {\bf Review of Vacuum Formalism}
\begin{itemize}
\setlength{\itemindent}{12pt}
\item[A]Overview
\item[B]The Nature of QCD Sum Rules
\item[C]Basic Strategy
\item[D]Field Conventions
\item[E]Ingredients of a QCD Sum-Rule Calculation
\item[F]Harmonic Oscillator Analogy
\item[G]Sum-Rule Results in Vacuum
\end{itemize}
 \item[\bf III] {\bf Nucleons at Finite Density}
\begin{itemize}
\setlength{\itemindent}{12pt}
\item[A]Overview
\item[B]Relativistic Nuclear Phenomenology
\item[C]The Chiral Condensate in Nuclear Matter
\item[D]Formalism: Nucleons in Medium
\begin{itemize}
\setlength{\itemindent}{38pt}
\item[1]Finite-Density Correlator
\item[2]Dispersion relation at fixed three-momentum
\item[3]Phenomenological Ansatz
\item[4]OPE at Fixed Three-Momentum
\item[5]Calculating Wilson Coefficients
\end{itemize}
\item[E]Estimating QCD Condensates
\begin{itemize}
\setlength{\itemindent}{38pt}
\item[1]Vacuum Condensates
\item[2]In-Medium Condensates
\end{itemize}
\item[F]Results and Qualitative Conclusions
\begin{itemize}
\setlength{\itemindent}{38pt}
\item[1]Borel Sum Rules
\item[2]Simplified Sum Rules
\item[3]Detailed Sum-Rule Analysis
\item[4]Qualitative Conclusions
\end{itemize}
\item[G]Other Approaches
\end{itemize}
 \item[\bf IV] {\bf Hyperons in Nuclear Medium}
\begin{itemize}
\setlength{\itemindent}{12pt}
\item[A]Overview
\item[B]$\Lambda $ Hyperons
\item[C]$\Sigma $ Hyperons
\item[D]Summary
\end{itemize}
 \item[\bf V] {\bf Vector Mesons in Medium}
\begin{itemize}
\setlength{\itemindent}{12pt}
\item[A]Introduction
\item[B]Formulation of the Sum Rule
\end{itemize}
 \item[\bf VI] {\bf Discussion}
\begin{itemize}
\setlength{\itemindent}{12pt}
\item[A]Critical Assessment of the QCD Sum-Rule Approach
\item[B]Simple Spectral Ansatz
\item[C]Does Dirac Phenomenology Make Sense?
\end{itemize}
\item[\bf VII] {\bf Summary and Outlook}
\item[\null]{\bf References}
\end{itemize}
\newpage

\section{Introduction}
\label{secti}

\subsection{Quantum Chromodynamics and Nuclear Physics}
\label{secti_i}

There is little disagreement today that quantum chromodynamics (QCD)
is the correct theory underlying strong-interaction physics.
Thus, the physics of nuclei is, in essence, an exercise in applied QCD.
Indeed, from a
fundamental perspective, {\em the\/} central problem in theoretical nuclear
physics is to develop connections between observed nuclear phenomena and
the interactions and symmetries of the
underlying quark and gluon degrees of freedom.

On the other hand, knowledge of the fundamental underlying
theory has had very little impact, to date, on the study of low-
and medium-energy nuclear phenomena.
Two general difficulties hinder the application of quantum chromodynamics
to nuclear physics.
The first is that, in this energy regime, the strong interaction is, in
fact, strong; straightforward
perturbation theory in the effective QCD coupling constant fails.
Therefore we need an alternative expansion scheme or an approach to
directly approximate QCD nonperturbatively.

The second difficulty is the mismatch of energy scales between hadronic
and nuclear physics.
Characteristic QCD scales for light-quark hadrons,
which are the building blocks of nuclei, are
hundreds of MeV up to several GeV.
The dynamics of nuclei, in contrast, is a
delicate and subtle enterprise involving physics at much smaller scales
and featuring sensitive cancellations.
Typical nuclear observables are on the scale of a few MeV or
perhaps tens of MeV.
Consider, for example, the phenomenon of
nuclear matter saturation.
The binding energy per nucleon is
approximately 16 MeV \cite{PRE75}, which is less than 2\% of the
nucleon's mass.
An error of a few percent in the binding could easily lead to
errors at the level of nuclear physics scales, which would render any
calculation useless.
Thus, an accurate
description of nuclear matter {\it saturation\/} or comparable properties
directly from QCD must entail very
precise calculations.

How, then, might one proceed to relate QCD and nuclear phenomena?
An {\it indirect\/} approach is to focus on the implications and
constraints of QCD symmetries
by developing effective field theories (EFT's) for nuclei.
The prototype EFT for strong-interaction physics is chiral perturbation theory
(ChPT), which provides a systematic expansion in
energy for low-energy scattering processes.
The degrees of freedom are the
Goldstone bosons of spontaneously broken chiral symmetry
(pions, {\it etc.\/}) and, when appropriate, nucleons.
This approach builds in constraints due to chiral symmetry without any
additional constraints on the dynamics or {\it ad hoc\/} model assumptions.
Physics beyond chiral symmetry is incorporated through constants
in the low-energy Lagrangian,
which are usually determined from experiment, although in principle they
could be determined directly from QCD.
Because additional constants are needed at each stage in the energy
expansion,
ChPT is predictive only
at sufficiently low energies, where the number
of parameters introduced does not overwhelm the data to be described.
The prospects for extending ChPT in a useful way to calculations at
finite density are unclear at present, although some progress has been made
on the few-body problem \cite{WEI90}.

Alternatively,
we can pursue a more {\it direct\/} approach by focusing on the properties
of hadrons rather than nuclear matter saturation.
If we can successfully describe hadronic resonances in the vacuum using QCD,
we might be able to extend this description to hadronic properties
at finite density.
The most complete approach to hadronic physics at zero density
uses Monte Carlo simulations of QCD on
a discrete (Euclidean) space-time lattice.
Significant and steady progress has been made on such calculations
in recent years.
(For a review of the current state of the art, see Ref.~\cite{LAT94}.)

The prospects for extending these lattice calculations to finite density are
unclear; there are certainly formidable obstacles.
One major difficulty is that
the functional determinant that arises
with a nonzero chemical potential is
not positive definite, and hence standard Monte Carlo techniques
cannot be applied in a straightforward manner \cite{HAS83}.
In addition,
 realistic nuclear matter calculations may require significantly larger
lattices than are required for single hadrons; there
must be enough room for a sufficiently large number of nucleons.
Even with significant progress on these problems,
useful lattice calculations at finite
density will not be available in the near future.

To make progress, we turn instead to QCD sum rules.
The QCD sum-rule approach has proven to be a
useful way of extracting qualitative and quantitative
information about hadronic physics (masses and coupling constants)
from QCD inputs.
The method was introduced
by Shifman, Vainshtein, and Zakharov in the late
1970's \cite{SHI79} and applied to describe mesonic properties.
An important
extension was made by Ioffe, who showed how the technique could be used
to describe baryons \cite{IOF81} (see also \cite{CHU81}).
There are several detailed
reviews of the subject \cite{REI85,NAR89,SHI92,DOS94}.

In this review, we will discuss how QCD sum-rule methods
can be generalized to describe hadrons propagating in nuclear matter.
The pioneering
attempts to apply sum rules to finite-density systems were made by
Drukarev and Levin \cite{DRU90}.
Subsequently, a
variety
\cite{HAT91,COH91,ADA91,DRU91a,DRU91,FUR92,HAT92,JIN93,SCH93,JIN94a,HEN93,%
ASA93,JIN94,DRU94,JIN94c} of approaches were
developed, differing in focus and
technical detail.
Here we will emphasize the treatment of baryons
in infinite nuclear matter, and we will closely follow the philosophy and
technical approach of
Refs.~\cite{COH91,FUR92,JIN93,JIN94}
for nucleons and Refs.~\cite{JIN94a,JIN94c}
for hyperons.
The properties of vector mesons in the nuclear
medium will be discussed following the treatment in
Refs.~\cite{HAT92,ASA93}.
We will also comment on
alternative formulations of the problem.

\subsection{QCD Sum Rules}
\label{secti_ii}

QCD sum rules focus on momentum-space
correlation functions (also called correlators) of local composite
operators [for example, see Eq.~(\ref{nucleon_corr_def})].
Each composite operator
is constructed using quark and/or gluon fields
so as to carry the quantum numbers of the hadron we wish to
study; we will refer to such operators as interpolating fields.
Correlation functions of interpolating fields are also the basic
objects of lattice investigations (but in Euclidean coordinate space).
But while the lattice correlators are calculated from first principles
in quantum chromodynamics with a precision that can be systematically improved,
sum-rule calculations rely on some phenomenological input and
have limited (but generally understood) accuracy \cite{SHI92}.

The basic idea of QCD sum rules is to
match a QCD description of an
appropriate correlation function with a phenomenological one.
The underlying concept is ``duality,''
which establishes a correspondence
between a description in terms of physical (hadronic) degrees of
freedom and one based on the underlying quark and gluon degrees of freedom.
A generic QCD sum-rule calculation consists of three main parts:
1) an approximate
description of the correlation function in terms of QCD degrees of
freedom via an {\it operator product expansion\/} (OPE),
2) a description of the same correlator in terms of physical intermediate
states through a Lehmann representation ({\it i.e.\/}
dispersion relation) \cite{LEH54}
that incorporates a simple ansatz for the spectral density,
and
3) a procedure for matching these two descriptions and
extracting the
parameters of the spectral function that characterize the
hadronic state of interest.

The concept of an operator product expansion
was introduced by Wilson in the 1960's
\cite{WIL64}; the OPE is reviewed in Ref.~\cite{COL84}.
As applied in QCD sum rules,
the OPE expresses a correlator of interpolating fields
as a sum of $c$-number coefficients (called Wilson
coefficients) times expectation values of composite local
operators constructed from quark and gluon fields.
These expectation
values are referred to as {\it condensates\/}.

The essence of the OPE is the separation of short-distance and
long-distance physics, or, in momentum space,
large and small (spacelike) momenta.
In QCD this separation, to a large extent, corresponds to a
separation between perturbative physics (the coefficients) and
nonperturbative physics (the condensates).
This suggests that the coefficients can be calculated from QCD via
perturbation theory \cite{SHI79}, while the nonperturbative physics is
isolated in the condensates.
In principle, the condensates are
calculable directly from QCD (for example, using lattice simulations),
but in practice they are usually determined phenomenologically
from one set of sum rules and applied to others.
The sum rules are predictive because a relatively small number
of condensates dominate the descriptions of many observables \cite{SHI92}.

The possibility of matching the OPE-based description with the phenomenological
description is often viewed by nonpractitioners with skepticism;
it seems like magic.
The OPE is essentially a
short-distance expansion, and at any finite order it can only
describe the correlation function accurately at sufficiently large
spacelike momenta.
On the other hand, we wish to learn about the
low-lying excitations in the spectral function, which one expects to
dominate the correlation function only at nearby timelike momenta or
the low-momentum spacelike regime.
How, then, is there any hope of matching the two?
The secret is that the phenomenological
spectral density ({\it i.e.\/}, the sum over physical states) can be smeared
in energy to a significant degree while preserving basic information
on well-defined low-energy excitations (hadrons).
On the other hand,
the smearing corresponds to short times so that the dual
description of the spectral density
in terms of the quark and gluon OPE is adequate.
The conversion from a momentum-space
correlator to a smeared spectral function is
conventionally achieved by applying a Borel transform \cite{SHI79}
(see Sec.~\ref{sectii_iv}).

As will be discussed in the course of this review, there are definite
limitations of the sum-rule approach, and one must take care in applying
sum rules to new domains, as we do here.
We wish to warn the reader at the outset that it is rather easy
to get totally incorrect results when using (or,
rather, abusing) QCD sum rules.
The basic dangers are that
the
phenomenological and OPE descriptions at the level of approximation
used for each simply may not match well enough to give reliable
information or that some new feature of the physics is omitted.
If, however, one nevertheless proceeds to
extract phenomenological information, this information will likely be spurious.
The need for internal and external consistency checks cannot be overstressed.

Keeping this in mind,
the extension of the sum rules to finite density is, in principle,
straightforward: simply take expectation values in a finite-density
ground state rather than in the vacuum.
Consequently, we study the propagation of hadrons in infinite nuclear
matter.
In nature, of course, infinite nuclear
matter does not exist.
Why then do we study this somewhat unphysical
system rather than real finite nuclei?
In fact, infinite
nuclear matter is in many ways a more general and fundamental
problem.
After all, the nuclear interaction saturates, and the centers
of  medium and large nuclei all look roughly like nuclear matter.
By working with an infinite system, we isolate the physics
that is related to QCD dynamics and not to details of the many-body physics
of finite systems.
Furthermore, we can exploit the similarities to the
vacuum formulation (including, for example, translational invariance).
The vacuum sum rules can be recovered in the zero-density limit, and we
can take ratios to divide out certain systematic errors.

The price we pay is that
we do not make direct predictions of experimental observables.
To do so one needs to make model-dependent assumptions,
such as local-density approximations, to relate
calculations of nuclear matter properties to experimental observables of
finite nuclei.
Thus we are ultimately forced to compare with phenomenology rather than
with experimental data.

One has to confront, in addition, a basic limitation of the
sum-rule approach; namely,
the spectral information that can be extracted is necessarily approximate
and cannot be systematically improved.
For example, the extraction of
hadronic masses with a reliability of better than $\sim$100\,MeV is
typically not possible \cite{REI85,NAR89,SHI92}.
Thus the approach is not sufficiently precise to make detailed
calculations of the nuclear matter ground state at the scale of a few
MeV, the natural scale for much of low-energy nuclear physics.
However, another limitation of the approach, the fact that it is not a
self-consistent dynamical theory, turns out, in this context, to be a
virtue.
In the sum-rule approach, one does not dynamically solve for
the ground state of the theory.
Rather, one {\em characterizes\/} the
ground state: the QCD description of the ground state is characterized
by various expectation values of composite
operators.
Thus the approach need not be precise enough to calculate
the ground state reliably; as long as one has a reasonable way to
extract these condensates  phenomenologically, one can proceed to study
the qualitative features of excitations without actually solving the
ground-state problem.

Although one can avoid solving for the nuclear
matter ground state, the approximate nature of the sum-rule
approach means that one must use extreme care if
using the method to extract quantities that are small on the scale of hadronic
physics.
Thus, the use of the technique to describe fine details of
low-energy nuclear physics must depend on the construction of reliable
sum rules for nuclear properties that are much more accurate (in
absolute terms) than the sum rules have proven to be at the hadronic level.
Whether or not this will prove to be possible remains an open
question, but we view any results for quantities at this scale with some
skepticism.
Indeed, the question
may be posed as to whether there are {\it any\/}
properties in low- or intermediate-energy nuclear physics whose natural
scale is sufficiently large as to make the sum-rule approach
reasonable.

\subsection{Dirac Phenomenology}
\label{secti_iii}

If we go beyond direct experimental observables
and turn to nuclear phenomenology, we {\em can\/} find
quantities in low- and intermediate-energy nuclear physics
whose natural scale is several hundred MeV.
The quantities
we have in mind are the Lorentz scalar and vector parts of the optical
potential for a baryon propagating in nuclear matter.
In the Dirac phenomenology of nucleon-nucleus scattering
\cite{CLA83a,CLA83b,CLA83c,MCN83,SHE83,TJO87,WAL87,HAM90,SER86,SER92},
the Lorentz
scalar piece of the real part of the optical potential for the nucleon
is typically several hundred MeV attractive while the (time component
of the) Lorentz vector piece is typically several hundred MeV repulsive.
We propose that QCD sum rules are a
natural and reasonable way to study these quantities, and we focus
our review on this goal.

The Dirac description of nucleons propagating in nuclear
matter appears to be quite different from the conventional low-energy
nuclear physics approach, in which the (central) optical potential is
a few tens of MeV.
The latter nonrelativistic approach has been quite
successful in describing cross sections in proton-nucleus
scattering.
Dirac phenomenology describes cross sections equally
well; the reason is that only the sum of the scalar and vector potentials
contributes to the cross section, and, because of large cancellations
between the scalar and vector optical potentials, the sum is only a
few tens of MeV.
Furthermore, most of the
phenomenological energy dependence of the nonrelativistic
central potential at low energies arises simply from the
nonrelativistic reduction of the Dirac potentials, even when
the latter are energy independent \cite{SER86}.

The most compelling support for the Dirac approach
comes from studies of spin
observables such as the analyzing power and the spin rotation.
These
observables depend on the {\em difference\/} between the scalar and
vector potentials, and, given the opposite sign, this difference is many
hundreds of MeV.
The phenomenological evidence in favor of this description is based on
direct fits of optical potentials to describe simultaneously  the spin
observables and cross sections \cite{CLA83a,CLA83b,CLA83c,HAM90}.
Moreover,
a relativistic impulse approximation (RIA) gives optical potentials that
are qualitatively similar to the empirical
fits \cite{MCN83,SHE83,TJO87,WAL87}.
Further, in relativistic structure calculations, there is evidence
that keeping track separately of scalar and vector parts leads to
phenomenologically desirable density dependences
and nonlocalities \cite{SER92}.

While the Dirac approach has been quite successful
phenomenologically, it has also been quite
controversial \cite{BRO84,THE85,COO86,BLE87,ACH88,JAR91,COH91a,WAL94}.
We believe that much of this
controversy is misplaced, involving issues of how
Dirac phenomenology is to be interpreted rather than the
fundamental underlying issues of physics (see Sec.~\ref{sec:diracsense}).
However, in light of this
controversy, it is important to ask whether there is any evidence from
QCD in favor of the qualitative picture of Dirac phenomenology.
{}From a phenomenological perspective, the question is not whether the
nucleon should be thought of as a point Dirac particle; it is clear
that the composite structure of the nucleon plays an essential role in
the dynamics.
The key  question, however, is whether
the optical potential for the nucleon has large and canceling pieces
that transform as scalars and vectors under Lorentz transformations.

If QCD sum rules can be implemented so as
to separately describe the scalar and vector parts of the optical
potential, then we can test
whether the scalar and vector parts of the optical potential have the
sign and scale suggested by Dirac phenomenology.

\subsection{The Chiral Condensate}
\label{secti_iv}

If the large-scale changes at nuclear matter densities
implied by Dirac phenomenology
are to be reconciled with QCD through sum rules,
we must expect to find correspondingly large changes in the QCD description.
Is this plausible?
An important clue comes from the chiral
condensate, which
is the expectation value of the operator
$\overline{q}q$, where $q$ is an up or down quark field.
The chiral condensate is often taken as the order
parameter for spontaneous chiral symmetry breaking.
As shown by Ioffe \cite{IOF81}, this
condensate plays a central role in the QCD sum-rule determination of
the nucleon mass.
This role is consistent with chiral models of the
nucleon and is  plausible from rather
general considerations; `t~Hooft's \cite{HOO80}
anomaly matching conditions require  massless
nucleons (for massless quarks) if chiral symmetry is not broken.
Thus, it comes
as no surprise that the amount of chiral symmetry breaking
(which is measured by the chiral condensate) is correlated with the
nucleon's mass.

So our question becomes:
Is the chiral condensate
in nuclear matter very different from its free space value?
In fact, the chiral condensate in medium can be estimated in a reasonably
reliable way
(for an overview of the current state of affairs, see the
review by Birse \cite{BIR94}).
The magnitude of the chiral
condensate in the interior of nuclei is reduced
from its vacuum value by $\sim 30$--40\% \cite{COH92}
(see Sec.~\ref{sec:chiralcon}).
To the extent that the
chiral condensate controls the nucleon mass, one might naively expect
the mass of a nucleon in the nuclear medium to be reduced by
$\sim 30$--40\%.
This may seem absurd; after all, the
empirical binding energy per nucleon in nuclear matter is approximately
16 MeV \cite{PRE75}, while $\sim 30$--40\% of the nucleon mass is several
hundred MeV.

In fact, however, in addition to a
shift in the nucleon mass (which is a Lorentz scalar), the in-medium
nucleon energy can depend on a self-energy that transforms as
a Lorentz four-vector.
For qualitative consistency with experiment and Dirac phenomenology,
one needs to find that the vector contribution is repulsive in nuclear
matter and largely
cancels the scalar attraction.
As we will show, under certain assumptions about
condensates, QCD sum rules predict this qualitative
behavior for the optical potential \cite{FUR92,JIN93,JIN94}.

\subsection{Other Hadrons in Medium}
\label{secti_v}

Of course, nucleons are not the only hadrons.
The same methods that
are used to describe nucleon self-energies in medium
can be directly adapted to describe optical potentials (self-energies)
of other baryons in nuclear matter.
Again, we
might expect that large changes in the chiral condensate
 (along with changes in other condensates) will imply large
effects.
This raises issues
similar to the ones in nucleon propagation:
Are large and
canceling Lorentz scalar and vector optical potentials predicted for other
baryons?
Are the sum-rule predictions consistent with experiment and phenomenology?
These calculations provide important consistency checks
on the sum-rule predictions for nucleons.
There are ongoing studies of various baryons;
in this review, we will consider
hyperon propagation through nuclear matter \cite{JIN94a,JIN94c},
for which there is  phenomenology to confront.

In addition to baryons,
one can consider meson propagation at finite density.
The case of vector mesons, is particularly interesting and important,
in part, because changes in vector-meson masses have
been proposed as an explanation for anomalies in electron
scattering \cite{BRO89,SOY93} and $K^+$ scattering data \cite{BRO88}, and,
in part, because measurements of dilepton pairs emerging
from a heavy-ion collision may provide a more-or-less
direct experimental probe \cite{DIL,MOS94}.
If vector mesons are created in the medium, they have a nonzero amplitude
for decay into two leptons.
Since the electromagnetic interaction is
weak, the dilepton pair has a good probability of leaving the nucleus
unscathed, providing essentially unfiltered information about the vector
meson in the nucleus.
A number of heavy-ion experiments have been
proposed to look for shifts in vector-meson masses via study of
dilepton pairs \cite{DIL}.
In addition, the
use of the virtual Compton process ($\gamma$,dilepton)
might provide an even cleaner probe of the vector meson in medium \cite{MOS94}.

\section{Review of Vacuum Formalism}
\label{sectii}

\subsection{Overview}
\label{sectii_i}

In this section, we review the QCD sum-rule approach as applied to the
calculation of hadronic masses in the vacuum, {\it i.e.\/}, at zero
density.
An overview of QCD sum-rule methods and results is provided by the
recent book edited by Shifman \cite{SHI92}, which includes a collection
of reprints along with an up-to-date commentary by the editor and
others.
Since thorough technical reviews exist, we will focus here on features
and analogies that make the generalization to the finite-density
problem more straightforward and plausible.

\subsection{The Nature of QCD Sum Rules}
\label{sectii_ii}

The designation ``QCD sum rules'' is often confusing to the nonexpert.
In quantum chromodynamics there are various relations referred to as ``sum
rules,'' which have nothing to do with the approach we are concerned
with here.
For example, there are many sum rules that originated with the parton model:
Bjorken sum rule, Ellis-Jaffe sum rule, Gottfried sum rule, {\it etc\/}.
These typically are relations that manifest a constraint
on properties of constituents, such as a conservation law.
So, for example, all of the momenta (or strangeness)
of the constituents must add up to
the total momentum (or strangeness).

In contrast,
the type of sum rule we are concerned with here is typified by the dipole
sum rules of atomic and nuclear physics.
Consider photo-absorption by nuclei.
The cross section for the excitation of a final state $|\nu\rangle$
by a photon of energy $E$ is (see, for example, Ref.~\cite{RIN80})
\begin{equation}
  \sigma_\nu(E) = {4\pi^2\e^2\over \hbar c}
        (E_\nu - E_0) |\langle\nu| D | 0 \rangle|^2
          \delta(E-E_\nu+E_0)  \ .  \label{eq:sigmanu}
\end{equation}
Here $D$ is the dipole operator for E1 radiation in the $z$ direction,
which can excite states that have
isovector, $J^\pi=1^-$ quantum numbers.
The total cross section for dipole absorption is found by summing
over all possible final states and integrating over the energy:
\begin{equation}
   \sigma_{\rm tot} =
     \int_0^\infty \! dE\, \sum_\nu \sigma_\nu(E)
       = {4\pi^2\e^2\over \hbar c} S_1(D)  \ ,
\end{equation}
where we have defined the {\it energy-weighted dipole sum rule\/}
\begin{equation}
  S_1(D) \equiv \sum_\nu (E_\nu-E_0) |\langle\nu | D | 0\rangle|^2 \ .
\end{equation}

On the other hand, one can use completeness
to rewrite $S_1(D)$ as a ground-state matrix element
involving $D$ and the Hamiltonian $H$ \cite{RIN80}:
\begin{equation}
   S_1(D) = \mbox{$1\over 2$}\langle 0 | [D, [H,D]] | 0 \rangle \ .
\end{equation}
The double commutator can be directly evaluated in a theoretical model.
For example, if we model the nucleon-nucleon interaction to have no exchange
mixtures and to be
velocity independent, we can evaluate the commutator immediately
(and obtain the TRK result) \cite{RIN80}.
More generally, we will have to do an approximate model calculation.

The sum rule consists of the equality between the ground-state
matrix element, which can be approximated in a theoretical model,
and the sum over excited states, which is a weighted average
of experimental observables.
The operator $D$ is the analog of the interpolating field in QCD sum rules,
while $\sum_\nu \sigma_\nu(E)$ is the analog of the spectral density.
Various general
observations can be made about these sum rules,
which will carry over to analogous QCD sum rules:
\begin{enumerate}
  \item  If one uses  different multipole operators, different
  intermediate states (corresponding to different quantum numbers)
  are excited.  Thus one
  can investigate iso\-scalar or iso\-vector states, and monopole, dipole,
  quadrupole excitations, and so on.
  For QCD sum rules, we choose an interpolating field that excites a
  particular channel of intermediate states which includes the
  hadron of interest ({\it e.g.\/}, nucleon or $\rho$ meson).
  \item  One can choose an operator that doesn't actually
  correspond to a physically realizable experiment.  The matrix
  elements in the sum over states may not be measured in a real
  experiment, but a sum rule can relate their values to a theoretical
  calculation.
  Furthermore, in QCD there is the additional possibility of a {\it
numerical\/}
  experiment to establish this sum, through lattice calculations.
  \item  One can choose different weighting functions for the sum rule
  and thereby
  change the weight in the sum over states ({\it cf}.\  dispersion relation).
  For example, given a single-particle operator $F$, we can derive
  the sum rule \cite{RIN80}
\begin{equation}
   S_k(D) \equiv \sum_\nu(E_\nu-E_0)^k |\langle\nu|F|0\rangle |^2
         =  \langle 0 | F(H-E_0)^k F | 0 \rangle \ .
\end{equation}
  Thus one has some freedom to adjust how important particular states are
  in the sum.
  In QCD sum rules, one finds that the weighting resulting from a {\em Borel
  transform\/} of the correlation function is particularly effective.
  \item  Certain states may dominate the sum.  The collective excitations
  known as giant resonances are a case in point.  If a giant resonance
  largely dominates (``saturates'') the sum over states of the strength,
  then one obtains from the sum rule a
  direct connection between some experimental properties of this resonance
  and a theoretical calculation.
  Furthermore, we can use the freedom of changing the
  weighting function to improve the dominance of the state
  of interest.
   Note that one does not learn everything about
  the states, just certain matrix elements.
\end{enumerate}

In  QCD
sum rules, we include an external parameter (the Borel mass) that adjusts the
weighting function.
We seek a range of parameter values for which we
find dominance by
the hadronic state of interest as well as a reasonable
theoretical approximation to
the ground-state expectation value.
That the latter goal is attainable in QCD is not obvious.
In the next section, we outline how it could be possible.

\subsection{Basic Strategy}

Shifman has compared the strategy of the
sum-rule approach to the quantum-mechanical many-body problem of some
external objects injected into a complicated medium \cite{SHI92b}.
(This is what a Green's function does, for example.)
The picture is that, over time, the external objects will interact with
themselves {\it and\/} with the medium and will eventually develop into a
stationary state or, more generally, an approximate stationary state with
a finite lifetime.
In many-body systems, the latter is known as a quasiparticle, while the
analog in particle physics is a hadronic resonance (such as the $\rho$
meson).
The objects can also develop into many other complicated states with the same
quantum numbers, but our goal is to isolate some properties of the
quasiparticle, such as its energy or amplitude for creation
({\it i.e.\/}, the spectral weight).%
\footnote{Note that the latter by itself is not generally an observable.}

The medium is too complicated to solve
microscopically.
At best we might know some coarse, averaged thermodynamic properties.
Ideally we would simply evolve the system for large times (but
still less than the quasiparticle lifetime),
so that the quasiparticle develops and
becomes dominant.
To do this, however,
we would need to calculate the interactions of the objects with the
medium and the reaction of the medium back on the objects.
By assumption we cannot calculate these details.

Suppose, however,
that the characteristic time for the medium to react is long
compared to the characteristic time for the external objects.
Then there is a window in time corresponding to some cycles of the
developing bound
state during which the medium is essentially frozen, and the objects see
only coarse properties.
Yet it is enough time that some basic features of the bound state are
established.
Then we have a chance to extrapolate from relatively short times!

The key feature is the distinction in time scales.
If certain aspects of the physics are established at short times, they
can persist to the measurement at long times, even though
the system appears very different.
This is reminiscent of the parton model, as applied to deep-inelastic
scattering.
In the infinite-momentum frame, one sees a clear separation of time scales:
the interaction of the photon with the constituent partons and the
interaction of the constituents among themselves.
One can calculate the cross section for inclusive scattering
using partons and get the right answer,
even though there are strong (confining) final-state interactions
and the partons end up in detectors as hadrons.

The analogy to QCD sum rules
is that the medium is the QCD vacuum and the external
objects are valence quarks (or gluons) with definite quantum numbers
created by external currents (which we specify).
We are able to calculate the behavior only at relatively short
times%
\footnote{Actually, we will look effectively at short {\em imaginary\/} times.}
using the operator product expansion.
The coarse properties of the medium seen by the valence quarks
are characterized by vacuum matrix elements of quark and gluon fields,
called condensates.
The question of whether there really
is a favorable relation of the time scale
characteristic of the valence quarks and that of the vacuum fluctuations
is not clearcut in QCD,
and seems to depend strongly on the channel \cite{NOV81}.

However,
for light-quark hadrons, the existence of two mass (or time) scales has
been postulated in various contexts (for example, Georgi and Manohar's
chiral quark model \cite{MAN84}).
The vacuum response is characterized by times of order
$1/\Lambda_{\rm QCD}$.
The shorter time scale for the light-quark hadrons might be set by the
chiral symmetry breaking scale.
So one is comparing roughly $\Lambda_{\rm QCD}$ to the mass of the
$\rho$ meson or the nucleon.
While this is not a huge difference,
it may be sufficient to establish some dominant features of the
ground state from the physics of chiral symmetry breaking
before confining interactions play a direct role in the bound state.
(Compare to a quantum-mechanical
system in a large box.
The box serves to discretize the spectrum, but does not
change the total cross section if we average over nearby levels \cite{SHI92}.)
This scenario
is consistent with a picture of constituent quarks forming first,
then becoming weakly bound at later times.
Indeed,
one of the basic physical predictions of QCD sum rules is
that chiral symmetry breaking, as manifested by the chiral condensate,
predominantly sets the scale for the masses of light-quark hadrons.

The generalization of our analogy
to finite density simply means that we inject valence
quarks into a modified medium, the nuclear matter ground state.
The background fields (condensates) seen by the injected quarks will
be somewhat different, which would then be reflected in the predictions for
self-energies at finite density.
If we focus on quantities that are not strongly dependent on physics
at large time scales, the successes of the vacuum sum rules should
carry over to finite density.

\subsection{Field Conventions}

In this section we review the basic QCD field operators
to establish our notation and conventions.
We start with the fields for the up, down, and strange quarks,
$u_{a\alpha}$, $d_{a\alpha}$, and $s_{a\alpha}$, where we use
$a,b,c,\ldots=\mbox{1--3}$ for quark color indices and
$\alpha,\beta,\gamma,\ldots=\mbox{1--4}$ for Dirac indices.
Due to isospin symmetry, we often do not distinguish between up and down
quark flavors; we use $q_{a\alpha}$ to refer to either an up or down
quark field.

The gluon field is denoted $A^A_\lambda$, where we use
$A,B,C,\ldots=\mbox{1--8}$ for gluon color indices and
$\lambda,\mu,\nu,\ldots=\mbox{0--3}$ for Lorentz indices.
A matrix form of the gluon field is obtained by multiplying the
color components of the gluon field by the generators of the SU(3)
Lie algebra:
\begin{equation}
{\cal A}^\mu_{ab}\equiv A^{A\mu}t^A_{ab}\ ,
\end{equation}
where $t^A\equiv\lambda^A/2$ are the SU(3) generators
in the fundamental representation
($\lambda^A$ are the Gell-Mann matrices \cite{ITZ80}).
They satisfy the following relations:
\begin{equation}
[t^A,t^B]=if^{ABC}t^C\ ,
\hspace{0.5in}
{\rm tr}(t^A)=0\ ,
\hspace{0.5in}
{\rm tr}(t^A t^B)=\mbox{$1\over 2$}\delta^{AB}\ ,
\label{Lie_algebra}
\end{equation}
where $f^{ABC}$ are totally antisymmetric structure
constants \cite{ITZ80} and tr denotes a trace over quark color indices.

It is useful to introduce the
gluon field tensor, which can be defined as
\begin{equation}
{\cal G}_{\mu\nu}\equiv G^A_{\mu\nu}t^A
\equiv D_\mu{\cal A}_\nu-D_\nu{\cal A}_\mu\ ,
\label{glue_tensor_def}
\end{equation}
where $D_\mu\equiv\partial_\mu-ig_s{\cal A}_\mu$
is the covariant derivative ($g_s$ is the quark-gluon coupling constant).
Useful identities that follow from
Eqs.~(\ref{Lie_algebra}) and (\ref{glue_tensor_def}) are
\begin{eqnarray}
{\cal G}_{\mu\nu}&=&{i\over g_s}[D_\mu,D_\nu]\ ,
\label{G_commutator}
\\*
G^A_{\mu\nu}&=&\partial_\mu A^A_\nu-\partial_\nu A^A_\mu
+g_s f^{ABC}A^B_\mu A^C_\nu\ .
\end{eqnarray}
We also define the dual of the gluon field tensor,
\begin{equation}
\widetilde{G}^{A\kappa\lambda}
\equiv\mbox{$1\over 2$}\epsilon^{\kappa\lambda\mu\nu}G^A_{\mu\nu}\ .
\label{dual_def}
\end{equation}

The quark fields are coupled to the gluon field by replacing ordinary
derivatives with covariant derivatives in the free quark Lagrangians.
Thus the quark fields $q$ and $\overline{q}$ (with current quark mass $m_q$)
satisfy the following equations of motion:
\begin{equation}
(i\rlap{\,/}{D}-m_q)q=0\ ,
\hspace{1in}
\overline{q}(i\DslashL+m_q)=0\ ,
\label{dirac}
\end{equation}
where we define
$\DL_\mu\equiv\partialL_\mu+ig_s{\cal A}_\mu$.
These equations of motion will be used in the subsequent discussion
to simplify
operator matrix elements.
In nucleon sum rules, the contributions from finite up and down current
quark
masses are numerically small; therefore, we neglect these contributions.

\subsection{Ingredients of a QCD Sum-Rule Calculation}
\label{sectii_iv}

In this section we review the ingredients of a QCD sum-rule calculation.
Since our primary focus in this review is on the nucleon sum rule
in medium,
we will use the nucleon sum rule in vacuum as our example here.
Particular details of other sum rules can be found in
Refs.~\cite{REI85,NAR89,SHI92}.

Conventional treatments of QCD sum rules start with a momentum-space
correlation function
(also called a correlator) of color-singlet currents.
For example, the analysis of the nucleon mass \cite{IOF81} starts with
the time-ordered correlation function $\Pi(q)$ defined by
\begin{equation}
\Pi_{\alpha\beta}(q)\equiv i\int d^4x\, \e^{iq\cdot x}
\langle{0}|T\eta_\alpha(x)\overline{\eta}_\beta(0)|{0}\rangle\ ,
\label{nucleon_corr_def}
\end{equation}
where $|{0}\rangle$ is the physical nonperturbative vacuum state,
and $\eta_\alpha$ is an interpolating field with the spin and isospin of a
nucleon, but with indefinite parity.
We have exhibited the Dirac indices $\alpha$ and $\beta$.
Here we choose to use the
interpolating field for the proton proposed by Ioffe \cite{IOF81,IOF83}:
\begin{equation}
\eta_p=\epsilon_{abc}(u_a^T C\gamma_\mu u_b)\gamma_5\gamma^\mu d_c\ ,
\label{int_field}
\end{equation}
where $u_a$ and $d_a$ are up and down quark fields ($a$ is a color index),
$T$ denotes a transpose in Dirac space,
and $C$ is the charge-conjugation matrix.

The choice of interpolating field for the nucleon is motivated by the
goals of maximizing the coupling to the nucleon intermediate state
relative to other (continuum) states while minimizing the contributions
of higher-order corrections from the operator product expansion.
Furthermore, we want the two invariant functions
[see Eq.~(\ref{nucleon_corr})] to
be more-or-less equally dominated by the nucleon contribution.
These goals plus the simplicity of an interpolating field with no
derivatives and the minimum number of quark fields motivate
$\eta_p$ \cite{IOF81,IOF83}.
%
%
In Sec.~\ref{result:nuc}
the sensitivity to this choice for finite-density sum rules is tested.

The fundamental physical object in the sum-rule analysis is the
{\it spectral density\/}
of the interpolating fields.
For our example here, we can identify spectral densities corresponding
to the two time orderings:
\begin{eqnarray}
  \rho_{\alpha\beta}(q)  &=& {1\over 2\pi}\int\! d^4x\, \e^{iq\cdot x}
\langle{0}|\eta_\alpha(x)\overline{\eta}_\beta(0)|{0}\rangle\  \ ,
\label{eq:rhoa}\\
  \widetilde\rho_{\alpha\beta}(q)  &=& {1\over 2\pi}
        \int\! d^4x\, \e^{iq\cdot x}
\langle{0}|\overline{\eta}_\beta(0)\eta_\alpha(x)|{0}\rangle\   \ .
\label{eq:rhob}
\end{eqnarray}
The spectral densities of the
correlator follow from the assumed dispersion relation
(ignoring, for simplicity, any subtractions, which
are not relevant here):
\begin{equation}
\Pi_{\alpha\beta}(q) = - \int_{-\infty}^{+\infty} dq_0'\,
   \biggl[ {\rho_{\alpha\beta}(q')\over q_0 - q_0' + i\epsilon} +
     {\widetilde\rho_{\alpha\beta}(q')\over q_0 - q_0' - i\epsilon} \biggr] \ ,
     \label{eq:disp}
\end{equation}
where $q'_\mu = (q_0',\vec{q})$.
{}From this expression, one sees that the spectral densities are equal,
up to overall constants,
to the discontinuity of the correlator across the
real axis.

After inserting intermediate states between the
currents in Eqs.~(\ref{eq:rhoa}) and (\ref{eq:rhob}),
we can extract the $x$ dependence of the $\eta$ field and
do the $x$ integral, yielding
\begin{equation}
  \rho_{\alpha\beta}(q) = (2\pi)^3 \sum_n \delta^{(4)}(q - P_n)
     \langle{0}|\eta_\alpha(0)|n \rangle \langle n| \overline\eta_\beta(0) | 0
     \rangle  \ ,
\end{equation}
and a similar expression for $\widetilde\rho_{\alpha\beta}(q)$
[but with $\delta^{(4)}(q+P_n)$].
Here $P_n^\mu$ is the four-momentum of the state $|n\rangle$.
For fixed three-momentum $\vec q$, the spectral function
measures the intensity at which
energy is absorbed from the current at different frequencies.
Thus it is analogous to the absorption cross section
$\sum_\nu \sigma_\nu(E)$ in Eq.~(\ref{eq:sigmanu}).
(In some finite-density sum rules,
we can also have scattering from the medium.)

Lorentz covariance and parity invariance imply that the Dirac structure
of $\rho_{\alpha\beta}(q)$ is of the form \cite{BJO65,ITZ80}
\begin{equation}
  \rho_{\alpha\beta}(q) = \rho_s(q^2)\delta_{\alpha\beta}
     + \rho_q(q^2)\rlap{/}{q}_{\alpha\beta} \ ,
\end{equation}
and similarly for $\widetilde\rho$, so that
$\Pi(q)$ has an analogous form:
\begin{equation}
\Pi_{\alpha\beta}(q) = \Pi_s(q^2)\delta_{\alpha\beta}
        +\Pi_q(q^2)\rlap{/}{q}_{\alpha\beta}\ ,
\label{nucleon_corr}
\end{equation}
where $\Pi_s$ and $\Pi_q$ are Lorentz scalar functions of $q^2$ only.
An independent sum rule for each scalar function, $\Pi_s(q^2)$ and
$\Pi_q(q^2)$, will be constructed.

Charge-conjugation invariance at zero density implies that
the invariant spectral functions $\rho_i(q^2)$ and $\widetilde\rho_i(q^2)$
(where $i = \{s,q\}$)
are, in fact, identical, up to an overall sign.
This will {\it not\/} be true at finite density.
Thus, for the vacuum case, one need integrate only over positive energy.
This means that the integral over energy in Eq.~(\ref{eq:disp})
can be transformed to one over $q'{}^2 \equiv s$.
The end result is the familiar dispersion relation for each invariant
function:
\begin{equation}
  \Pi_i(q^2) =  \int_0^\infty
        ds\, {\rho_i(s) \over s-q^2}  + \mbox{polynomial}  \ .
        \label{eq:dispp}
\end{equation}
At finite density we will keep the dispersion relations as integrals
over energy.

The dispersion relation in Eqs.~(\ref{eq:disp}) or (\ref{eq:dispp})
embodies a particular weighting of the
spectral density, integrated over all energies.
When the correlator is evaluated for spacelike external momenta ($q^2<0$),
the weighting function
is analytic and serves to smear smoothly the spectral density in energy.
It is also clear that one could devise a different
weighting function that goes to
zero sufficiently fast at large energy that the weighted integral is
assured to be finite (and no subtraction is needed).
This weighted sum over states is one half of our sum rule.

The other half of the sum rule is a direct calculation of the correlator
using
the operator product expansion (OPE), which provides a QCD approximation to
the correlator that is applicable at large spacelike $q^2$.
The OPE is a useful tool for extracting
phenomenological information from renormalizable quantum field theories.
The central idea is that the time-ordered product of two local (elementary or
composite) operators at short distances can be expanded in terms of
a complete set of regular local operators $\widehat O_n(0)$
\cite{WIL64,COL84}:
\begin{equation}
TA(x)B(0)\stackrel{x\rightarrow 0}{=}
\sum_n C^{AB}_n(x) \widehat O_n(0)\ .
\end{equation}
The $c$-number coefficients $C^{AB}_n$ of the expansion are called
Wilson coefficients.
In this expansion, the singularities at short distances are factored
out from the regular operators, and the terms in the expansion are
organized in decreasing order of singularity.
Formally, the OPE has only been proven in perturbation theory;
the validity of the OPE in the presence of
nonperturbative effects is a complex issue.

There has been a series of papers discussing the nature of the OPE beyond
perturbation theory \cite{VAI85,NOV85,DAV86}.
It was shown that, to apply unambiguously the OPE, one must define
the coefficient functions and composite operators
by introducing an auxiliary parameter, the normalization point $\mu$:
\begin{equation}
TA(x)B(0)\stackrel{x\rightarrow 0}{=}
\sum_n C^{AB}_n(x,\mu) \widehat O_n(0,\mu)\ .
\label{blib1}
\end{equation}
Physics from
momenta above $\mu$ are put in the coefficients $C^{AB}_n$, while
physics from momenta below $\mu$ are put into the operators $\widehat O_n$.
Both the coefficient functions and the operators contain, in general,
nonperturbative as well as perturbative contributions.
However, in practical applications of the OPE
(in particular, in the QCD sum-rule formalism),
one usually applies a simplified version.
Namely, the Wilson coefficients are evaluated in perturbation theory,
while the nonperturbative effects are contained entirely in the
vacuum expectation values of composite operators, which are assumed
to contain no perturbative contributions.

This simplified version of the OPE is justified in part by the phenomenological
success of QCD sum rules.
In Ref.~\cite{NOV85},
the reason behind this success is attributed to the following:
There seems to be a range in $\mu$ in which $\mu$ is large enough with
respect to $\Lambda_{\rm QCD}$ (the QCD scale parameter) so that
nonperturbative corrections to the coefficients are small and can be
neglected, but small enough so that the values of the
condensates are quite insensitive to variations in $\mu$.
In other words, there seems to be a ``window'' in QCD where the simplified
version of the OPE applies.
Perturbative corrections to the
operators can be taken into account in the leading-logarithmic
approximation through anomalous-dimension factors \cite{SHI79}
(these will be implicit in the rest of our discussion).

In momentum space, the OPE correlator takes the general form
\begin{equation}
 \Pi(Q^2) = \sum_n C^i_n(Q^2) \langle \widehat O_n \rangle_{\rm vac} \ ,
\end{equation}
where the $C^i_n(Q^2)$ are $c$-number functions, calculated in QCD
perturbation theory, and the $\langle\widehat O_n\rangle_{\rm vac}
\equiv \langle 0 | \widehat O_n | 0 \rangle $ are vacuum expectation
values of QCD operators---the condensates.
The expansion is in inverse powers of $Q^2$ and thus is useful in
the deep spacelike region,
for which $Q^2 \equiv -q^2$ is large and positive.

Equating the OPE correlator and the spectral expansion, each
evaluated at spacelike $q^2$, gives us a sum rule.
The equivalence exploits the
underlying principle of {\it duality\/}.
This term is used in various contexts;
for our purposes, it is most clearly expressed in terms of smeared
spectral densities.
On the one hand, the spectral density is a sum of matrix elements with
physical intermediate states, including the hadron of particular interest.
On the other hand, one could calculate the spectral density in perturbation
theory with a basis of quarks and gluons.%
\footnote{The OPE effectively
gives corrections to the perturbative spectral density
due to nonperturbative effects.}
Superficially these have little in common;
nevertheless, if these spectral densities are sufficiently smeared, one
should obtain the same result with either representation.
(See the harmonic oscillator example in Sec.~\ref{sec:sho}.)

Smearing broadly with an analytic function can be thought of as focusing
on the short Euclidean time parts of the spectral density,
which makes it amenable to approximation by an OPE.
The Euclidean correlator on the lattice in imaginary time is
a clear example [see Eq.~(\ref{eq:latt})].
We note that the choice of smearing function is rather arbitrary and,
furthermore, that some choices are more useful than others.
The experience with QCD sum rules is that a near optimal choice
corresponds to a Gaussian weighting function (in energy).
We can achieve this weighting starting from the correlator by applying
a Borel transform.

Given a function $f(Q^2)$, the Borel transform
${\cal B}[f(Q^2)] \equiv \widehat{f}(M^2)$ can be
defined in practice by \cite{SHI79}
\begin{equation}
\widehat{f}(M^2)\equiv
\lim_{\stackrel{\scriptstyle Q^2,n\rightarrow\infty}
{\scriptstyle Q^2/n=M^2}}
{(Q^2)^{n+1}\over n!}\left(-{d\over dQ^2}\right)^n f(Q^2)\ ,
\label{borel_def_appendix}
\end{equation}
which depends on the ``Borel mass'' $M$.
Table~\ref{borel_table} lists the Borel transforms of
the most commonly encountered functions in sum-rule applications.%
\footnote{Some additional useful results can be derived
by using integral representations of functions that isolate
the $Q^2$ dependence in the integrand in the form
$f(Q^2)=\e^{-zQ^2}$ with $z>0$.
The Borel transform is found using
$\widehat{f}(M^2)=M^2\delta(zM^2-1)$.}
Expressions involving the
running coupling
$\alpha_s$ are derived using the one-loop expression:
\begin{equation}
\alpha_s(Q^2)\simeq{4\pi\over b\ln(Q^2/\Lambda_{\rm QCD}^2)}\ ,
\label{alphas}
\end{equation}
and are valid to lowest order in $\alpha_s$.

One notes from Table~\ref{borel_table} that any simple polynomial in
$Q^2$ is eliminated by the Borel transform.
This has two useful (and related) consequences:
The subtraction terms accompanying the dispersion relation and any
divergent (or renormalized) polynomials from the OPE are simultaneously
eliminated.
It is also evident that the higher-order terms in the OPE, which contain
inverse powers of $Q^2$, are factorially suppressed by the Borel transform.

After applying the Borel transform to the dispersion
relation of each invariant function, we obtain the desired weighting:
\begin{equation}
  \widetilde\Pi_i(M^2) = \int_0^\infty
       ds\, \e^{-s/M^2} \rho_i(s)  \ .  \label{eq:borelint}
\end{equation}
For $M$ near the mass of the nucleon, higher-mass contributions to
the integral are exponentially suppressed.

\setlength{\tabcolsep}{20pt}
\begin{table}[tb]
\centering
\caption{Borel transforms of common functions, where $k=1,2,3,\ldots$,
$m=0,1,2,\ldots$, and $\epsilon$ is not necessarily integral.
Transforms involving  $\alpha_s$ are given to
leading order.}
\vspace{12pt}
\begin{tabular}{cc}
\hline\hline
  $f(Q^2)$  &  $\widehat{f}(M^2)$ \\
\hline
  &  \\[-8pt]
  ${\displaystyle{1\over(Q^2)^k}}$  &
    ${\displaystyle{1\over(k-1)!(M^2)^{k-1}}}$\\[12pt]
  ${\displaystyle{\alpha_s(Q^2)\over(Q^2)^k}}$  &
    ${\displaystyle{\alpha_s(M^2)\over(k-1)!(M^2)^{k-1}}}+\cdots$\\[12pt]
   $(Q^2)^m$  &  0  \\[12pt]
  $(Q^2)^m\ln Q^2$  &
    $(-1)^{m+1}m!(M^2)^{m+1}$\\[12pt]
  $\alpha_s(Q^2)(Q^2)^m\ln Q^2$  &
    $\alpha_s(M^2)(-1)^{m+1}m!(M^2)^{m+1}+\cdots$\\[12pt]
 ${\displaystyle 1\over \displaystyle (Q^2+\mu^2)^\epsilon}$ &
   $   {\displaystyle 1\over
   \displaystyle \Gamma(\epsilon)(M^2)^{\epsilon-1}}
   \e^{-\mu^2/M^2} $
    \\[10pt]
\hline\hline
\end{tabular}
\label{borel_table}
\end{table}

The correlator in spectral form can be evaluated by introducing a
phenomenological model for the spectral density.
The lowest-energy contribution to the spectral function is from the
nucleon pole.
Its contribution can be constructed from the matrix element
\begin{equation}
  \langle 0 | \eta(0) | q \rangle  = \lambda_N u(q)  \ , \label{eq:mel}
\end{equation}
where $| q \rangle$ is a one-nucleon state with four-momentum $q^\mu$
(with $q^2 = M_N^2$) and $u(q)$ is a Dirac spinor for the nucleon.
Contributions to the spectral functions from higher-mass states are
roughly approximated using the leading terms in the OPE, starting at a
threshold $s_0$ [see Eqs.~(\ref{eq:phens}) and (\ref{eq:phenq})].

Equating the Borel transforms of the OPE and phenomenological
descriptions yields relations---sum rules---for each invariant function.
To summarize, the general recipe for a QCD sum-rule calculation is:
\begin{enumerate}
	\item  identify an appropriate interpolating field for the hadron of
	interest and construct a correlator;

	\item  identify the tensor structure and invariant functions
	of the correlator;

	\item  write dispersion relations for each invariant function
	         with a spectral ansatz;

	\item  construct the OPE for each invariant function;

	\item  convert to Borel weighting;

	\item  match and extract parameters of the ansatz.
\end{enumerate}

Examples of this procedure are given in the next two sections.

\subsection{Harmonic Oscillator Analogy}
\label{sec:sho}

To illustrate the QCD sum-rule approach for nonexperts, it is  useful to
look at a simple, solvable problem from ordinary quantum mechanics.
Indeed, it is difficult to believe that the approach is feasible without
some direct evidence.
Illustrative calculations for various three-dimensional quantum mechanical
potentials are given in the literature (see, for example,
Refs.~\cite{NOV81} and \cite{PAS84}).
There are also examples in model field theories, which provide
additional verifications of
the methods.

Here we will go through the three-dimensional harmonic oscillator.
The harmonic oscillator is a particularly nice example, because, besides
the advantage of being exactly solvable, it has
features analogous to QCD:  confinement and asymptotic freedom.
The confinement analogy is obvious: the potential goes to infinity so
that there are only bound states, no continuum states.
(One should imagine the one-body harmonic oscillator problem as arising
from a two-particle problem, with a harmonic oscillator potential between them.
This, of course, can be reduced to a one-body problem in an
external potential.)
The analog of asymptotic freedom is less obvious, but
follows because the potential is
nonsingular at the origin, which means that the kinetic
energy will dominate the potential energy at short times.

The Hamiltonian for the harmonic oscillator is
\begin{equation}
  H =  {1 \over 2m }{\vec p}^2 + {m \omega^2 \over 2} {\vec x}^2 \ ,
\end{equation}
where $m$ is the mass and $\omega$ is the oscillator parameter.
Complete information about the system is given by the energies $E_n$
and wave
functions $\psi_n({\vec x})$ of the bound states, but this is far more
than we can calculate with sum-rule methods.
Instead we focus on the ground-state energy and
wave function at the origin.
The latter is the analog to something calculated in QCD sum rules:  an
amplitude for finding quarks in the bound state on top of each other.
Thus we will deal with $s$ states only in the sequel.

Our goal with the sum rules is to calculate the energy [$E_0=(3/2)\omega$]
and wave function
at the origin [$|\psi_0(0)|^2=(m\omega/\pi)^{3/2}$] of the lowest bound state.
We start with the coordinate-space Green's operator or resolvent \cite{PAS84}:
\begin{equation}
 G(\vec x,\vec y;z) \equiv \langle \vec x | {1\over z-H} | \vec y \rangle
   = \sum_n {\psi_n(\vec x) \psi^\ast_n(\vec y) \over
         z - E_n}  \ ,
\end{equation}
where the second equality is obtained by inserting a complete set of
eigenstates of $H$.
In the complex $z$ plane, there are poles at $z=E_n$.
We will evaluate $G$ for $z = -E < 0$, away from the poles (``spacelike'').
The analog of the QCD correlator follows by setting $\vec x=\vec y=0$:
\begin{equation}
  M_1(E) \equiv -G(\vec x,\vec y;-E)\bigg|_{\vec x=\vec y=0}
      = \sum_n {|\psi_n(0)|^2 \over E + E_n}  \ .
\end{equation}
But this doesn't converge ({\it cf}.\ the need for subtractions),
so we look at the {\em derivative\/} instead:
\begin{equation}
  M_2(E) \equiv {d\over dE}G(\vec x,\vec y;-E)\bigg|_{\vec x=\vec y=0}
      = \sum_n {|\psi_n(0)|^2 \over (E + E_n)^2}  \ . \label{eq:deriv}
\end{equation}
For the harmonic oscillator, $E_n$ and $|\psi_n(0)|^2$ are given by
\begin{equation}
   E_n = (\mbox{$3\over 2$} + 2n)\omega  \qquad \mbox{and} \qquad
     |\psi_n(0)|^2 = {(2n+1)!!\over 2^n n!}
       \biggl({m\omega\over\pi}\biggr)^{3/2}  \ .
\end{equation}
Equation~(\ref{eq:deriv})
is half of our candidate sum rule: a sum over states.

We can also evaluate $M_2(E)$ directly in perturbation theory after
constructing the Born series for $G$.
The first-order result, for a general central potential $V(r)$, is
\begin{equation}
  M_2(E) = \biggl({m^3\over 8\pi^2 E}\biggr)^{1/2}
          [ 1 - 4m\int_0^\infty \! dr\, r\,\e^{-\sqrt{8mE}\,r}\, V(r) +
           \cdots\
           ] \ ,
\end{equation}
and for the harmonic oscillator is
\begin{equation}
  M_2(E) = \biggl({m^3\over 8\pi^2 E}\biggr)^{1/2}
           \biggl( 1 - {3\over 16}{\omega^2\over E^2} +
           \cdots\
           \biggr) \ .
          \label{eq:pert}
\end{equation}
Note that, for large $E$, the free term dominates and the contribution
from the potential is a controlled
correction.
Equating the two representation of $M_2(E)$ gives us our sum rule.

Can we find values of $E$ for which Eq.~(\ref{eq:pert}) is accurately
calculated with one or two corrections {\em and\/} Eq.~(\ref{eq:deriv}) is
dominated by the $n=0$ term?
For the harmonic oscillator, taking $E\simeq\omega$ in Eq.~(\ref{eq:pert})
gives the exact answer to within a few percent.
However, the contribution of the ground state in Eq.~(\ref{eq:deriv}) is only
about 1/3 of the total and even summing the first eight states only
gives about 3/4!
It would appear that our sum rule is useless.

The solution is to use a better weighting function in the sum, such
as an exponential, to create a new, improved sum rule.
We can reach such a sum rule by applying a Borel transform
with respect to $E$
to each representation of $M_2(E)$.%
\footnote{Use Eq.~(\ref{borel_def_appendix}) with $Q^2 \rightarrow E$
and $M^2 \rightarrow \epsilon$.}
The resulting sum rule is \cite{PAS84}:
\begin{eqnarray}
\sum_{n=0}^{\infty} |\psi_n(0)|^2 \e^{-E_n/\epsilon}
  &=& \biggl({m\epsilon\over 2\pi}\biggr)^{3/2}
           [ 1 - 4m\int_0^\infty \! dr\, r\,\e^{-2m\epsilon r^2} V(r) + \cdots
           ] \nonumber \\
     &=&
 \biggl({m\epsilon\over 2\pi}\biggr)^{3/2}
           \biggl( 1 - {\omega^2\over 4\epsilon^2} + \cdots
           \biggr) \ .
                    \label{eq:borelrule}
\end{eqnarray}
This function can be recognized as a special case of the
time-dependent Green's function, continued to imaginary
time $-i/\epsilon$.
Note that for large $\epsilon$ a large number of levels make an important
contribution to the sum, whereas as we go to small $\epsilon$ the lowest
level dominates increasingly.

Now we try the harmonic oscillator again, with $\epsilon=\omega$,
and again we find that first-order perturbation theory gives the exact
answer to within five percent.
But now the ground state contribution is 80\% of the exact sum, and the
first two levels together give 97\%!
So it is plausible that this sum rule will let us extract information
about the ground state from low-order perturbation theory.

It is instructive to introduce the spectral density
\begin{equation}
    \rho^{\rm osc}(E) = \pi\sum_{n=0}^\infty |\psi_n(0)|^2 \delta(E-E_n)   \ ,
      \label{eq:spect}
\end{equation}
where the sum is over a complete set of energy eigenstates.
We can write the left-hand side of Eq.~(\ref{eq:borelrule}) as
\begin{equation}
\sum_{n=0}^{\infty} |\psi_n(0)|^2 \e^{-E_n/\epsilon}
  = {1\over\pi}
      \int_{-\infty}^{+\infty} dE\, \e^{-E/\epsilon}\, \rho^{\rm osc}(E) \ .
\end{equation}
The free spectral density follows from substituting the free Hamiltonian
and plane-wave states normalized in volume $\Omega$:
\begin{equation}
  \rho^{\rm free}(E) = \pi \sum_{\vec k} {1\over \Omega}
            \delta(E - {\vec k}^2/2m)  \longrightarrow
           \biggl({m^3\over 2\pi^2}\biggr)^{1/2} E^{1/2}\theta(E)  \ .
            \label{eq:rhofree}
\end{equation}
It would not appear from comparing Eqs.~(\ref{eq:spect}) and (\ref{eq:rhofree})
that $\rho^{\rm free}$ and $\rho^{\rm osc}$ have much in
common.

However, if we smear the spectral densities in energy, we find
some similarities.
If we rewrite Eq.~(\ref{eq:borelrule}) in terms of integrals
over spectral densities and transfer the free contribution
to the left-hand side, it takes the form
\begin{equation}
  {1\over \pi} \int_{-\infty}^{\infty}\!  dE\,
     [\rho^{\rm osc}(E) - \rho^{\rm free}(E)]
       \e^{-E/\epsilon}  =
       {1\over\sqrt{\epsilon}}\, \sum_{n=0}^{\infty} {A_n\over \epsilon^{2n}}
          \ .
\end{equation}
The terms on the right-hand side, which are suppressed for large
$\epsilon$ by powers of $1/\epsilon$, are usually referred to as ``power
corrections.''
If we take the limit $\epsilon \rightarrow \infty$, we find that the
integrated spectral densities are equal:
\begin{equation}
  {1\over \pi} \int_{-\infty}^{\infty}\!  dE\,
     [\rho^{\rm osc}(E) - \rho^{\rm free}(E)] = 0  \ .
\end{equation}
This is referred to as a ``global duality'' relation.

This similarity between the spectral densities can be extended to more
local relations.
That is, if we integrate from zero to the midpoint between the first
and second levels, we get quite close to the same result:
\begin{equation}
   {1\over \pi} \int_0^{5\omega/2}\! dE\, \rho^{\rm free}(E)
   \simeq
   {1\over \pi} \int_0^{5\omega/2}\! dE\, \rho^{\rm osc}(E)
   = \biggl({m\omega\over\pi}\biggr)^{3/2}
    \ .
\end{equation}
Similar results are obtained for other levels and if the integrands
include a factor of $E$.
This is called ``local duality'' between the bound and free states.
Each bound state
eats up its part of the free spectral density, leaving
the magnitude of the integral unchanged.

The trick of the sum rules is to exploit this duality.
In particular, we approximate the true spectral density as the
contribution from the lowest state, which is what we seek to determine,
and a contribution from higher states approximated by duality.
Thus we use $\rho^{\rm free}$ starting at some threshold $s_0$, which becomes
a parameter to be determined as well.
Then we have
\begin{equation}
  |\psi_0(0)|^2 \e^{-E_0/\epsilon} +
  {1\over\pi}\int_{s_0}^{\infty} \! dE\,\rho^{\rm free}(E)\, \e^{-E/\epsilon}
  = \biggl({m\epsilon\over 2\pi}\biggr)^{3/2}
  \biggl(1 - {\omega^2\over 4\epsilon^2}
           + {19\over 480}{\omega^4\over \epsilon^4}
     + \cdots \biggr)\ ,
\end{equation}
or, putting the continuum contribution on the right hand side,
\begin{equation}
  |\psi_0(0)|^2 \e^{-E_0/\epsilon} =
  {1\over\pi}\int_0^{s_0} \! dE\,\rho^{\rm free}(E)\, \e^{-E/\epsilon}
  + \biggl({m\epsilon\over 2\pi}\biggr)^{3/2}
  \biggl(- {\omega^2\over 4\epsilon^2} + {19\over 480}{\omega^4\over
\epsilon^4}
     + \cdots \biggr)  \ .
\end{equation}

The parameters to be extracted from the sum rule are $E_0$,
$|\psi_0(0)|^2$, and $s_0$.
One notes that $\epsilon$ is an auxiliary parameter in the sum rule.
If the sum rule were perfect, we could pick any value of $\epsilon$.
However, each side is approximate, so at best we expect a {\it region\/}
in $\epsilon$ for which each side is well approximated.
This fiducial
region is determined in practice by insisting that the errors due to
the rough model of the higher states and the contributions of
omitted higher
power corrections are both small.

There are various methods for extracting the parameters.
The ratio method can be useful if we have two sum rules that we can divide to
remove the dependence on the wave function.
In the present case, differentiating with respect to $-1/\epsilon$
generates another sum rule.
Dividing the two sum rules yields an estimate for $E_0$:
\begin{equation}
  E_0  =  {
       {1\over\pi} \int_0^{s_0} \! dE\,\rho^{\rm free}(E)\, E\e^{-E/\epsilon}
  + \bigl({m\epsilon\over 2\pi}\bigr)^{3/2}\epsilon
  ({3\over 2}
  + {\omega^2\over 8\epsilon^2} - {19\over 192}{\omega^4\over \epsilon^4}
     + \cdots )
     \over
     {1\over\pi}  \int_0^{s_0} \! dE\,\rho^{\rm free}(E)\, \e^{-E/\epsilon}
  + \bigl({m\epsilon\over 2\pi}\bigr)^{3/2}
  (- {\omega^2\over 4\epsilon^2} + {19\over 480}{\omega^4\over \epsilon^4}
     + \cdots ) } \ .
\end{equation}
We can plot this for various values of $s_0$ and
choose the flattest curve in the fiducial interval
(say $0.6 < {\omega/\epsilon} < 1.2$) to determine $E_0$.
With just two or three terms, results close to
$E_0 = (3/2)\omega$ are found, with $s_0 \simeq (5/2)\omega$.
We can then return to the original sum rule and extract $|\psi_0(0)|^2$.

There are still some subjective elements in this ratio analysis because
$s_0$ is not determined very well.
An alternative approach is simply to treat the sum rule as an
optimization problem for $E_0$,
$|\psi_0(0)|^2$, and $s_0$ in a specified interval where the sum rule
should work (results should be relatively insensitive to the precise
boundaries).
One finds that accurate and more stable results are obtained this way,
particularly if the continuum estimate is improved by adding
the first-order correction to $\rho^{\rm free}$.
The QCD analog of the optimization method
is described in Sec.~\ref{result:nuc} and will be our method of
choice for quantitative analysis of the QCD sum rules.

The moral of the harmonic oscillator analogy is that
it {\it is\/} possible to construct a sum rule that can be accurately
calculated in low-order perturbation theory, but which is also largely
saturated by the lowest physical state.
The key underlying principle is that of duality between the two
representations.
In QCD sum rules, we exploit the duality between the hadronic
description (the physical spectral density) and the quark-gluon description
from the OPE.
Note that the main source of error for the harmonic oscillator
example
is the rough model of the spectral density for the higher states.
In the QCD case, the situation is better:  Only
the first resonance is narrow, with higher-state contributions broadened.
Therefore a simple continuum model may be quite reasonable.

\subsection{Sum-Rule Results in Vacuum}

Now we return to the nucleon correlator in vacuum and put together
a QCD sum rule, following the recipe of Sec.~\ref{sectii_iv}
and the harmonic oscillator example.
We start with
the OPE for the functions $\Pi_s$ and $\Pi_q$
of Eq.~(\ref{nucleon_corr}).
(A method to calculate Wilson coefficients is described in
Sec.~\ref{wilson:coe}.)
With the interpolating field
in Eq.~(\ref{int_field}) and keeping operators up through
mass dimension six,%
\footnote{Recall that terms with explicit dependence on $m_q$ are
omitted here.}
 one obtains \cite{IOF81}
\begin{eqnarray}
\Pi_s^{\rm OPE}(q^2)
&=&{q^2\over 4\pi^2}\ln(-q^2)\langle\overline{q}q\rangle_{\rm vac}\ ,\\*
\Pi_q^{\rm OPE}(q^2)
&=&-{(q^2)^2\over 64\pi^4}\ln(-q^2)
-{1\over 32\pi^2}\ln(-q^2)\left<{\alpha_s\over\pi}G^2\right>_{\rm vac}
-{2\over 3q^2}\langle\overline{q}q\rangle_{\rm vac}^2\ ,
\label{nucleon_corr_qcd}
\end{eqnarray}
where all polynomials in $q^2$, which vanish under the Borel transform,
have been omitted.
In Eq.~(\ref{nucleon_corr_qcd}), the four-quark condensates have
already been estimated in terms of the square of
$\langle\overline{q}q\rangle_{\rm vac}$;
this factorization assumption will be discussed in detail later.
The Borel transform can be calculated using the
formulas in Table~\ref{borel_table} to obtain
\begin{eqnarray}
\widehat\Pi_s^{\rm OPE}(M^2)
&=&-{1\over 4\pi^2}M^4\langle\overline{q}q\rangle_{\rm vac}\ ,\\*
\widehat\Pi_q^{\rm OPE}(M^2)
&=&{1\over 32\pi^4}M^6+{1\over 32\pi^2}M^2\left<{\alpha_s\over\pi}
       G^2\right>_{\rm vac}
+{2\over 3}\langle\overline{q}q\rangle_{\rm vac}^2\ .
\end{eqnarray}

Now consider the phenomenological side of the sum rule.
Given the form of the correlator in Eq.~(\ref{nucleon_corr_def}), one
can simply write down the phenomenological form
of a nucleon pole plus higher-energy continuum states:
\begin{equation}
\Pi^{\rm phen}(q)=-\lambda_N^2{\rlap{/}{q}+M_N\over q^2-M_N^2+i\epsilon}
+{\rm continuum}\ ,
\end{equation}
where $M_N$ is the nucleon mass and $\lambda_N$ was introduced in
Eq.~(\ref{eq:mel}).
In terms of the scalar functions of Eq.~(\ref{nucleon_corr}), one has
\begin{eqnarray}
\Pi_s^{\rm phen}(q^2)
&=&-\lambda_N^2{M_N\over q^2-M_N^2+i\epsilon}+{\rm continuum}\ ,\\*
\Pi_q^{\rm phen}(q^2)
&=&-\lambda_N^2{1\over q^2-M_N^2+i\epsilon}+{\rm continuum}\ .
\end{eqnarray}

The continuum contribution is approximated by the equivalent OPE
spectral densities, starting at a sharp threshold $s_0$,
yielding
\begin{eqnarray}
\rho_s^{\rm phen}(s)
&=&\lambda_N^2 M_N\delta(s-M_N^2)
-{1\over 4\pi^2}s\langle\overline{q}q\rangle_{\rm vac}\theta(s-s_0)\ ,
  \label{eq:phens} \\*
\rho_q^{\rm phen}(s)
&=&\lambda_N^2\delta(s-M_N^2)
+\left[{1\over 64\pi^4}s^2+{1\over 32\pi^2}\left<{\alpha_s\over\pi}
        G^2\right>_{\rm vac}\right]
\theta(s-s_0)\ .  \label{eq:phenq}
\end{eqnarray}
The same continuum threshold $s_0$ is assumed for simplicity
to be relevant for each
spectral density.
Substituting into Eq.~(\ref{eq:borelint}), one obtains
the Borel transform of the phenomenological correlation
function:
\begin{eqnarray}
\widehat\Pi_s^{\rm phen}(M^2)&=&\lambda_N^2 M_N \e^{-M_N^2/M^2}
-{M^4\over 4\pi^2}\langle\overline{q}q\rangle_{\rm vac}
\left(1+{s_0\over M^2}\right)\e^{-s_0/M^2}\ ,\\*
\widehat\Pi_q^{\rm phen}(M^2)&=&\lambda_N^2 \e^{-M_N^2/M^2}
+{M^6\over 32\pi^4}\left(1+{s_0\over M^2}+{s_0^2\over 2M^4}\right)
\e^{-s_0/M^2}
\nonumber
\\*
\hspace*{1.5cm}& &
+{M^2\over 32\pi^2}\left<{\alpha_s\over\pi}
     G^2\right>_{\rm vac}\e^{-s_0/M^2}\ .
\label{pinuc_phen}
\end{eqnarray}

At this point, one can make the identifications
$\widehat\Pi_s^{\rm phen}(M^2)=\widehat\Pi_s^{\rm OPE}(M^2)$ and
$\widehat\Pi_q^{\rm phen}(M^2)=\widehat\Pi_q^{\rm OPE}(M^2)$ and solve for
$M_N$ by dividing the equations, with the following result:
\begin{equation}
M_N = { 2aM^4\left[1-\left(1+{s_0\over M^2}\right)\e^{-s_0/M^2}\right] \over
   M^6\left[1-\left(1+{s_0\over M^2}+{s_0^2\over 2M^4}\right)
\e^{-s_0/M^2}\right]
+b M^2\left(1-\e^{-s_0/M^2}\right)
+{4\over 3}a^2   }   \ ,
\label{nucleon_mass}
\end{equation}
where $a=-(2\pi)^2\langle\overline{q}q\rangle_{\rm vac}$ and
$b=\pi^2\langle(\alpha_s/\pi)G^2\rangle_{\rm vac}$.
Perturbative corrections $\sim\alpha_s^n$ can be taken into account
through anomalous-dimension factors (see Ref.~\cite{IOF84} for formulas).

The principal physical content of the full sum rule, that the
scale of the nucleon mass is largely determined by the quark
condensate, is manifest in a highly
simplified version of the nucleon sum rule.
Detailed sum-rule analyses of the nucleon mass have been made by Ioffe
\cite{IOF81} and many others \cite{REI84,REI85}.
Ioffe concluded that the contributions of higher-dimensional
condensates and the continuum tend to cancel for values of the Borel
mass in the vicinity of $M_N$, such that meaningful predictions can be
made.
If one neglects contributions from the continuum,
anomalous dimensions, and the gluon
and four-quark condensates in Eq.~(\ref{nucleon_mass}), the following
simple result is obtained:
\begin{equation}
M_N=-{8\pi^2\over M^2}\langle\overline{q}q\rangle_{\rm vac}\ ,
\label{nucleon_mass_Ioffe}
\end{equation}
which is to be evaluated for $M^2\sim 1\,\mbox{GeV}^2$.
This is known as the Ioffe formula, which we generalize to finite density
in Sec.~\ref{result:nuc}.

In Fig.~\ref{nucleon_mass_fig} we plot the nucleon mass
predicted by the ratio sum rule [Eq.~(\ref{nucleon_mass})] as a function
of Borel $M^2$ for several different continuum thresholds.
We also plot the higher-order sum rule of Ioffe and Smilga \cite{IOF84}.
We expect that the sum rule should be valid in a region near
$M^2=1\,\mbox{GeV}^2$ \cite{IOF81}.
That is to say, the overlap between the region where the sum rule is
dominated by the nucleon contribution and the region where the truncated
operator product expansion is reliable is around $1\,\mbox{GeV}^2$.

\begin{figure}
\vbox to 4.in{\vss\hbox to 5.625in{\hss
{\includegraphics{ratio_fig1f.ps}}\hss}}
\caption{Ratio sum-rule predictions for the nucleon mass $M_N$
at zero density.  The lower curves are from the ratio sum rule
up to dimension-six condensates only
[Eq.~(\protect{\ref{nucleon_mass}})],
and the upper group of curves show the predictions of the sum rule
from Ref.~\protect\cite{IOF84},
which includes higher-dimensional operators.
Anomalous-dimension corrections are neglected.
In both cases, the three curves correspond to continuum thresholds
$s_0=2.0\,\mbox{GeV}^2$ (solid),
$s_0=2.5\,\mbox{GeV}^2$ (dashes), and
$s_0=3.0\,\mbox{GeV}^2$ (dot-dashes). }
\label{nucleon_mass_fig}
\end{figure}

The sum rules for the nucleon are less than ideal for a number of
reasons.
First, we expect that the ratio sum-rule prediction for the
nucleon, if well-satisfied, should have a flat region as a function of
the Borel mass, which is just an auxiliary parameter.
In fact, the nucleon sum rule is not close to flat
if we truncate the OPE at dimension-six condensates, as we will do at
finite density.
On the other hand,
the sum rule of Ioffe and Smilga \cite{IOF84},
which includes various higher-dimensional condensates, becomes
considerably flatter over a larger region.
We interpret the change in the ratio within
the fiducial region as a measure of the theoretical error bar for the mass.
Second, perturbative corrections to many of the Wilson
coefficients are large, although the net effect on the sum-rule
predictions seems to be small \cite{OVC88}.
Third, the continuum contribution is unavoidably large because the gap
between the nucleon and higher-mass states is only of order 500 MeV.
This is generally undesirable, because we only use a simple model to
account for the continuum states; however, lattice tests \cite{LEI94}
find the approximation to be quite reasonable (see Sec.~\ref{sec:ansatz}).
Finally,
it has been suggested that instanton effects
are also important for the nucleon sum rules \cite{DOR90,FOR93} and
that inclusion of instanton effects help to stabilize the sum rule.

The vacuum sum rules for the nucleon (and other baryons) appear
to be phenomenologically successful \cite{SHI92}.
However, there are a variety of ways in which
the sum rules could be imprecise (or even fail).
While this observation might make one hesitate before including the
further complications of finite density, we think that the situation is
actually quite favorable.
Indeed, since we focus on {\em changes\/} in spectral properties with
density, we are less sensitive to details that affect the
{\em absolute\/} predictions of vacuum properties.
We will rely on the cancellation of systematic
discrepancies (including those from the truncation to dimension-six operators)
by normalizing all finite-density self-energies
to the zero-density prediction for the mass \cite{FUR92}.
Moreover, even predictions with large uncertainties ({\it e.g.,} 50\%)
will be useful in assessing relativistic phenomenology.
So we will proceed to generalize the nucleon
sum-rule formalism, assuming that the
zero-density limit is valid.

QCD sum-rule methods have also been widely
applied to the calculation of
baryons other than the nucleon
and to meson masses, coupling constants, and form factors.
Results for almost every low-spin ($J<3$) meson have been calculated with only
a handful of parameters.
These calculations are well documented elsewhere; the book by Shifman
\cite{SHI92} and
the reviews by Reinders {\it et al.} \cite{REI85}
and Narison \cite{NAR89} provide good guides to the literature.

\section{Nucleons at Finite Density}

\subsection{Overview}

In this section, we extend the sum-rule approach to the finite-density problem
of nucleons propagating in infinite nuclear matter.
We start with a review of relativistic phenomenology, which identifies
the quantities to calculate and the physics we hope to test
with the sum rules.
Next we focus on changes in the chiral condensate at finite density,
which suggests that nucleon propagation should be strongly
modified.
Then we put together these inputs through a generalization of the
sum-rule approach, which encompasses several subsections.
Finally we present results and mention alternative calculations.
We have omitted some technical details in places
so that we can focus on the physics, but complete details are
available in the cited references.

\subsection{Relativistic Nuclear Phenomenology}
\label{sec:dirac}

In this section, we review some elements of relativistic nuclear
physics phenomenology.
Our discussion follows that of Ref.~\cite{FUR92}.
Our goal is twofold: to explain the basic
physics underlying the approach and to give insight into the
appropriate way to model the spectral density for QCD sum rules at finite
density.
We also will introduce notation that will be used for the sum rules.


The most significant success of relativistic nuclear physics
has been the economical description of proton-nucleus spin observables
at intermediate energies for a wide range of target nuclei.
This
problem has been studied both purely phenomenologically using a global
parameterization of the scattering data \cite{CLA83a,CLA83b,CLA83c,HAM90}
and in a
meson-theoretical framework based on a relativistic impulse
approximation \cite{MCN83,SHE83,TJO87,WAL87}.
The key to the successes of both
approaches is that nucleon propagation in the nuclear medium is
described by a Dirac equation featuring large Lorentz scalar and vector
optical potentials.
That is, the nucleon wave function $\psi$ satisfies
\begin{equation}
  (E\gamma_0 - \vec\gamma
  \cdot \vec{q}- M_N - U) \psi = 0 \ ,
  \qquad\qquad U \simeq S + V\gamma_0 \ ,
\end{equation}
where $S$ and $V$ are the scalar and
vector optical potentials, $M_N$ is the nucleon mass, and $E$ is the
nucleon energy.

Although the two approaches give slightly different optical
potentials, the qualitative features are quite similar
\cite{CLA83a,CLA83b,CLA83c,MCN83,SHE83,TJO87,WAL87,HAM90}:
\begin{itemize}
\item[$\bullet$]
Attractive scalar ($\mbox{Re\ }S<0$) and repulsive vector ($\mbox{Re\ }V>0$)
potentials are found,
with magnitudes reaching several hundred MeV at nuclear matter
saturation density.
\item[$\bullet$] Significant cancellation between the
potentials occurs, so
that the effective nonrelativistic central potential is only
tens of MeV in magnitude.
\item[$\bullet$] The scalar and vector imaginary parts also exhibit
significant cancellation; the imaginary parts are significantly
smaller than the real parts.
\item[$\bullet$] The real parts of the
potentials have relatively weak energy dependence.
\end{itemize}

These qualities naturally suggest that a nucleon at intermediate
energies can be regarded as a quasiparticle with large scalar and
vector self-energies (corresponding to the optical potentials).
The imaginary parts of the optical potential
indicate that the width in energy of the quasinucleon excitation is
relatively small on hadronic scales; that is, small compared to the
spacing between the free-space nucleon and the Roper resonance.
As discussed below, a similar picture emerges for nucleons below the Fermi
sea in mean-field QHD \cite{SER86,SER92} or relativistic
Brueckner calculations \cite{SER86,DEJ91,AMO92,LI93}.

This picture is different from conventional
descriptions based on nonrelativistic $NN$ potentials.
It is worth asking why one might expect such a picture.
The empirical low-energy $NN$ scattering amplitudes are conventionally
parameterized using Galilean invariants;
one might, instead, choose to express these
amplitudes using Lorentz invariants.
In this case, one
finds Lorentz scalar and vector components that are much larger than
the amplitudes deduced from the nonrelativistic decomposition.
In ordinary, spin-saturated nuclear matter, all of the other
Lorentz components arising from the $NN$ interaction largely average
out, leaving only small effects; these include terms arising from
one-pion exchange.
Thus, the dynamics of neutral  scalar and vector
components are the most important for describing nucleons in bulk
nuclear matter \cite{SER86}.

Relativistic hadronic field theories of nuclear phenomena
such as quantum hadrodynamics (QHD) provide
a qualitative description of this physics.
In this description, large
scalar and vector self-energies emerge at the mean-field level from
the interaction of a nucleon with all other nucleons in the Fermi sea
via the exchange of isoscalar scalar and vector mesons.
This simple picture has many phenomenological successes;
for example, relativistic mean-field
models provide a quantitatively accurate description of many bulk
properties of nuclei \cite{SER86}.

It is important to stress that the QCD sum rules we construct will not
test this underlying dynamical picture of meson exchange as the origin
of the quasinucleon self-energies.
Concerns about issues such as the
use of the Dirac equation to describe composite nucleons (such as the
role of Z-graph physics) have dominated the past discussion of
connections between
relativistic nuclear physics and
QCD \cite{BRO84,THE85,COO86,BLE87,ACH88,JAR91,COH92,WAL94}.
We will briefly address some of these issues at the end of this article,
but they do not affect our discussion.
Instead, we consider the
{\em spectral\/} properties themselves, which can be
studied outside the context of a hadronic model.
Having done such a
study, one can compare to the predictions of Dirac phenomenology or QHD.
The critical questions are the magnitudes and signs of the
Lorentz scalar and vector parts of the optical potential.

To put this more concretely, we consider the nucleon propagator in a
QHD theory
\begin{equation}
   G(q) =
      -i \int \, d^4{x}\, \e^{iq\cdot x} \,
	\langle \Psi_0 |T\psi(x) \overline{\psi}(0)|{\Psi_0}
	\rangle
	     \ ,  \label{eq:nuclprop}
\end{equation}
where $|{\Psi_0} \rangle $ is the nuclear matter
ground state and $\psi(x)$ is a nucleon field \cite{SER86}.
The analytic structure of $G$ in the mean-field (and more sophisticated)
approximations can suggest what we should expect to find
for an analogous QCD correlator.
The nucleon self-energy $\Sigma$ can
be defined formally in terms of the solution of Dyson's equation for
the inverse propagator:
\begin{equation}
   [G(q)]^{-1} = \qslash  - M_N - \Sigma(q) \ .
\end{equation}
The self-energy can be identified directly from the analytic properties of
the propagator $G(q)$.
In particular, the discontinuities of $G$
across the real $q_0$ axis, which are proportional to the spectral
densities, can
be used to extract the on-shell self-energy.

Since Lorentz transformation properties are at the heart of this
problem, it is useful to identify the various Lorentz
vectors in the problem.
We will use these vectors to classify the possible terms in $G(q)$.
There are only two independent vectors:
$q_\mu$ itself and $u_{\mu}$, the four-velocity of the matter.
Since $u_{\mu}$ will play an essential role in what follows, it is
useful to give some feeling for what $u_{\mu}$ means.
The key point is that infinite nuclear matter has a natural rest frame
specified by $u_\mu$: the frame in which the expectation value of the
baryon current is zero for all spatial components; the time component
of the current in this frame is simply the baryon density $\rho_N$.
In the rest frame of nuclear matter, the components of
the four-velocity are $(1, \vec 0)$.
An alternative way to understand $u_{\mu}$
is to consider infinite nuclear matter as the limit of a finite nucleus
when the mass and particle number go to infinity.
The four-velocity is
simply given by $u_{\mu} = p_{\mu}/M$ where $p_{\mu}$ is the four-momentum
of the nucleus and $M = \sqrt{p \cdot p}$ is the invariant mass.
(In general, any four-vector thermodynamic quantity in an equilibrium
system, such as the momentum or baryon densities, is proportional
to $u_\mu$.)

We start with a general decomposition of $G(q)$:
\begin{equation}
 G(q) =  G_s(q^2,q\cdot u) + G_q(q^2,q\cdot u)\rlap/{q}
	+ G_u(q^2,q\cdot u)\rlap/u   \ .   \label{eq:Gofq}
\end{equation}
This form is determined by Lorentz covariance and the
assumed invariance of the ground state under parity and time-reversal
symmetries \cite{RUS94}.
The self-energy can be decomposed similarly; this will define the
notation used in subsequent sections.
The nucleon self-energy is written as
\begin{equation}
\Sigma(q)=\widetilde\Sigma_s(q^2,q \cdot u)
+\widetilde\Sigma_v^\mu(q)\gamma_\mu \ ,
\end{equation}
where
\begin{equation}
\widetilde\Sigma_v^\mu(q)
=\Sigma_u(q^2,q \cdot u)u^\mu+\Sigma_q(q^2,q \cdot u)q^\mu  \ .
\end{equation}
We also define an in-medium scalar self-energy
\begin{equation}
\Sigma_s\equiv M_N^*-M_N\ , \ \ \
M_N^*\equiv {M_N+\widetilde\Sigma_s\over 1-\Sigma_q}\ ,
\label{eq:MNstar}
\end{equation}
and an in-medium vector self-energy
\begin{equation}
\Sigma_v \equiv
{\Sigma_u\over 1-\Sigma_q}\ ,
\label{eq:SigmaV}
\end{equation}
The combinations in Eqs.~(\ref{eq:MNstar}) and (\ref{eq:SigmaV}) appear
naturally when one solves for the nucleon pole.
In a certain sense, the ``scalar self-energy''
of the nucleon in the medium is $M_N^*$; this is the scalar quantity
we will calculate using the sum rules.
However, we will follow the
conventional nuclear physics nomenclature and refer to $\Sigma_s$ as
the scalar self-energy in the medium.

In the mean-field approximation, $\Sigma_s$ and $\Sigma_v$ are found to
be real and independent of momentum, and $\Sigma_q$ is taken to be
identically zero.
In this description, nucleons of any three-momentum
appear as stable quasiparticles with
self-energies that are nearly linear in the density up to nuclear
matter density \cite{SER86}.
For mean-field models that
provide quantitative fits to bulk properties of finite nuclei, the
self-energies are typically several hundred MeV in magnitude at
nuclear matter saturation density: $\Sigma_s \sim -350$ MeV and
$\Sigma_v \sim +300$ MeV.
These correspond to real, energy-independent
optical potentials, $S$ and $V$, and are similar in magnitude
to the real parts of the optical potentials
used to describe proton-nucleus scattering
\cite{CLA83a,CLA83b,CLA83c,MCN83,SHE83,TJO87,WAL87,HAM90}.
We wish to stress that the effective nucleon mass
$M_N^*$ defined in Eq.~(\ref{eq:MNstar}) is {\em not\/} equivalent to
the usual nonrelativistic effective nucleon mass that is connected with
the momentum dependence of the optical potential \cite{MAH85}.
It comes from
different physics and does not reduce to the nonrelativistic $M_N^*$ in
the nonrelativistic reduction of the model.
We refer the reader to
Ref.~\cite{JAM89} for further discussion of the relations between
different effective masses.

Clearly, mean-field physics is not the entire story.  We
know {\it a priori\/} that the optical potentials are complex
and energy dependent.
Thus, although mean-field models
successfully describe a wide range of phenomena, their simplicity leads
us to question whether the basic physics survives in a more
sophisticated analysis.
The most sophisticated relativistic calculations of nuclear
matter have been performed in what is usually known as the Dirac
Brueckner Hartree-Fock (DBHF) approximation \cite{SER86,DEJ91,AMO92,LI93}.
This approximation incorporates
effects from short-range correlations, which are critical in the
nonrelativistic description of nuclear matter saturation.
While these
calculations involve some untested assumptions, they  provide a unified
and quantitative description of $NN$ scattering observables and nuclear
matter saturation properties.

Dirac Brueckner calculations generally find that the on-shell
self-energies are only weakly  dependent on the three-momentum
$\vec{q}$.
(This corresponds to a weak energy dependence for the real
parts of the relativistic scalar and vector optical potentials seen by
a scattered nucleon.)
Here, ``on-shell'' means that the self-energies
are evaluated at the $q_0$ corresponding to the pole position, which is
found by solving a transcendental equation for the self-consistent
single-particle energy \cite{SER86}.
The self-energies $\Sigma_s$
and $\Sigma_v$ are found to be essentially
similar in magnitude, sign, and density
dependence to those from  mean-field calculations.
Furthermore, the
magnitude of the dimensionless $\Sigma_q$, which is zero in the
mean-field approximation, is typically much less than one in DBHF
calculations (see, however, Ref.~\cite{AMO92}).
Thus the mean-field
quasiparticle picture is qualitatively unchanged in the DBHF approximation.
We use this picture to guide us in formulating our QCD
sum-rule spectral ansatz.

In the mean-field approximation, the propagator with real self-energies
in the rest frame of nuclear matter is
\begin{equation}
 G(q) = {1 \over \rlap/q - M_N - \Sigma(q)}
    \longrightarrow \lambda^2 {\rlap/q + M_N^* - \rlap/u \Sigma_v \over
    (q_0 - E_q)(q_0 - \overline{E}_q)} \ ,  \label{eq:mftprop}
\end{equation}
where $E_q$ and $\overline{E}_q$,
the positions of the positive- and negative-energy poles,
are defined in
Eqs.~(\ref{eq:Eq}) and (\ref{eq:Eqbar}) below.
We have introduced a
common residue factor $\lambda^2$, which is unity here, but
includes the factor  $(1-\Sigma_q)^{-1}$
in more general approximations.
(Infinitesimals are not needed in the present discussion, so we
suppress them.)
Although, we have given the result in a particular
frame, it is straightforward to go to any other frame via a Lorentz boost.

We can pick out the functions $G_q$, $G_s$, and $G_u$ directly
from Eq.~(\ref{eq:mftprop}) using Eq.~(\ref{eq:Gofq}).
A Lehmann
representation, obtained  by inserting a complete set of intermediate
states between the $\psi$ and $\overline\psi$ in
Eq.~(\ref{eq:nuclprop}), shows that $G_q$, $G_s$, and $G_u$
have the same singularity structure.
In general, the entire real
$q_0$ axis is cut (except for a small gap at the
chemical potential), but in the mean-field approximation there are only
two simple poles.
The discontinuities of the $G_i$'s across the real $q_0$
axis are proportional to the spectral functions.
There are no singularities elsewhere in the complex $q_0$ plane.

In the mean-field approximation, the positive- and negative-energy
poles in $q_0$ are at
\begin{eqnarray}
  E_q &=& \Sigma_v + \sqrt{\vec{q}^2 + {M_N^*}^2}
      \equiv \Sigma_v + E_q^* \ ,  \label{eq:Eq} \\[4pt]
  \overline{E}_q&=& \Sigma_v -  \sqrt{\vec{q}^2 + {M_N^*}^2}
		\equiv \Sigma_v - E_q^*       \label{eq:Eqbar}
\end{eqnarray}
The discontinuities of the propagator functions across
the real $q_0$ axis, for real, fixed $|\vec{q}|$, are simply $\delta$
functions in this approximation:
\begin{eqnarray}
    \Delta G_s(q_0) &=& -2\pi i{M_N^*\lambda^2\over 2 E_q^*}
	    [\delta(q_0-E_q) - \delta(q_0-\overline{E}_q)]  \ ,
	    \\[4pt]
    \Delta G_q(q_0) &=& -2\pi i{\lambda^2\over 2 E_q^*}
	    [\delta(q_0-E_q) - \delta(q_0-\overline{E}_q)]  \ ,
	    \\[4pt]
    \Delta G_u(q_0) &=& +2\pi i{\Sigma_v\lambda^2\over 2 E_q^*}
	    [\delta(q_0-E_q) - \delta(q_0-\overline{E}_q)]      \ .
\end{eqnarray}
These define the mean-field spectral densities up to an
overall factor of $2\pi i$.
It is clear from Eq.~(\ref{eq:mftprop}) that the {\em relative\/} residues of
the $\rlap/{q}$, scalar, and $\rlap/u$ poles
directly determine $M_N^*$ and $\Sigma_v$.
These are the two independent quantities we wish
to extract; when we generalize to the QCD case, these will be  the
quantities we want to extract via our spectral ansatz for the QCD
nucleon correlator.
Note, however, that in the
mean-field approximation, the on-shell self-energies are
independent of $\vec{q}$, while the self-energies we
extract from the sum rules
will depend explicitly (but weakly) on $\vec{q}$.%
\footnote{The momentum dependence means that the residues of the
positive- and negative-energy poles will differ.
This is an important effect in some other formulations of finite-density
nucleon sum rules \cite{FUR94}.
Because we will explicitly
suppress negative-energy contributions, it will not be an important
consideration in our sum rules.\/}

Evidently, to match the empirical fact that quasinucleon
single-particle energies are mostly unchanged at nuclear matter
density, we must find significant cancellation between $\Sigma_s$ and
$\Sigma_v$ to keep $E_q$ roughly constant.
In contrast, the mean-field
ansatz predicts a significant shift of the negative-energy
``pole'' position $\overline{E}_q$ with increasing density, since the
scalar and vector self-energies are both attractive in this channel.
In reality, we  expect a broad distribution of strength rather than a
narrow excitation, so the simple ansatz is far more realistic for the
positive-energy quasinucleon than for the antinucleon.
Thus we find
that, although a simple real momentum-independent description of the
self-energies yields a plausible description of the
nucleon in nuclear matter, it does a rather poor job of describing
antinucleons in nuclear matter.
It should be noted that this spread
in the antinucleon strength occurs rather naturally in Dirac
phenomenologically.
While the imaginary parts of the scalar and vector
self-energies tend to cancel for the nucleon, they add  for the
antinucleon, implying a large imaginary self-energy and, hence, a large width.
Thus, while the neglect of the imaginary part of the self-energy may be
justified in descriptions of the nucleon, it is
certainly not justified in descriptions of the antinucleon.
Accordingly, it is important when using such an overly simple
description to ensure that calculated quantities do not depend strongly
on contributions coming from the antinucleons.

One might imagine that, given the large self-energies predicted in
relativistic models, one could find clear experimental signatures.
This is not the case.
The {\em individual\/}  self-energies are not directly
observed in nuclei; only combinations of scalar and vector  appear.
Moreover,
given the large cancellations between scalar and vector in many
observables, one finds that the best evidence of
a modified relativistic effective mass $M_N^*$, for example, is rather
indirect, coming from fits to spin observables in intermediate-energy
scattering.

QCD sum rules, however, are not able to predict directly such spin observables.
As a result, in our QCD sum-rule studies, we adopt an
indirect approach.
Instead of attempting to make direct predictions of
experimental observables, we will use QCD sum rules to predict  the
individual scalar and vector self-energies themselves.
To do so, we consider the analytic structure of
something like the nucleon propagator in QHD models.
This leads us to
the natural analog in QCD of the QHD propagator:  a correlator of
interpolating fields with nucleon quantum numbers.

Our aim is to study  the QCD correlator at finite density.
In many ways the problem resembles the simple QHD problem discussed in
this section.
For example, we can exploit the fact that  the
correlator can be  decomposed as in Eq.~(\ref{eq:Gofq}).
We will also
use the Lehmann representation to relate the discontinuities in $q_0$
across the real axis to  the spectral densities, which, in turn,
determine the correlator everywhere in the complex $q_0$ plane.
Phenomenologically, we
will assume a quasinucleon model to describe the region of the cut
corresponding to the energy of a nucleon in nuclear matter.
That is,
we will take Eq.~(\ref{eq:mftprop}) as our ansatz for the quasinucleon
contribution to the correlator, although  we will
allow the self-energies to depend on the three-momentum $\vec{q}$.
We  stress that
all of this can be done without making the assumptions that QHD and
the mean-field approximation are valid.
All we assume is that the nucleon in medium is well approximated as a
reasonably well-defined quasiparticle.

Lorentz covariance and the assumed invariance  of the nuclear matter
ground state under time-reversal and parity symmetries constrain the form of
the spectral functions for the nucleon propagator in QHD models.
In particular, they imply that only Lorentz scalar and vector
self-energies ({\it i.e.\/},  no pseudoscalar or tensor components) are
associated with a quasiparticle pole \cite{RUS94}.
These same constraints also
apply to the QCD correlator of nucleon interpolating fields.
They {\it do not\/} involve further assumptions about hadronic degrees of
freedom or other aspects of relativistic phenomenology.
Therefore, the
principal issue we address here is not whether there are
scalar and vector self-energies that characterize the nucleon-like
excitation in medium; this is a given once we believe that a
quasiparticle approximation is reasonable.
The real question is: What
are the magnitudes, signs, and density dependencies of the
self-energies?
Phenomenologically, the quasiparticle energy of a nucleon in
nuclear matter is only fractionally shifted from the free nucleon
energy, and the excitation is only tens of MeV wide at most.
This small shift will not be a built-in constraint, but  should be
predicted by the sum rules as well.

Of course the quasiparticle ansatz is overly simple:  the
true QCD correlator at finite density will not
have simple poles on the real axis.
However, we believe the
width of the positive-energy quasinucleon excitation to be small on
hadronic scales (and also  compared to the energy over which we average
in the sum rule), so we are justified in making a pole ansatz.
Moreover, since this ansatz does not explicitly include
a background or continuum at these energies, the effective
self-energies we extract will account for {\em all\/} of the strength
in the nuclear domain.
In this sense our in-medium nucleon ``pole''
is like the $\rho$-meson ``pole'' in the vacuum sum-rules; it
summarizes spectral strength over a narrow region.
On the other hand,
the spectral density on the negative-energy side, which corresponds to
an antinucleon ``pole,''  describes the spectral function quite
badly; we expect a broad distribution as the density of the nuclear
system is increased.  Therefore, we will minimize our sensitivity to
this part of the spectral density by constructing a sum rule that
suppresses this contribution relative to the positive-energy side.

\subsection{The Chiral Condensate in Nuclear Matter}
\label{sec:chiralcon}

Perhaps the  most compelling
evidence that light-quark hadron properties are significantly altered in
matter stems from two basic facts: 1) The chiral condensate
$\langle\overline q q\rangle$ plays
an essential role in the structure of these hadrons, and 2)
in the nuclear medium, the magnitude of the chiral condensate is
substantially reduced relative to its vacuum value.
This decrease of the chiral condensate in medium is often
referred to as ``partial restoration of chiral symmetry.''
In this section, we will review some of the important issues concerning
the chiral condensate at finite density.
Additional details may be found in the recent review by Birse \cite{BIR94}.

To leading order, the reduction in magnitude
of the  chiral condensate in nuclear matter
from its vacuum value is  simply the reduction per nucleon times the
density of nucleons \cite{DRU90}.
The essential observation is that
the net amount a single isolated  nucleon reduces the integrated
scalar  density is closely related to an experimental observable.
In particular, the nucleon $\sigma$ term, defined by
\begin{equation}
\sigma_N\equiv 2m_q\int d^3 x\,
(\langle\widetilde{N}|\overline{q}q|\widetilde{N}\rangle
-\langle{0}|\overline{q}q|{0}\rangle)\ ,
\end{equation}
with $m_q$ the average of the up and down quark masses
and $|\widetilde{N}\rangle$ a box-normalized nucleon state
(see Sec.~\ref{condensates}),
is extractable from
$\pi$-$N$ scattering data after some (nontrivial) theoretical extrapolations.

Ignoring $NN$ interactions, one can
express the value of the chiral condensate at finite nuclear matter
density as%
\footnote{The condensate $\langle \overline{q} q \rangle$ is defined
with an implicit normalization scale, typically taken to be
0.5--1.0\,GeV.
The ratio in Eq.~(\ref{model-ind}), however, is
renormalization-group invariant.}
\begin{equation}
\frac{\langle\overline{q}q\rangle_{\rho_N}}
{\langle\overline{q}q\rangle_{\rm vac}}
=1-\frac{\sigma_N\rho_N}{m_\pi^2 f_\pi^2}+\cdots\ ,
\label{model-ind}
\end{equation}
where we have introduced the notation
$\langle{\widehat O_n}\rangle_{\rho_N}
\equiv\langle\Psi_0 |{\widehat O_n}|\Psi_0\rangle$
and used the
Gell-Mann--Oakes-Renner (GMOR) relation \cite{GEL68},
\begin{equation}
   2 m_q \langle
  \overline{q} q \rangle_{\rm vac} = - m_\pi^2 f_\pi^2 \ .
  \label{eq:gmor}
\end{equation}

One can use Eq.~(\ref{model-ind}) to get
a first approximation to how much one expects the chiral condensate to
change in the nuclear medium.
For definiteness, we take nuclear matter saturation
density to be $\rho_N = (110\,{\rm MeV})^3 \simeq 0.17\,{\rm fm}^{-3}$.
The extraction of the
$\sigma$ term requires a sophisticated analysis and
there remain significant uncertainties.
A recent analysis gives a value for $\sigma_N$
of approximately 45 MeV with an uncertainty in the 7--10\,MeV range
\cite{GAS91};  future experiments and theoretical work should
tighten the limits further.
Taking $\sigma_N=45\,\mbox{MeV}$ and using Eq.~(\ref{model-ind})
gives $\langle\overline{q}q\rangle_{\rho_N}/
\langle\overline{q}q\rangle_{\rm vac}\simeq 0.64$.
Thus, this simple analysis suggests
that the chiral condensate at nuclear matter density is substantially
reduced from its free-space value.

The expression in Eq.~(\ref{model-ind}) is, in a very real sense, model
{\em independent\/} \cite{COH92}.
It is valid for any description of nuclear
matter so long
as the density is sufficiently low that one can ignore the effects of
interactions.
There has been considerable
discussion \cite{DRU90,DRU91,COH92,BIR93,CEL93,CHA93,ERI93,NYF93,ERI94,BIR94}
about corrections to Eq.~(\ref{model-ind}),
which are model {\em dependent\/}.
Fortunately, there is
good reason to believe that these corrections are small ($\leq 20$\%)
up to nuclear matter density \cite{COH92}.

There is a very general way to relate the in-medium chiral condensate
to the quark-mass dependence of the nuclear matter energy density.
As shown in Ref.~\cite{COH92}, one can use the Hellmann-Feynman
theorem \cite{MER70} to express the in-medium condensate as%
\footnote{One can avoid renormalization subtleties in applying the
Hellmann-Feynman theorem by working with bare quantities.
Since Eq.~(\ref{model-ind1}) is renormalization-group invariant, the bare
quantities can be replaced with analogous scale-dependent
renormalized quantities.}
\begin{equation}
{\langle\overline{q}q\rangle_{\rho_N}\over
\langle\overline{q}q\rangle_{\rm vac}}
=1-{1\over m_\pi^2 f_\pi^2}
\left(\sigma_N\rho_N+m_q{d\delta{\cal E}\over dm_q}\right)\ ,
\label{model-ind1}
\end{equation}
where $\delta{\cal E}$, which is of higher order in $\rho_N$, is
the contribution to the energy density from interactions and
Fermi motion.

The derivation of Eq.~(\ref{model-ind1}) is quite simple.
The Hellmann-Feynman theorem \cite{MER70} states that, given a Hermitian
operator
$H(\lambda)$, depending on the real parameter $\lambda$, with a set of
normalized eigenstates $|\psi_i(\lambda)\rangle$, one obtains
\begin{equation}
\langle\psi_i(\lambda)|{d\over d\lambda}H(\lambda)|\psi_i(\lambda)\rangle
={d\over d\lambda}\langle\psi_i(\lambda)|H(\lambda)|\psi_i(\lambda)\rangle\ .
\end{equation}

Now consider the application of this theorem to QCD.
The operator $H$ can be taken to be the QCD Hamiltonian,
\begin{equation}
H_{\rm QCD}=H_{\rm QCD}^\chi
+\int d^3 x\,m_q(\overline{u}u+\overline{d}d)
+\mbox{$1\over 2$}\int d^3 x\,(m_u - m_d)(\overline{u}u-\overline{d}d)
+H_{\rm QCD}^{\rm heavy}\ ,
\end{equation}
where $H_{\rm QCD}^\chi$ is the QCD Hamiltonian in the chiral limit,
$H_{\rm QCD}^{\rm heavy}$ contains the quark mass terms for strange and
heavier quarks, and $m_q$ is the average of the up and down quark masses.
One can apply the Hellmann-Feynman theorem to this system with
$m_q$ playing the role of $\lambda$ and $|\Psi_0\rangle$ and $|0\rangle$
playing the role of $|\psi_i(\lambda)\rangle$.
After using the fact that the system is translationally invariant
to remove an overall volume factor, one obtains
\begin{equation}
2m_q(\langle\overline{q}q\rangle_{\rho_N}
-\langle\overline{q}q\rangle_{\rm vac})
=m_q{d{\cal E}\over dm_q}\ ,
\label{FH2}
\end{equation}
where ${\cal E}=M_N\rho_N+\delta{\cal E}$ is the nuclear matter energy density.

Upon using this expression for the energy density, Eq.~(\ref{FH2}) becomes
\begin{equation}
2m_q(\langle\overline{q}q\rangle_{\rho_N}
-\langle\overline{q}q\rangle_{\rm vac})
=m_q\left({dM_N\over dm_q}\rho_N+{d\delta{\cal E}\over dm_q}\right)\ ,
\label{FH3}
\end{equation}
A second application of the Hellmann-Feynman theorem, this time using
the nucleon and vacuum eigenstates of $H_{\rm QCD}$, yields
$\sigma_N=m_q(dM_N/dm_q)$.
This result plus the GMOR relation [Eq.~(\ref{eq:gmor})] gives us
Eq.~(\ref{model-ind1}).

{}From Eq.~(\ref{FH3}), one might expect the corrections to
Eq.~(\ref{model-ind}) to be small at nuclear matter densities and
below.
The essential point is that the interaction energy density is much
smaller than the energy density associated with the nucleon masses.
(The interaction energy per nucleon is about 16 MeV, while the nucleon
mass is nearly two orders of magnitude larger.)
Thus, one expects the first term on the right-hand side of
Eq.~(\ref{FH3}) to dominate the second.

Of course, this argument is somewhat naive:
$\delta{\cal E}\ll M_N\rho_N$ does not necessarily imply that
$d\delta{\cal E}_\rho/dm_q\ll(dM_N/dm_q)\rho_N$.
Nevertheless, the argument is quite suggestive.
To proceed
beyond this qualitative level, one needs to consider explicit models.
The magnitude and even the sign of the correction to
Eq.~(\ref{model-ind}) have been subject to considerable recent discussion
\cite{DRU90,DRU91,COH92,BIR93,CEL93,CHA93,ERI93,NYF93,ERI94,BIR94}.
At present there are probably no models
that can be used to calculate these corrections {\em reliably\/}.
It is
encouraging, however, to note that all of the models on the market,
whether based on explicit quark degrees of freedom or hadronic
degrees of freedom, give
small corrections to the model-independent relation at least up to
nuclear matter saturation density \cite{BIR94}.
We are,
therefore, reasonably confident that the in-medium condensate is reduced from
its free space value by a substantial amount $\sim 35$\%.

However, one must ask whether the in-medium chiral condensate is a
dynamically important quantity.
This question was raised by Ericson \cite{ERI94} in the following
context:
The chiral condensate is a spatially averaged quantity.
In any realistic description of nuclear matter there are spatial
correlations in the local value of $\overline{q}q$; one might
expect that inside nucleons and in their immediate neighborhood the
condensate is very different from its vacuum value, while between
nucleons the value of $\overline{q}q$ may not be much altered
from the vacuum value.
Given this picture, it is reasonable to ask whether this average
quantity should play any special role.
The condensate apparently represents the average of two very different
kinds of physics, and one might imagine that the average could change
substantially by simply increasing the number density of regions where
$\overline{q}q$ differs from the vacuum value without anything
special happening elsewhere.

As noted by
Birse \cite{BIR94}, however, this picture is not realistic: the range of
the spatial correlations for $\overline{q}q$ is fairly long; it
is typically given by two-pion exchange.
Thus, at nuclear densities, any given nucleon is influenced by the change
in the condensate induced by its neighbors, since the range (and the
physical origin) is similar to that of the scalar meson of
one-boson-exchange phenomenology.
Moreover, if the picture of localized bits of altered
condensate were correct and the chiral condensate itself were of no
dynamical significance, as suggested
by Ericson, then one could have dense matter with
the condensate going to zero and eventually changing signs without
anything special happening.
As noted recently \cite{COH94}, however, this is
not possible; the chiral condensate can be shown to be negative semi-definite.

What are the consequences of a decreasing chiral condensate?
As stressed earlier,
the mass of the nucleon is strongly tied to chiral symmetry breaking.
This relationship is seen both in simple chiral models \cite{BHA88}
and in Ioffe's QCD sum-rule treatment of the nucleon \cite{IOF81}.
The reduction of the chiral condensate by about $35\%$ in
the nuclear medium suggests, but does not prove, that the
nucleon mass (more precisely, its scalar self-energy)
in the medium should be significantly reduced.
To be consistent with the empirical fact that nucleons are very weakly
bound in energy, there must also be a strong vector repulsion.
This is the picture suggested by the Dirac phenomenology.
We next turn to the formulation of
QCD sum rules for nucleons in medium to see whether this
qualitative picture plays out.

\subsection{Formalism: Nucleons in Medium}

\subsubsection{Finite-Density Correlator}

The generalization of the vacuum QCD sum rules to finite density starts
with the same correlation function of interpolating fields
as in Eq.~(\ref{nucleon_corr_def}).
The only immediate
difference is that, instead of a vacuum expectation value,
matrix elements are evaluated in the finite-density ground state.
Thus we focus on%
\footnote{As explained later, it is not important whether we
start with the time-ordered or retarded correlator.}
\beq
\Pi(q)\equiv i\int d^4 x\, \e^{iq\cdot x}
\Psibra T\eta(x)\overline{\eta}(0)\Psiket\ ,
\label{corr_def}
\eeq
where $\eta(x)$ is a colorless interpolating field made up of
quark fields with the quantum numbers of a nucleon.
The ground
state of nuclear matter $\Psiket$ is characterized by the rest frame
nucleon density $\rhonuc$ and by the four-velocity $u^\mu$; it is
assumed to be invariant under parity and time reversal.
Formally, we work at fixed volume and baryon number until the
end, when we take the thermodynamic limit.

We can choose to work either at fixed density, with interpolating
fields in the Heisenberg picture as at zero density, or with a chemical
potential and grand canonical Heisenberg operators.
Since
we work exclusively at zero temperature here, the distinction between
working with the density and with a chemical potential is not critical.
However, there are several reasons to work in the Heisenberg picture:
\begin{enumerate}
\item  We wish to exploit the zero-density sum rules as a means of
normalizing the finite-density sum rules, so we want a smooth limit to zero
density.

\item  Our estimates for the finite-density matrix elements (condensates)
will give them as
functions of density, not chemical potential.

\item  The physics of the chemical potential will not be
reproduced by the sum rules.  That is,
the singularity structure signaled
by the chemical potential will not be reflected on the OPE side;
more simply, the sum rules don't ``know'' about the Fermi surface.

\end{enumerate}

We generalize the discussion of Sec.~\ref{sectii_iv} by
considering all
nucleon interpolating fields (currents) that contain no
derivatives and couple to spin-$1\over 2$
and isospin-$1\over 2$ states only.
There are two
linearly independent fields with these features corresponding to a
scalar or pseudoscalar up-down
diquark coupled to the remaining quark.
For the proton these
two independent interpolating fields are
\begin{eqnarray}
\eta_1(x)=\epsilon_{abc}[u_a^T(x) C \gamma_5 d_b(x)]u_c(x)\ ,
\\*
\eta_2(x)=\epsilon_{abc}[u_a^T(x) C d_b(x)]\gamma_5 u_c(x)\ ,
\end{eqnarray}
where $T$ denotes a transpose in Dirac space and $C$ is the charge
conjugation-matrix.
The analogous fields for the neutron follow by interchanging the
up and
down quark fields and changing the overall sign.

Here we define a linear combination of these two fields:
\begin{equation}
\eta(x)=2\left[t\eta_1(x) + \eta_2(x)\right] ,\label{def_ge}
\end{equation}
where $t$ is an arbitrary real parameter. The current with $t=-1$
corresponds to Ioffe's choice \cite{IOF81}, which was considered
in the discussion of the vacuum sum rule.
If the correlator were
calculated to arbitrary accuracy from both the QCD expansion
and the phenomenological
description, one would expect that the sum-rule
predictions of physical observables would be independent
of the choice of $t$. In practice, however, the QCD expansion is
truncated and the phenomenological description is represented crudely.
The criterion of choosing the interpolating field in QCD sum-rule
applications is to maximize the coupling of the interpolating field to
the desired physical intermediate state relative to other
(continuum) states, while minimizing the contributions of higher-order
terms in the OPE. These goals cannot be simultaneously realized. The
optimal choice of the nucleon interpolating field seems to be close to
Ioffe's choice.  However,
interpolating fields with $t\simeq -1.1$ have also been used in
nucleon sum-rule studies \cite{NAR89}, in particular, in studying
direct small-scale instanton effects in nucleon sum
rules \cite{DOR90,FOR93}. We
shall consider values in the range $-1.15\leq t\leq -0.95$ here
to evaluate the sensitivity of our predictions.

It is often remarked in the literature that one cannot work covariantly
at finite density or temperature because of the existence of a preferred
frame of reference, \ie\ the rest frame of nuclear matter.
This is a misconception.
While the ground state is not {\it invariant\/}
under all Lorentz transformations (unlike the vacuum state), matrix
elements in this state do have well-defined Lorentz transformation properties.
So two observers in different frames can still compare calculations or
observations as prescribed by special relativity.
The new feature is the additional four-vector $u_\mu$ that must
be transformed when making the comparisons
and must be included when building tensors  or identifying
invariant functions.
(The situation here
is analogous to considering diagonal matrix elements
of a spin-averaged proton state, which is also characterized by a
single four-vector, the four-momentum of the proton.)

The correlation function $\Pi(q)$
is a $4\times 4$ matrix in Dirac space, so
we can expand it in the usual complete set of Dirac matrices.
Using the transformation properties of $\eta(x)$
and keeping in mind the role of $u_\mu$,
we can constrain the form of $\Pi(q)$.
The arguments are analogous to those in Chapter~16 of Ref.~\cite{BJO65}.
Lorentz covariance dictates that the general form of the correlator
is
\begin{equation}
 \Pi(q) = \Pi_s   +  \Pi_q \qslash  + \Piu \uslash
   + \Pi_1 \gammafive
          + \Pi_2 \qslash\gammafive
          + \Pi_3 \uslash\gammafive
  + \Pi_4(q_\mu u_\nu - q_\nu u_\mu)\sigmamunu
   + \Pi_5 \epsilon_{\mu\nu\kappa\lambda}
       q^\kappa u^{\lambda} \sigmamunu \ ,
\end{equation}
where the $\Pi_i$'s are scalar functions of the invariants
$q^2$ and $\qdotu$.
We assume that the nuclear matter ground state is invariant under parity
and time reversal in its rest frame.
In a general frame, we must take
$u^\mu \rightarrow u_\mu$ as well as $q^\mu \rightarrow q_\mu$
under these transformations; thus $q^2$ and
$\qdotu$ are unchanged.
The parity constraint implies $\Pi_1 = \Pi_2 = \Pi_3 = 0$.
In the vacuum, a term proportional to $\sigmamunu$ can be excluded,
because it can only be contracted with the symmetric combination
$q_\mu q_\nu$.  In finite-density nuclear matter, this argument is no longer
sufficient, but the
assumed parity and time-reversal invariance implies
$\Pi_4 = \Pi_5 = 0$.
(See Ref.~\cite{RUS94} for a more complete discussion based on spectral
functions.)

Thus
Lorentz covariance, parity, and time reversal
imply that $\Pi(q)$ has the form
\begin{equation}
  \Pi(q)  \equiv  \Pi_s(\qsq,\qdotu)
         + \Pi_q(\qsq,\qdotu) \qslash
        +  \Pi_u(\qsq,\qdotu) \uslash
        \ .   \label{eq:decompose}
\end{equation}
%
There are {\it three\/} distinct structures---scalar,
$\qslash$, and $\uslash$---and thus
three invariant functions of the two scalars
$\qsq$ and $\qdotu$  (or any convenient
combination).
Recall that in the vacuum there are only two structures:
scalar and $\qslash$.
In the zero-density limit, $\Pi_u \rightarrow 0$,
and $\Pi_s$ and $\Pi_q$ become functions of $q^2$ alone.
For simplicity, we specialize the sum rules
to the rest frame of the nuclear medium, where the variables
$\qzero$ and $\vec q^2$ are most useful.
A covariant form can be recovered in general by repeating the
analysis with
$\qzero \rightarrow \qdotu$ and $-\vec q^2 \rightarrow
\widetilde q^2 \equiv \qsq - (\qdotu)^2$.

We can project out the individual invariant functions by taking
traces:
\begin{eqnarray}
      \Pi_s &=& \mbox{$1\over 4$}\Tr(\Pi)\ , \label{eq:tracei} \\[4pt]
    \Pi_q &=& {1\over q^2 - (\qdotu)^2}
        \Bigl[
    \mbox{$1\over 4$}\Tr(\qslash\Pi) - {\qdotu\over 4}\Tr(\uslash\Pi)
        \Bigr]  \ ,  \label{eq:traceii} \\[4pt]
    \Piu &=& {1\over q^2 - (\qdotu)^2}
        \Bigl[
    {q^2\over4}\Tr(\uslash\Pi) - {\qdotu\over 4}\Tr(\qslash\Pi)
        \Bigr] \ . \label{eq:traceiii}
\end{eqnarray}
These projections require that $q^2 - (\qdotu)^2$ be nonzero;
this means $\vec{q} \neq 0$ in the rest frame.
If this is not the case, there are only two functions, and the
second is projected by a trace with $\gammazero$.

\subsubsection{Dispersion Relation at Fixed Three-Momentum}

As discussed in Sec.~\ref{sectii_iv}, the fundamental physical
objects in the sum rules are the spectral
functions
\begin{eqnarray}
	\rho_{\alpha\beta}(q,u)&= & {1\over 2\pi}\int\!d^4x\ \e^{iq\cdot x}
	\langle\Psi_0|\eta_\alpha(x)\overline\eta_\beta(0)
		|\Psi_0\rangle \label{eq:rhodefinition} \ ,\\
 	\tilde\rho_{\alpha\beta}(q,u)&=&  {1\over 2\pi}
 	      \int\!d^4x\  \e^{iq\cdot x}
		\langle\Psi_0|\overline\eta_\beta(0)\eta_\alpha(x)
		|\Psi_0\rangle  \ .
		\label{eq:rhotildedefinition}
\end{eqnarray}
Because  the spectral functions are expectation values in an equilibrium
state, they are characterized by the density in the rest frame of the
medium and the medium
four-velocity $u^\mu$.%
\footnote{Relativistic thermodynamics
dictates that expectation values in a uniform medium in equilibrium are
characterized by a single four-velocity.
Since all equilibrium four-vectors such as the baryon current or the
four-momentum of the ground state will be proportional to $u^\mu$, we
could choose one of them as well.
However, since in the end we wish to take the thermodynamic limit, it is
most convenient to work with the four-velocity.}
The sum rules consist of weighted integrals in energy of the spectral
functions (with an analytic weighting function),
obtained from  the correlator through Borel transforms.
All of the various correlators (time-ordered, advanced, retarded)
have the same spectral functions, so it doesn't matter which correlator
we start with.

The spectral functions can be written as a sum over
a complete set of energy-momentum eigenstates.
For example (recall that $\eta$ is a Heisenberg picture operator),
\begin{eqnarray}
	\rho_{\alpha\beta}(q,u)&=&
	{1\over 2\pi}\sum_n\int\! d^4x\ \e^{iq\cdot x}\langle\Psi_0
		|\eta_\alpha(x)|n\rangle \langle n |
		\overline\eta_\beta(0) |\Psi_0 \rangle  \nonumber  \\
	& = &
	{1\over 2\pi}\sum_n\int\! d^4x\ \e^{iq\cdot x} \langle\Psi_0
		|\e^{i\widehat P\cdot x}\eta_\alpha(0)
		\e^{-i\widehat P\cdot x}|n \rangle
		\langle n |\overline\eta_\beta(0) |\Psi_0 \rangle
	\nonumber  \\
	& = &
	{1\over 2\pi}\sum_n\int\! d^4x\
	\e^{i(q\cdot x+P_0\cdot x-P_n\cdot x)}
		 \langle\Psi_0
		 |\eta_\alpha(0) |n \rangle
		 \langle n |\overline\eta_\beta(0) |\Psi_0 \rangle
       \nonumber \\
	& = &(2\pi)^3\sum_n\delta^{(4)}(q+P_0-P_n)
		 \langle\Psi_0
		 |\eta_\alpha(0) |n \rangle
		 \langle n |\overline\eta_\beta(0) |\Psi_0 \rangle
	\label{eq:rhodistribution}\ .
\end{eqnarray}
Here, $P_0^\mu$ is the ground state four-momentum, equal to the product of
the ground state
mass with the four-velocity $u^\mu$, and $P_n^\mu$ is the
four-momentum of the state $|n\rangle$.
Recall that we work at fixed volume and baryon number until the end,
when the thermodynamic limit is to be taken.

At zero density, the spectral densities $\rho$ and $\widetilde\rho$
(which include the nucleon and antinucleon states, respectively)
are
related by discrete space-time symmetries,
because the vacuum is invariant under these operations \cite{BJO65}.
In contrast, the finite-density ground state $|\Psi_0 \rangle$
is not invariant
under charge conjugation, so
there is no definite relationship between the functions
$\rho$ and $\widetilde\rho$ and, in particular,
the spectral densities for nucleon and antinucleon quasiparticles are not
simply related.

The spectral representation
of the correlator $\Pi(q)$ starts as an integral in
energy over the spectral densities at fixed three-momentum.
In the vacuum, where the invariant functions depend only on $q^2$,
the separation of $\qzero$ and $|\qvec|$ dependence is not
necessary (or particularly useful).
Also, the discrete symmetries require that the spectral density for
negative-energy states (corresponding, for example, to the antiparticle)
is equal (up to a sign) to the spectral density for positive-energy states.
These properties allow us to change the integration over energy to one over
$q^2$.
In contrast, the four-velocity of the medium $u_\mu$ makes the distinction
important in our case.

At finite density, we keep the dispersion relations as integrals over
$q_0$, with the three-momentum $|\qvec|$ held fixed.
This provides a clean identification of the intermediate quasinucleon
states, which are naturally labeled by $|\qvec|$.
The contribution from negative-energy quasinucleons (antinucleons)
is clearly separated, which lets us isolate to a large degree the
positive-energy quasinucleon contribution in the subsequent sum rule
by adopting an appropriate weighting function.

We can write a dispersion relation for
each of the Lorentz structures
$\Pi_i$ [$i=\{s,q,u\}$],
\beq
\Pi_i(q_0,|\vec{q}|)={1\over 2\pi
i}\int^{\infty}_{-\infty}d\omega{\Delta \Pi_i(\omega,|\vec{q}|)\over
\omega-q_0}\ ,  \label{des_re}
\eeq
where we have omitted possible polynomials arising from subtractions,
which will be eliminated by a subsequent Borel transform. We have also
omitted the infinitesimals (which distinguish time-ordered
and retarded correlators), since we will
not evaluate $q_0$ on the real axis. The discontinuity,
defined by
\beq
\Delta\Pi_i(\omega,|\vec{q}|)\equiv \lim_{\epsilon\rightarrow 0^+}
[\Pi_i(\omega+i\epsilon,|\vec{q}|)-\Pi_i(\omega-i\epsilon,|\vec{q}|)]\ ,
\eeq
contains the spectral information on the quasiparticle, quasihole, and
higher-energy states.
It is simply proportional to the sum  of the spectral functions
in Eqs.~(\ref{eq:rhodefinition}) and (\ref{eq:rhotildedefinition}) [taking
into account the tensor structure].
The correlator is analytic
except for cuts on the  real $q_0$ axis (which vanish at the chemical
potential) \cite{FET71}.
For $q_0$ off the real axis, one has
\begin{equation}
\Pi_i(q_0^*,|\qvec|)   =
   [\Pi_i(q_0,|\qvec|)]^*
   \ ,
\end{equation}
which relates the function in the upper and lower half planes.
The deep
spacelike (Euclidean) limit with $\qvec$ fixed takes $q_0 \rightarrow
i\infty$.

\subsubsection{Phenomenological Ansatz}

We  generalize the
usual zero-density ansatz for the spectral functions
by assuming a quasiparticle pole
for the nucleon,%
\footnote{Note that the sum rules do not distinguish between quasiparticles
and quasiholes.  The other nucleons only provide background fields to
be seen by the injected quarks at relatively short times, so
Pauli-principle correlations do not arise.}
with real self-energies (dependent on $\qvec$);
all higher-energy excitations are included in a
continuum contribution.
We use the notation of Sec.~\ref{sec:dirac}.
For infinite nuclear matter, where in the rest frame we can label a
quasinucleon state by its three-momentum, energy dependence of optical
potentials translates into three-momentum dependence of the self-energies.
At the mean-field level, there is no dependence; all quasinucleons see
the same self-energies.
But even in the most sophisticated Dirac Brueckner Hartree-Fock (DBHF)
calculations, the three-momentum dependence is weak
(at least for bound and low-lying continuum states).%

The quasiparticle-pole
contribution to the correlator is
\begin{equation}
  \Pi(q) \propto
   {1\over
     (q^\mu - \widetilde\Sigmav^\mu)\gammamul - (\MN + \widetilde\Sigmas) }
   \ ,
\end{equation}
where
$\widetilde\Sigmav^\mu$ and $\widetilde\Sigmas$
are  the in-medium self-energies.
In the language of the hadronic theories discussed in Sec.~\ref{sec:dirac},
these are the on-shell self-energies for a quasinucleon with
three-momentum $\qvec$.
The representations of the individual
invariant functions  are  (in the nuclear matter rest frame)
\begin{eqnarray}
  \Pi_s(q_0,|\qvec|) & = &
  -\lamNsq {\MNstar  \over
    (q_0 - E_q)(q_0 - \Eqbar)}
   + \cdots \ ,  \label{eq:tom} \\[4pt]
  \Pi_q(q_0,|\qvec|) & = &
  -\lamNsq {1  \over
    (q_0 - E_q)(q_0 - \Eqbar)}
    + \cdots \ ,   \label{eq:dick}   \\[4pt]
  \Pi_u(q_0,|\qvec|) & = &
  \lamNsq {\Sigmav \over
    (q_0 - E_q)(q_0 - \Eqbar)}
 + \cdots     \ ,  \label{eq:harry}
\end{eqnarray}
where we have defined $\MNstar$, $\Sigmav$, $E_q$, and $\Eqbar$ as in
Eqs.~(\ref{eq:MNstar}), (\ref{eq:SigmaV}), (\ref{eq:Eq}), and (\ref{eq:Eqbar})
and introduced an overall residue $\lamNsq$
that denotes the coupling of the interpolating field to the physical
quasinucleon state.
The positive- and negative-energy quasinucleon poles are explicit,
and $\cdots$ denotes the  contribution from higher-energy states,
which will be included later.
Recall that we expect this to be a reasonable ansatz for representing the
nucleon self-energy, but a poor ansatz for the antinucleon.
Differences in the residues between positive- and negative-energy states
are not important in our formalism, but can be critical in other
analyses \cite{FUR94}.

\subsubsection{OPE at Fixed Three-Momentum}

The next ingredient in our sum rule is an operator product
expansion (OPE)
of the
time-ordered product in Eq.~(\ref{corr_def}) at short distances.
The correlator is studied in the limit that $q_0$
becomes large and imaginary while $|\vec q|$ remains fixed
(in the nuclear matter rest frame).
This limit takes $q^2 \rightarrow -\infty$ with
$|q^2/q\cdot u|\rightarrow \infty$, which satisfies the conditions
discussed in Ref.~\cite{COL84} for a short distance expansion.%
\footnote{
Note that in Refs.~\cite{DRU90,DRU91,DRU94} the finite-density correlator
is studied using kinematics analogous to that of deep-inelastic scattering
({\it i.e.}, \ $q^2/q\cdot u$ is fixed and finite).
This ensures that only the light cone is probed, but does not imply a
short-distance expansion.
Furthermore, the identification of physical
quasinucleon intermediate states is
obscured.
}
It might appear that in this limit long-distance correlations in the spatial
distance are possible.
However, the singularities of the correlator lie on the light cone, so to
the extent that we are dominated by these singularities
(which are still in $q^2$ at finite density),
we have short
distance as well as short time in this limit.
We also apply Borel transforms in $q_0^2$ (or, equivalently, in $q^2$
with fixed $\vec q^2$),
which implies that only terms in the expansion that are discontinuous
across the
real $q_0$ axis contribute to the sum rules (for example, polynomials are
eliminated).

At finite density, the OPE for the invariant functions of the
nucleon correlator
take the general form
\begin{equation}
 \Pi_i(q^2,q\cdot u)=\sum_n C^i_n(q^2,q\cdot u)
       \langle{\widehat O_n}\rangle_{\rho_N}\ .
\label{blib2}
\end{equation}
(Recall
$\langle{\widehat O_n}\rangle_{\rho_N}
\equiv\langle\Psi_0 |{\widehat O_n}|\Psi_0\rangle$.)
The $C^i_n(q^2,q\cdot u)$ ($i=\{s,q,u\}$)
are the Wilson coefficients, which depend on QCD Lagrangian
parameters such as the quark masses and the strong coupling constant.
%
The Wilson coefficients in the OPE only depend on $q^\mu$, and
the ground-state expectation values of the operators are proportional to
tensors constructed from the nuclear matter four-velocity $u^\mu$,
the metric $g^{\mu\nu}$, and the antisymmetric tensor
$\epsilon^{\kappa\lambda\mu\nu}$.
In Eq.~(\ref{blib2}) we incorporate the contraction of $q^\mu$
(from the OPE) and $u^\mu$ (from the ground-state expectation values of the
operators) into the definition of the Wilson coefficients
$C^i_n(q^2,q\cdot u)$.
Thus the dependence on $q \cdot u$ is solely in the form of polynomial factors.
%
We have suppressed the dependence on the normalization point $\mu$.

The $\widehat O_n$
are local composite operators constructed from quark and gluon fields;
examples of such operators are $\overline q q$ and $(\alpha_s/\pi)G^2$.
The ground-state expectation values of these operators are the
in-medium condensates.
The operators are defined so that the density dependence of the correlator
resides solely in the in-medium condensates;
thus, the only substantial difference from
the vacuum calculations is that more composite operator matrix
elements are nonzero in nuclear matter.
The operators $\widehat O_n$ are ordered by dimension (measured as a power of
mass) and the $C^i_n(q^2,q\cdot u)$ for higher-dimensional operators
fall off by corresponding powers of $Q^2\equiv -q^2$.
Therefore, for sufficiently large $Q^2$, the operators of lowest dimension
dominate, and the OPE can be truncated after a small number of
lower-dimensional operators.

In the next subsections we consider in turn the calculation of  the
Wilson coefficients and the estimation of finite-density
condensates.

%
%
%
\subsubsection{Calculating Wilson Coefficients}
\label{wilson:coe}

We define the composite operators in the operator product expansion at
finite density using the same renormalization prescription and
subtractions as at zero density.
This means that the coefficients of the OPE,
the Wilson
coefficients, will be independent of density. One can thus calculate
the Wilson coefficients following the same methods developed for
the vacuum sum rules (see, for example, Ref.~\cite{REI85}).

Note, however, that expectation values of local operators with any integer
spin can be nonzero in medium due to the appearance of the additional
four-vector $u^\mu$. Consequently, there are a large number
of new condensates, which, while vanishing in vacuum due to Lorentz
invariance, will survive in medium.
Thus  one needs to calculate
the Wilson coefficients of the operators corresponding to
these new condensates.

Here we discuss some simple rules for the calculation
of Wilson coefficients using the fixed-point gauge and background-field
techniques, which have proven to be convenient for light-quark
systems. We will  assume that working to  lowest order
in perturbation theory is sufficient
to calculate the Wilson coefficients with acceptable accuracy.

By construction, the hadron interpolating fields are colorless; thus the
correlators of these interpolating fields
are gauge invariant. Therefore, one can evaluate the correlators in any
desired gauge. Here we adopt the fixed-point gauge, which was
introduced for use in electrodynamics in Refs.~\cite{FOC37,SCH70},
and reintroduced
for use in QCD in Ref.~\cite{CRO80}.  The fixed-point gauge condition is
\begin{equation}
x_\mu{\cal A}^\mu(x)=0\ ,
\label{gauge_condition}
\end{equation}
where ${\cal A}^\mu\equiv A^{A\mu}t^A$, with
$A^{A\mu}$ the gluon field and $t^A\equiv\lambda^A/2$
the color SU(3) generators in the fundamental representation.
In this gauge, the gluon field ${\cal A}_\nu$ can be expressed directly in
terms of the gluon field tensor
${\cal G}_{\mu\nu}$ \cite{CRO80,SMI82,HUB82,NOV84b,SHI80,REI85}:
\begin{equation}
{\cal A}_\nu(x)=\int_0^1 d\alpha\,\alpha x^\mu{\cal G}_{\mu\nu}(\alpha x)
=\mbox{$1\over 2$}x^\mu{\cal G}_{\mu\nu}(0)
+\mbox{$1\over 3$}x^\lambda x^\mu(D_\lambda{\cal G}_{\mu\nu})_{x=0}
+\cdots\ ,
\label{gluon_expand}
\end{equation}
where ${\cal G}_{\mu\nu}\equiv G^A_{\mu\nu}t^A\equiv D_\mu
{\cal A}_\nu-D_\nu{\cal A}_\mu$, with $D_\mu\equiv\partial_\mu
-ig_s{\cal A}_\mu$ the covariant derivative.
This allows one to obtain manifestly gauge-invariant results in a relatively
simple way.

In the background-field method \cite{SMI82,HUB82,NOV84b,SHI80,REI85},
the presence of nonperturbative quark and
gluon condensates is parameterized by Grassmann background quark fields,
$\chi^q_{a\alpha}$ and $\overline{\chi}^q_{a\alpha}$, and a classical
background gluon field $F^A_{\mu\nu}$.
It is most convenient to first work
in coordinate space and then transform to momentum space.
The coordinate-space quark propagator for light quarks in the presence of
the background fields takes the following form in the
fixed-point gauge\cite{REI85}:
\begin{eqnarray}
S^q_{ab,\alpha\beta}(x,0)
&\equiv&\langle Tq_{a\alpha}(x)\overline{q}_{b\beta}(0)\rangle_{\rho_N}
  \nonumber\\*[7.2pt]
&=&{i\over 2\pi^2}\delta_{ab}{1\over(x^2)^2}[\rlap{/}x]_{\alpha\beta}
-{im_q\over 4\pi^2}\delta_{ab}{\delta_{\alpha\beta}\over x^2}
\nonumber\\*[7.2pt]
& &\hspace{0.125in}\null
+\chi^q_{a\alpha}(x)\overline{\chi}^q_{b\beta}(0)
-{ig_s\over 32\pi^2}F^A_{\mu\nu}(0)t^A_{ab}
{1\over x^2}[\rlap{/}x\sigma^{\mu\nu}+\sigma^{\mu\nu}\rlap{/}x]_{\alpha\beta}
+\cdots\ ,
\label{bfprop}
\end{eqnarray}
where the first and  second terms are the expansion of the
free quark propagator to first order in the quark mass, and the third and
fourth
terms are the contributions due to the background quark and gluon fields,
respectively.
%
%
%
%
The gluonic contribution to Eq.~(\ref{bfprop}) comes from a single
gluon insertion, retaining only the leading term in the short-distance
expansion of the gluon field [see Eq.~(\ref{gluon_expand})].
Contributions from derivatives of the gluon field tensor,
which are less singular at $x\rightarrow 0$,
and additional gluon insertions are not included.
These refinements are not needed given the level of truncation in the OPE
considered in this review.

Products of Grassmann background quark fields and classical background
gluon fields obtained in Eq.~(\ref{bfprop}) correspond to ground-state matrix
elements of the corresponding quark and gluon operators:
\begin{equation}
\begin{array}{l}
\chi^q_{a\alpha}(x)\overline{\chi}^q_{b\beta}(0)
=\langle q_{a\alpha}(x)\overline{q}_{b\beta}(0)\rangle_{\rho_N}\ ,
%
\qquad
F^A_{\kappa\lambda}F^B_{\mu\nu}
=\langle G^A_{\kappa\lambda} G^B_{\mu\nu}\rangle_{\rho_N}\ ,
\\[14.4pt]
\chi^q_{a\alpha}\overline{\chi}^q_{b\beta}F^A_{\mu\nu}
=\langle q_{a\alpha}\overline{q}_{b\beta}G^A_{\mu\nu}\rangle_{\rho_N}\ ,
%
\qquad
\chi^q_{a\alpha}\overline{\chi}^q_{b\beta}\chi^q_{c\gamma}
\overline{\chi}^q_{d\delta}
=\langle q_{a\alpha}\overline{q}_{b\beta}q_{c\gamma}
\overline{q}_{d\delta}\rangle_{\rho_N}\ ,
\\
\end{array}
\label{bg_to_me}
\end{equation}
where the fields are evaluated at $x=0$ unless otherwise noted.
In Eq.~(\ref{bg_to_me}), we have only shown those matrix elements that are
needed in order to carry out the OPE to the level we are considering.
Thus we evaluate the fields in the higher-dimensional operators at the same
point, since nonlocalities would only introduce condensates that are higher
in dimension than those we wish to consider.
The composite operators in Eq.~(\ref{bg_to_me}) are implicitly
normal-ordered with respect to the perturbative vacuum at zero density;
we can write these matrix elements in terms of scalar local
condensates by  projecting out the Dirac, Lorentz, and color structure and
performing a short-distance expansion if necessary.
We discuss this procedure in detail; it is through this
discussion that we introduce the relevant condensates for sum-rule
calculations.

The Dirac and color structure of the matrix element
$\langle q_{a\alpha}(x)\overline{q}_{b\beta}(0)\rangle_{\rho_N}$
can be projected out to obtain
\begin{equation}
\langle q_{a\alpha}(x)\overline{q}_{b\beta}(0)\rangle_{\rho_N}
=-{\delta_{ab}\over 12}
\left[\langle \overline{q}(0)q(x)\rangle_{\rho_N}\delta_{\alpha\beta}
+\langle\overline{q}(0)\gamma_\lambda q(x)\rangle_{\rho_N}
\gamma^\lambda_{\alpha\beta}\right]\ ,
\label{fierz}
\end{equation}
since nuclear matter is colorless and
the ground state is (assumed to be) invariant under parity and time reversal.
(Other matrix elements of the form $\langle\overline{q}(0)\Gamma
q(x)\rangle_{\rho_N}$ do not
contribute due to parity and/or time reversal.)
We evaluate Eq.~(\ref{fierz}) at short distances by
expanding the quark field $q(x)$ in a Taylor series:
\begin{equation}
q(x)=q(0)
+x^\mu (\partial_\mu q)_{x=0}
+\mbox{$1\over 2$}x^\mu x^\nu (\partial_\mu\partial_\nu q)_{x=0}
+\cdots\ .
\label{quark_taylor}
\end{equation}
However, since the correlator is gauge invariant, the ordinary
derivatives in
Eq.~(\ref{quark_taylor}) must ultimately become covariant derivatives.
In standard calculations, gluon fields in higher-order terms of the OPE
combine with the
ordinary derivatives in lower-order terms to form covariant derivatives.

The situation is much more straightforward in the fixed-point gauge;
the ordinary derivatives can be replaced with covariant derivatives
immediately.
We follow the discussion of Ref.~\cite{NOV84b}.
Using the fixed-point gauge condition in Eq.~(\ref{gauge_condition}),
and expanding the gluon field, one obtains
\begin{equation}
x^\nu{\cal A}_\nu(0)
+x^\mu x^\nu (\partial_\mu{\cal A}_\nu)_{x=0}
+\mbox{$1\over 2$}x^\lambda x^\mu x^\nu
(\partial_\lambda\partial_\mu{\cal A}_\nu)_{x=0}
+\cdots
=0\ .
\label{expand}
\end{equation}
Since $x$ is arbitrary, the individual terms of Eq.~(\ref{expand}) must vanish;
using this fact, one can readily show
\begin{equation}
x^\mu (D_\mu q)_{x=0}
=x^\mu (\partial_\mu q)_{x=0}\ ,
\hspace{0.3in}
x^\mu x^\nu (D_\mu D_\nu q)_{x=0}
=x^\mu x^\nu (\partial_\mu\partial_\nu q)_{x=0}\ ,
\end{equation}
and so on.
Combining this result with Eq.~(\ref{quark_taylor}),
one derives the following
covariant Taylor expansion:
\begin{equation}
q(x)=q(0)
+x^\mu (D_\mu q)_{x=0}
+\mbox{$1\over 2$}x^\mu x^\nu (D_\mu D_\nu q)_{x=0}
+\cdots\ .
\end{equation}
An analogous expansion of the gluon field tensor at short
distances [used in Eq.~(\ref{gluon_expand})] is proved to all orders
using mathematical induction in Ref.~\cite{SHI80}.
Therefore, we obtain
\begin{eqnarray}
\langle q_{a\alpha}(x)\overline{q}_{b\beta}(0)\rangle_{\rho_N}
&=&-{\delta_{ab}\over 12}
[(\langle\overline{q}q\rangle_{\rho_N}
+x^\mu\langle\overline{q}D_\mu q\rangle_{\rho_N}
+\mbox{$1\over 2$}
x^\mu x^\nu\langle\overline{q}D_\mu D_\nu q\rangle_{\rho_N}
+\cdots)\delta_{\alpha\beta}
\nonumber\\*[7.2pt]
& &\null
+(\langle\overline{q}\gamma_\lambda q\rangle_{\rho_N}
+x^\mu\langle\overline{q}\gamma_\lambda D_\mu q\rangle_{\rho_N}
+\mbox{$1\over 2$}
x^\mu x^\nu\langle\overline{q}\gamma_\lambda D_\mu D_\nu q\rangle_{\rho_N}
+\cdots)\gamma^\lambda_{\alpha\beta}]\ ,
\label{contract}
\end{eqnarray}
where all fields and field derivatives in the condensates are evaluated
at $x=0$.

It is useful to note one particular calculational convenience:
Since the Dirac matrices involved in calculating the Wilson coefficient of
$\langle\overline{q}D_{\mu_1}\cdots D_{\mu_n}q\rangle_{\rho_N}$
[$\langle\overline{q}\gamma_\mu D_{\mu_1}
\cdots D_{\mu_n}q\rangle_{\rho_N}$]
are the same as those involved in
calculating the Wilson coefficient of $\langle\overline{q}q\rangle_{\rho_N}$
[$\langle\overline{q}\gamma_\mu q\rangle_{\rho_N}$],
we conclude that the coordinate-space coefficients are related as
follows:
\begin{eqnarray}
C_{\overline{q}D_{\mu_1}\cdots D_{\mu_n}q}(x)
&=&{1\over n!}x^{\mu_1}\cdots x^{\mu_n}C_{\overline{q}q}(x)\ ,
\\*[7.2pt]
C_{\overline{q}\gamma_\mu D_{\mu_1}\cdots D_{\mu_n}q}(x)
&=&{1\over n!}x^{\mu_1}\cdots x^{\mu_n}C_{\overline{q}\gamma_\mu q}(x)\ .
\end{eqnarray}
This implies that the momentum-space Wilson coefficients are related by
\begin{eqnarray}
C_{\overline{q} D_{\mu_1}\cdots D_{\mu_n}q}(q)
&=&{(-i)^n\over n!}
\left({\partial\over\partial q_{\mu_1}}\cdots
      {\partial\over\partial q_{\mu_n}}\right)
C_{\overline{q}q}(q)\ ,
\\*[7.2pt]
C_{\overline{q}\gamma_\mu D_{\mu_1}\cdots D_{\mu_n}q}(q)
&=&{(-i)^n\over n!}
\left({\partial\over\partial q_{\mu_1}}\cdots
      {\partial\over\partial q_{\mu_n}}\right)
C_{\overline{q}\gamma_\mu q}(q)\ .
\end{eqnarray}

We now proceed to evaluate the condensates appearing in Eq.~(\ref{contract})
in terms of expectation values of scalar operators multiplied by quantities
that contain the Lorentz structure of the original condensates.
In vacuum, these condensates can only be expressed in terms of the metric
tensor $g^{\mu\nu}$ and the antisymmetric tensor
$\epsilon^{\kappa\lambda\mu\nu}$;
thus condensates with an odd number of uncontracted
Lorentz indices must vanish in the vacuum.
In-medium condensates, however, can also be expressed in terms of the nuclear
matter four-velocity $u^\mu$, which leads to new condensates
and new Lorentz structures.
The general procedure for evaluating the condensates in Eq.~(\ref{contract})
is to write each as a sum of all possible Lorentz structures
with unknown coefficients.
These coefficients, which will turn out to be expectation values of
scalar operators, can then be determined by taking appropriate traces over
the Lorentz indices. Following this procedure, one can express various
condensates in Eq.~(\ref{contract}) as \cite{JIN93,JIN93b}
\begin{eqnarray}
& &\langle\overline{q}\gamma_\mu q\rangle_{\rho_N}=\langle\overline{q}
      \rlap{/}u
q\rangle_{\rho_N}u_\mu\ ,
\\*[7.2pt]
& &\langle\overline{q}D_\mu q\rangle_{\rho_N}
   =\langle\overline{q}u\cdot Dq\rangle_{\rho_N}u_\mu
  = -im_q \langle\overline{q}
      \uslash q\rangle_{\rho_N}u_\mu\ ,  \label{eq:2nd}
\\[7.2pt]
& &\langle\overline{q}\gamma_\mu D_\nu q\rangle_{\rho_N}
=\mbox{$4\over 3$}\langle\overline{q}\rlap{/}u u\cdot D q\rangle_{\rho_N}
(u_\mu u_\nu-\mbox{$1\over 4$}g_{\mu\nu})
+{i\over 3}m_q\langle\overline{q}q\rangle_{\rho_N}(u_\mu u_\nu-g_{\mu\nu})\ ,
\label{dim4_2index}
\\[7.2pt]
& &\langle\overline{q}D_\mu D_\nu q\rangle_{\rho_N}
=\mbox{$4\over 3$}\langle\overline{q} u\cdot D\, u\cdot D
q\rangle_{\rho_N}
(u_\mu u_\nu-\mbox{$1\over 4$}g_{\mu\nu})
-\mbox{$1\over 6$}\langle g_s \overline{q}\sigma\cdot{\cal G}q\rangle_{\rho_N}
(u_\mu u_\nu-g_{\mu\nu})\ ,
\label{dim5_2index}
\\[7.2pt]
& &\langle\overline{q}\gamma_\lambda D_\mu D_\nu q\rangle_{\rho_N}
=2\langle\overline{q}\rlap{/}u u\cdot D\, u\cdot Dq\rangle_{\rho_N}
[u_\lambda u_\mu u_\nu
-\mbox{$1\over 6$}
(u_\lambda g_{\mu\nu}+u_\mu g_{\lambda\nu}+u_\nu g_{\lambda\mu})]
\nonumber\\*[7.2pt]
& &\phantom{\langle\overline{q}\gamma_\lambda D_\mu D_\nu q\rangle_{\rho_N}=}
\null
-\mbox{$1\over 6$}
\langle g_s\overline{q}\rlap/{u}\sigma\cdot{\cal G}q\rangle_{\rho_N}
(u_\lambda u_\mu u_\nu-u_\lambda g_{\mu\nu})\ ,
\end{eqnarray}
where the equations of motion have been used,%
\footnote{To obtain the second equality in Eq.~(\ref{eq:2nd}), we use the
identity $D_\mu\equiv\mbox{$1\over 2$}
(\gamma_\mu\rlap{\,/}{D}+\rlap{\,/}{D}\gamma_\mu)$
and translation invariance, which implies
$\langle\overline{q}i\rlap{\,/}{D}\rlap/{u}q\rangle_{\rho_N}
=-\langle\overline{q}i\DslashL\rlap/{u}q\rangle_{\rho_N}$.}
and $O(m_q^2)$
terms have been neglected.
Thus the expansion of $\langle q_{a\alpha}(0)\overline{q}_{b\beta}(x)
\rangle_{\rho_N}$ up to dimension
five includes quark condensates and quark-gluon condensates.

Another source of quark-gluon condensates is from contributions of the form
$\chi^q_{a\alpha}\overline{\chi}^q_{b\beta}F^A_{\mu\nu}$ in
Eq.~(\ref{bfprop}).
The corresponding matrix element can be decomposed as
\begin{eqnarray}
\langle g_s q_{a\alpha}\overline{q}_{b\beta}
G^A_{\mu\nu}\rangle_{\rho_N}
&=&-{t^A_{ab}\over 96}
\biggl\{
\langle g_s\overline{q}\sigma\cdot{\cal G}q\rangle_{\rho_N}
\Bigl[\sigma_{\mu\nu}
+i(u_\mu\gamma_\nu-u_\nu\gamma_\mu)\rlap{/}u\Bigr]_{\alpha\beta}
\nonumber\\*[7.2pt]
& &\hspace*{-1in}\null
+\langle g_s\overline{q}\rlap/{u}\sigma\cdot{\cal G} q\rangle_{\rho_N}
\Bigl[\sigma_{\mu\nu}\rlap{/}u
+i(u_\mu\gamma_\nu-u_\nu\gamma_\mu)\Bigr]_{\alpha\beta}
\nonumber\\*[7.2pt]
& &\hspace*{-1in}\null
-4\Bigl(\langle\overline{q} u\cdot D\, u\cdot D q\rangle_{\rho_N}
+im_q\langle\overline{q}\rlap{/}u u\cdot Dq\rangle_{\rho_N}\Bigr)
\Bigl[\sigma_{\mu\nu}
+2i(u_\mu\gamma_\nu-u_\nu\gamma_\mu)\rlap{/}u\Bigr]_{\alpha\beta}
\biggr\}\ ,
\label{big_pain}
\end{eqnarray}
which is obtained by projecting out the color, Dirac, and Lorentz structure
by taking appropriate traces.
Many of the ``condensates'' encountered in the derivation of
Eq.~(\ref{big_pain}) vanish due to the assumed parity and time-reversal
invariance of the nuclear matter ground state.

The dimension-four
gluon condensates arise from factors of
$F^A_{\kappa\lambda}F^B_{\mu\nu}$ in Eq.~(\ref{bfprop}).%
\footnote{Here, in the calculation of the Wilson coefficients
multiplying the gluon condensates, we neglect the quark masses.
Note that the factor $F^A_{\kappa\lambda}F^B_{\mu\nu}$
comes from single
gluon insertions in two of the quark propagators.
One must also consider  two gluon lines
emanating from the same quark propagator, which gives contributions
proportional
to $\langle(\alpha_s/\pi)(\vec{E}^2+\vec{B}^2)\rangle_{\rho_N}$
in the massless quark limit. Since this condensate
only gives very small contributions to the baryon sum rules,
for simplicity we omit these terms entirely.
Note that the corresponding coefficients in Refs.~\cite{JIN93}
and \cite{JIN94}
are missing the two-gluon-line contributions.}
The matrix element $\langle G^A_{\kappa\lambda}
G^B_{\mu\nu}\rangle_{\rho_N}$
can be written in terms of two independent gluon condensates,
which, in the rest frame, are proportional to
$\langle\vec{E}^2-\vec{B}^2 \rangle_{\rho_N}
=-{1\over 2}\langle G^2\rangle_{\rho_N}$
and
$\langle\vec{E}^2+\vec{B}^2\rangle_{\rho_N}$,
where $\vec{E}^A$ and $\vec{B}^A$ are the color-electric
and color-magnetic fields.
Contributions from
$\langle\vec{E}^2+\vec{B}^2\rangle_{\rho_N}$
are small and are neglected in the sequel.
Thus, we are left with
\begin{equation}
\langle G^A_{\kappa\lambda}
G^B_{\mu\nu}\rangle_{\rho_N}
={\delta^{AB}\over 96}
[\langle G^2\rangle_{\rho_N}
(g_{\kappa\mu}g_{\lambda\nu}-g_{\kappa\nu}g_{\lambda\mu})
  + O(\langle\vec{E}^2+\vec{B}^2)\rangle_{\rho_N})]
\ .
\label{little_pain}
\end{equation}

At dimension six we only consider the four-quark condensates.
The leading-order four-quark condensate contributions to the baryon correlators
arise at tree level; thus they do not carry the suppression factors associated
with loops. Such contributions appear as terms proportional to
$\chi^u_{a\alpha}\overline{\chi}^u_{b\beta}
\chi^d_{c\gamma}\overline{\chi}^d_{d\delta}$, for example.
The Lorentz, Dirac, and color structure of these matrix elements can be
projected out in a manner similar to that discussed above. There are,
in general,
a large number of four-quark condensates contributing to
the baryon correlators.
If one assumes in-medium factorization (or ground-state saturation),
these four-quark condensates can be expressed
in terms of a few dimension-three quark
condensates.  The basic factorization formulas
are  \cite{JIN93,JIN93b}
\begin{eqnarray}
& &\langle\overline{u}_{a\alpha}u_{b\beta}
\overline{u}_{c\gamma}u_{d\delta}\rangle_{\rho_N}
\simeq\langle\overline{u}_{a\alpha}u_{b\beta}\rangle_{\rho_N}
\langle\overline{u}_{c\gamma}u_{d\delta}\rangle_{\rho_N}
-\langle\overline{u}_{a\alpha}u_{d\delta}\rangle_{\rho_N}
\langle\overline{u}_{c\gamma}u_{b\beta}\rangle_{\rho_N}\ ,
\label{factorize1}
\\*[7.2pt]
& &\langle\overline{u}_{a\alpha}u_{b\beta}\
\overline{d}_{c\gamma}d_{d\delta}\rangle_{\rho_N}
\simeq\langle\overline{u}_{a\alpha}u_{b\beta}\rangle_{\rho_N}
\langle\overline{d}_{c\gamma}d_{d\delta}\rangle_{\rho_N}\ .
\label{factorize2}
\end{eqnarray}
{\em However,  factorization has not
been justified for nuclear matter and may well be wrong\/}
\cite{FUR92,JIN94a}.

Summarizing the  rules for calculating Wilson coefficients:
\begin{enumerate}
\item Apply Wick's theorem to the time-ordered product in
the correlator, retaining only those contributions in which the
quark fields are fully contracted;
\item use the background quark propagator [Eq.~(\ref{bfprop})],
rather than the free quark propagator, for each contraction;
\item express the various condensates in terms of scalar operators
multiplied by appropriate tensors.
\end{enumerate}
In order to transform the expression to momentum space one can use
the formulas \cite{NOV84b,REI85}
\begin{eqnarray}
\int{d^4 x\over x^2}\e^{iq\cdot x}&=&-{4\pi^2 i\over q^2}\ ,
\\*
\int{d^4 x \over (x^2)^n}\e^{iq\cdot x}
&=&{i(-1)^n 2^{4-2n}\pi^2\over\Gamma(n-1)\Gamma(n)}
(q^2)^{n-2}\ln(-q^2)
+P_{n-2}(q^2)
\hspace{0.5in}(n\geq 2)\ ,
\end{eqnarray}
and their derivatives with respect to $q^\mu$.
$P_m(q^2)$ is a polynomial in $q^2$ of degree $m$ with divergent
coefficients.
The precise forms of the polynomials are not important, since they do not
contribute to the Borel transformed sum rules. Following this procedure,
one gets the whole OPE expression for the correlator,
from which one can easily identify the Wilson coefficients.%
%
%

One can now apply these simple rules to evaluate the nucleon correlator.
For convenience we separate the invariant functions into pieces that
are even and odd in $q_0$:
\begin{equation}
\Pi_i(q_0,|\vec{q}|)=\Pi_i^{\mbox{\tiny\rm E}}
(q_0^2,|\vec{q}|)+q_0\Pi_i^{\mbox{\tiny\rm O}}(q_0^2,|\vec{q}|)\ .
\label{inv_sap}
\end{equation}
The full results for the general interpolating field of Eq.~(\ref{def_ge})
are given in Ref.~\cite{JIN94}; here we list the dominant terms:
\begin{eqnarray}
\Pi_s^{\mbox{\tiny\rm E}}
&=&{c_1\over 16\pi^2}q^2\ln(-q^2)\langle\overline{q}q\rangle_{\rho_N}
   +{3c_2\over 16\pi^2}\ln(-q^2)
   \langle g_s\overline{q}\sigma\!\cdot\!{\cal G}q\rangle_{\rho_N}
\nonumber\\*
& &\null
   +{2c_3\over 3\pi^2}{q_0^2\over q^2}
   (\langle\overline{q} iD_0 iD_0 q\rangle_{\rho_N}+\mbox{$1 \over 8$}
   \langle g_s\overline{q}\sigma\!\cdot\!{\cal G}q\rangle_{\rho_N})
     + \cdots \ ,
\label{PisE_ope}
\\*
\Pi_s^{\mbox{\tiny\rm O}}
&=&-{c_1\over 8\pi^2}\ln(-q^2)\langle\overline{q} iD_0 q\rangle_{\rho_N}
   -{c_1\over 3q^2}\langle\overline{q}q\rangle_{\rho_N}
   \langle q^\dagger q\rangle_{\rho_N} + \cdots \ ,
\label{PisO_ope}
\\
\Pi_q^{\mbox{\tiny\rm E}}
&=&-{c_4\over 512\pi^4}(q^2)^2\ln(-q^2)
   +{c_4\over72\pi^2}\left(5\ln(-q^2)-{8q_0^2\over q^2}\right)
   \langle q^\dagger iD_0 q\rangle_{\rho_N}
\nonumber\\*
& &\null
   -{c_4\over 256\pi^2}\ln(-q^2)
   \left<{\alpha_s\over\pi}G^2\right>_{\rho_N}
   -{c_1\over 6q^2}\langle\overline{q}q\rangle_{\rho_N}^2
   -{c_4\over 6q^2}\langle q^\dagger q\rangle_{\rho_N}^2 + \cdots \ ,
\label{PiqE_ope}
\\*
\Pi_q^{\mbox{\tiny\rm O}}
&=&{c_4\over 24\pi^2}\ln(-q^2)\langle q^\dagger q\rangle_{\rho_N}
   +{c_5\over 72\pi^2 q^2}
   \langle g_s q^\dagger\sigma\!\cdot\!{\cal G}q\rangle_{\rho_N}
\nonumber\\*
& &\null
   -{c_4\over 12\pi^2 q^2}\left(1+{2q_0^2\over q^2}\right)
   (\langle q^\dagger iD_0 iD_0 q\rangle_{\rho_N}+\mbox{$1 \over 12$}
   \langle g_s q^\dagger\sigma\!\cdot\!{\cal G}q\rangle_{\rho_N})
        + \cdots \ ,
\label{PiqO_ope}
\\
\Pi_u^{\mbox{\tiny\rm E}}
&=&{c_4\over 12\pi^2}q^2\ln(-q^2)\langle q^\dagger q\rangle_{\rho_N}
   -{c_5\over 48\pi^2}\ln(-q^2)
   \langle g_s q^\dagger\sigma\!\cdot\!{\cal G}q\rangle_{\rho_N}
\nonumber\\*
& &\null
   +{c_4\over2\pi^2}{q_0^2\over q^2}
   (\langle q^\dagger iD_0 iD_0 q\rangle_{\rho_N}+\mbox{$1 \over 12$}
   \langle g_s q^\dagger\sigma\!\cdot\!{\cal G}q\rangle_{\rho_N})
     + \cdots \ ,
\label{PiuE_ope}
\\*
\Pi_u^{\mbox{\tiny\rm O}}
&=&-{5c_4\over 18\pi^2}\ln(-q^2)\langle q^\dagger iD_0 q\rangle_{\rho_N}
   -{c_4\over 3q^2}\langle q^\dagger q\rangle_{\rho_N}^2 + \cdots \ ,
\label{PiuO_ope}
\end{eqnarray}
where all polynomials in $q^2$ and $q_0^2$, which vanish under the
Borel transform, have been omitted, and the terms proportional
to quark masses have been neglected. Note that the contributions
from four-quark condensates are included in factorized
form. We work in
the rest frame of the nuclear matter and have defined
\begin{eqnarray}
c_1 &=& 7t^2-2t-5\ ,  \qquad\quad
c_2 = 1-t^2\ ,  \qquad\quad
c_3 = 2t^2-t-1\ ,
\\*[4pt]
c_4&=&5t^2+2t+5\ , \qquad\quad
c_5 = 7t^2+10t+7\ .
\label{t_dep}
\end{eqnarray}

\subsection{Estimating QCD Condensates}
\label{vacuum}

\subsubsection{Vacuum Condensates}

Before considering finite-density condensates, we review
standard estimates of vacuum condensates.
The Lorentz invariance of the vacuum state $|0\rangle$ dictates that
only spin-0 operators can have nonvanishing vacuum expectation values;
the lowest-dimensional vacuum condensates are
\begin{equation}
\begin{array}{ll}
\langle\overline{q}q\rangle_{\rm vac}\ ,
\langle\overline{s}s\rangle_{\rm vac}\ ,&
\hspace{1in}d=3\ ,\\[14.4pt]
\left<{\displaystyle{\alpha_s\over\pi}}G^2\right>_{\rm vac}\ ,&
\hspace{1in}d=4\ ,\\[14.4pt]
\langle g_s\overline{q}\sigma\cdot{\cal G}q\rangle_{\rm vac}\ ,
\langle g_s\overline{s}\sigma\cdot{\cal G}s\rangle_{\rm vac}\ ,&
\hspace{1in}d=5\ ,\\[14.4pt]
\langle\overline{u}\Gamma_1 u\,\overline{u}\Gamma_2 u\rangle_{\rm vac}\ ,
\langle\overline{u}\Gamma_1\lambda^A u\,
\overline{u}\Gamma_2\lambda^A u\rangle_{\rm vac}\ ,
\ldots\ ,&
\hspace{1in}d=6\ ,\\[14.4pt]
\langle g_s^3 fG^3\rangle_{\rm vac}\ ,&
\hspace{1in}d=6\ ,\\
\label{condensate_list}
\end{array}
\end{equation}
where $d$ denotes the mass dimension of the condensate.
We take $\sigma_{\mu\nu}\equiv i[\gamma_\mu,\gamma_\nu]/2$;
we have used the notation
$\langle\widehat{O}\rangle_{\rm vac}
\equiv\langle{0}|\widehat{O}|{0}\rangle$,
$G^2\equiv G^A_{\mu\nu}G^{A\mu\nu}$,
$\sigma\cdot{\cal G}\equiv\sigma_{\mu\nu}{\cal G}^{\mu\nu}$, and
$f G^3\equiv f^{ABC}G_\lambda^{A\mu}G_\mu^{B\nu}G_\nu^{C\lambda}$.
Other condensates, such as $\langle\overline{q}D^2 q\rangle_{\rm vac}$,
can be related to those
listed in Eq.~(\ref{condensate_list}) by using the field equations.
We will omit discussion of
$\langle g_s\overline{s}\sigma\cdot{\cal G}s\rangle_{\rm vac}$ and
$\langle g_s^3 fG^3\rangle_{\rm vac}$,
which will not play a role in our sum rules.

We first consider the quark condensate
$\langle\overline{q}q\rangle_{\rm vac}$.
Due to isospin symmetry one has
\begin{equation}
\langle\overline{q}q\rangle_{\rm vac}
\simeq\langle\overline{u}u\rangle_{\rm vac}
\simeq\langle\overline{d}d\rangle_{\rm vac}\ .
\end{equation}
The numerical value of $\langle\overline{q}q\rangle_{\rm vac}$ can be
determined from the Gell-Mann--Oakes--Renner relation,
\begin{equation}
(m_u+m_d)\langle\overline{q}q\rangle_{\rm vac}
=-m_\pi^2 f_\pi^2\,[1+O(m_\pi^2)]\ ,
\label{GMOR}
\end{equation}
where $m_\pi$ and $f_\pi$ are the pion mass and pion decay constant, and
$m_u$ and $m_d$ are the up and down current quark masses.
Both sides of Eq.~(\ref{GMOR}) are renormalization-group invariant
\cite{TAR82};
therefore, given the current quark masses at a particular renormalization
scale, one can determine the quark condensate at that same scale.
We take $m_\pi=138\,\mbox{MeV}$ and $f_\pi=93\,\mbox{MeV}$;
using the standard values of
the light quark masses, one obtains $m_u+m_d=14\pm 4\,\mbox{MeV}$
at a renormalization scale of $1\,\mbox{GeV}$ \cite{GAS82}.
Thus one has
\begin{equation}
\langle\overline{q}q\rangle_{\rm vac}
\simeq -(0.225\pm 0.025\,\mbox{GeV})^3
\end{equation}
at a renormalization scale of $1\,\mbox{GeV}$ \cite{GAS82}.
The value of the strange quark condensate is usually specified in terms
of the up and down quark condensate; we take
\begin{equation}
\langle\overline{s}s\rangle_{\rm vac}
=f_s\langle\overline{q}q\rangle_{\rm vac}\ ,
\end{equation}
with $f_s\simeq 0.8$ \cite{BEL82,REI85,LEI90}.

The gluon condensate was first estimated from an analysis of leptonic decays
of $\rho^0$ and $\phi^0$ mesons \cite{VAI78} and from a sum-rule
analysis of the charmonium spectrum \cite{SHI79}.
Its numerical value is taken to be \cite{SHU88}
\begin{equation}
\left<{\alpha_s\over\pi}G^2\right>_{\rm vac}
\simeq(0.33\pm 0.04\,\mbox{GeV})^4\ .
\label{gluoncon_num}
\end{equation}
(Also see Ref.~\cite{SHU88} for a discussion of lattice QCD extractions
of the gluon condensate.)
Note that the product $(\alpha_s/\pi)G^2$ is approximately
renormalization-group
invariant;
violations of renormalization-group invariance are of higher order
in $\alpha_s$ \cite{TAR82}.

The quark-gluon condensate
$\langle g_s\overline{q}\sigma\cdot{\cal G}q\rangle_{\rm vac}$
is expressed in terms of the quark
condensate $\langle\overline{q}q\rangle_{\rm vac}$:
\begin{equation}
\langle g_s\overline{q}\sigma\cdot{\cal G}q\rangle_{\rm vac}
=2\langle\overline{q}D^2 q\rangle_{\rm vac}
\equiv 2\lambda_q^2\langle\overline{q}q\rangle_{\rm vac}\ ,
\label{mix_trick}
\end{equation}
where we have used Eqs.~(\ref{G_commutator}) and (\ref{dirac})
to obtain the first equality.
Thus $\lambda_q^2$ parameterizes the average vacuum gluon field strength and
the average virtuality (momentum squared) of the quarks in the QCD vacuum.
The standard QCD sum-rule estimate of this quantity is
$\lambda_q^2=0.4\pm 0.1\,\mbox{GeV}^2$ \cite{BEL82,OVC88}.
Somewhat larger values have been obtained from a lattice calculation
($\lambda_q^2=0.55\pm 0.05\,\mbox{GeV}^2$) \cite{KRE87}
and a QCD sum-rule analysis of the
pion form factor using nonlocal quark and gluon
condensates ($\lambda_q^2=0.7\pm 0.1\,\mbox{GeV}^2$) \cite{BAK91}.
A much larger value for $\lambda_q^2$ is obtained in a sum-rule
analysis of the
pion wave function using nonlocal condensates \cite{MIK92}.
The value suggested by this analysis is
$\lambda_q^2\simeq 1.2\,\mbox{GeV}^2$,
which
agrees with the value obtained with an instanton liquid model \cite{SHU89}.

In QCD sum-rule applications, higher-dimensional condensates are usually
approximated in terms of $\langle\overline{q}q\rangle_{\rm vac}$ and
$\langle(\alpha_s/\pi)G^2\rangle_{\rm vac}$.
For example, the four-quark condensates
are frequently estimated in terms of $\langle\overline{q}q\rangle_{\rm vac}^2$
by using the
factorization, or vacuum-saturation, approximation.
This approximation corresponds to inserting a complete set of intermediate
states in the middle of the four-quark matrix element,
but retaining only the dominant vacuum intermediate state.
An analogous approximation is commonly used in many-body physics \cite{FET71}.
The factorization approximation has been justified in
large-$N_c$ QCD \cite{NOV84};
in QCD with $N_c=3$, it has been argued that the contribution to four-quark
condensates from single-pion intermediate states
(the lowest excitations of the vacuum)
is small compared to that of
the vacuum intermediate state \cite{SHI79}.
Four-quark condensates in vacuum are thus estimated as
\begin{eqnarray}
& &\langle\overline{u}\Gamma_1 u\,\overline{u}\Gamma_2 u\rangle_{\rm vac}
=\mbox{$1\over 16$}\langle\overline{u}u\rangle_{\rm vac}^2
[{\rm Tr}(\Gamma_1){\rm Tr}(\Gamma_2)
-\mbox{$1\over 3$}{\rm Tr}(\Gamma_1\Gamma_2)]\ ,
\\*[4pt]
& &\langle\overline{u}\Gamma_1\lambda^A u\,
\overline{u}\Gamma_2\lambda^A u\rangle_{\rm vac}
=-\mbox{$1\over 9$}\langle\overline{u}u\rangle_{\rm vac}^2
{\rm Tr}(\Gamma_1\Gamma_2)\ ,
\end{eqnarray}
and so on,
where $\Gamma_1$ and $\Gamma_2$ are Dirac matrices, and Tr denotes a trace
over Dirac indices.
A more detailed discussion of the factorization approximation and the
estimation of four-quark condensates at finite density,
including those of mixed flavor (not shown here), is in
Ref.~\cite{JIN93}.
Note that phenomenological studies \cite{ZHI85,GOV87}
and instanton liquid models \cite{SHU88}
suggest strong deviations from the factorized results in some cases.

Thus there are a small number of condensates up to dimension six that could
contribute to nucleon sum rules in the vacuum.
To generalize the sum rules to finite density, the density dependence of
these condensates must be estimated.
In addition, there are a number of new condensates that vanish in
the vacuum, but are nonzero in nuclear matter.


\subsubsection{In-Medium Condensates}
\label{condensates}

To calculate the nucleon correlator at finite density, we need to know the
condensates in nuclear matter.
A detailed discussion  of how these can be estimated is given in
Ref.~\cite{JIN93}.
Here we focus on the dominant condensates and outline how others can be
determined.

We work in the rest frame, where
$u^\mu\rightarrow u^{\prime\mu}\equiv(1,\vec{0})$,
and expand the in-medium
condensates in terms of the rest-frame nucleon density.
To first order in the nucleon density, we have
\begin{equation}
\langle\widehat{O}\rangle_{\rho_N}=\langle\widehat{O}\rangle_{\rm vac}
+\langle\widehat{O}\rangle_N\rho_N+\cdots\ ,
\label{rho_exp}
\end{equation}
where $\cdots$ denotes correction terms that are of higher order
in the nucleon
density.
Note that this expansion is {\it not\/} a Taylor series
expansion in $\rho_N$,
since the next term in the expansion is not $O(\rho_N^2)$.
The spin-averaged nucleon matrix element is
\begin{equation}
\langle\widehat{O}\rangle_N=\int_V d^3 x\,
(\langle{\widetilde N}|\widehat{O}|{\widetilde N}\rangle
-\langle{0}|\widehat{O}|{0}\rangle)\ ,
\label{spock}
\end{equation}
where $|\widetilde{N}\rangle$ is the state vector for a nucleon at rest
normalized to unity
($\langle\widetilde{N}|\widetilde{N}\rangle=1$) in a box of volume $V$.
Using more conventional notation, the nucleon matrix element is given by
\begin{equation}
\langle\widehat{O}\rangle_N=\langle N|\widehat{O}|N\rangle\ ,
\end{equation}
where $|N\rangle$ is once again the state vector for a nucleon at rest.
In this case, the connected matrix element is implied, which is equivalent to
making a vacuum subtraction as in Eq.~(\ref{spock}), and
the nucleon plane-wave states are normalized as follows:
\begin{equation}
\langle N(p)|N(p^\prime)\rangle
={\omega_p\over M_N}(2\pi)^3\delta^3(\vec{p}-\vec{p}^\prime)\ ,
\label{normal}
\end{equation}
with $\omega_p=p_0=\sqrt{\vec{p}^2+M_N^2}$\ .

For a general operator $\widehat{O}$, there is not a systematic
way to study contributions to $\langle\widehat{O}\rangle_{\rho_N}$ that are
of higher order in $\rho_N$.
In the case of $\langle\overline{q}q\rangle_{\rho_N}$, however, higher-order
corrections can be
systematically
studied with an application of the Hellmann-Feynman theorem
(see Sec.~\ref{sec:chiralcon}), although the
corrections are necessarily model dependent.
Estimates of
$\langle\overline{q}q\rangle_{\rho_N}$ in Ref.~\cite{COH92} imply that the
linear approximation is reasonably good
(higher-order corrections $\sim 20\%$ of the linear term) up to nuclear matter
saturation density, although this cannot be considered a definitive conclusion.
Without further justification,
 we assume that the first-order approximation for
{\it all\/}
the condensates is good up to nuclear matter saturation density.
We now proceed to estimate the in-medium condensates.

The most important condensates in finite-density QCD sum rules for baryons
are the dimension-three quark condensates
$\langle\overline{q}q\rangle_{\rho_N}$, $\langle\overline{s}s\rangle_{\rho_N}$
$\langle q^\dagger q\rangle_{\rho_N}$, and
$\langle s^\dagger s\rangle_{\rho_N}$.
The quark condensates alone contribute to the leading-order
sum-rule results for the baryon self-energies
\cite{COH91,FUR92,JIN94a,JIN94c}.

The simplest dimension-three quark condensates are
$\langle q^\dagger q\rangle_{\rho_N}$
and $\langle s^\dagger s\rangle_{\rho_N}$.
Since the baryon current is conserved, these condensates are
proportional to the net nucleon and strangeness densities, respectively:
\begin{equation}
\langle q^\dagger q\rangle_{\rho_N}=\mbox{$3\over 2$}\rho_N\ ,\
\langle s^\dagger s\rangle_{\rho_N}=0\ .
\label{qdaggerq}
\end{equation}
These are exact results.

The in-medium quark condensate $\langle\overline{q}q\rangle_{\rho_N}$
can be expanded in terms of the
nucleon density as
\begin{equation}
\langle\overline{q}q\rangle_{\rho_N}=\langle\overline{q}q\rangle_{\rm vac}
+\langle\overline{q}q\rangle_N\rho_N+\cdots\ ,
\end{equation}
where we have used Eq.~(\ref{rho_exp}).
This condensate has been discussed in Sec.~\ref{sec:chiralcon}
and more extensively
in Refs.~\cite{COH92,DRU90,DRU91};
for completeness, we simply quote the result again here.
The nucleon matrix element $\langle\overline{q}q\rangle_N$ is related to the
nucleon $\sigma$
term $\sigma_N\equiv(m_u+m_d)\langle\overline{q}q\rangle_N$,
where $m_u$ and $m_d$ are the up and down current quark masses.
Combined with the Gell-Mann--Oakes--Renner relation [Eq~(\ref{GMOR})],
one obtains
\begin{equation}
{\langle\overline{q}q\rangle_{\rho_N}
\over\langle\overline{q}q\rangle_{\rm vac}}
=1-{\sigma_N\rho_N\over m_\pi^2 f_\pi^2}+\cdots\ .
\label{qbar-im}
\end{equation}
The most recent estimate of the $\sigma$ term is
$\sigma_N\simeq 45\pm 10\,\mbox{MeV}$ \cite{GAS91}; thus the in-medium quark
condensate is 30--45\% smaller than its vacuum value at nuclear matter
saturation density.

The strange quark condensate $\langle\overline{s}s\rangle_{\rho_N}$ is
expanded in a similar manner:
\begin{equation}
\langle\overline{s}s\rangle_{\rho_N}=\langle\overline{s}s\rangle_{\rm vac}
+\langle\overline{s}s\rangle_N\rho_N+\cdots\ .
\end{equation}
The nucleon matrix element $\langle\overline{s}s\rangle_N$
is commonly parameterized by the dimensionless quantity
$y\equiv\langle\overline{s}s\rangle_N/\langle\overline{q}q\rangle_N$.
Calculations that analyze the mass spectrum of the baryon octet in the context
of SU(3) flavor symmetry indicate that the $\sigma$ term is
related to the strangeness content $y$ in the following manner \cite{GAS91}:
\begin{equation}
\sigma_N={\sigma_N^0\over 1-y}\ ,
\label{wagner}
\end{equation}
where $\sigma_N^0$ is the $\sigma$ term in the limit of vanishing strangeness
content.
Thus one obtains
\begin{equation}
{\langle\overline{s}s\rangle_{\rho_N}
\over\langle\overline{s}s\rangle_{\rm vac}}
=1-{(\sigma_N-\sigma_N^0)\rho_N\over f_s m_\pi^2 f_\pi^2}+\cdots\ .
\end{equation}
An analysis of $\sigma_N^0$ based on second-order perturbation
theory in $m_s-m_q$ yields $\sigma_N^0\simeq 35\pm 5\,\mbox{MeV}$
\cite{GAS91};
hence, the in-medium strange quark condensate is 0--25\% smaller than its
vacuum value at nuclear matter saturation density.

Next we consider
dimension-four quark condensates of the form
$\langle q^\dagger iD_0 q\rangle_{\rho_N}$
and
$\langle(\alpha_s/\pi)G^2\rangle_{\rho_N}$
[note that $\langle\overline{q}iD_0 q\rangle_{\rho_N}$ follows from
Eq.~(\ref{eq:2nd}) and
$\langle(\alpha_s/\pi)(\vec{E}^2+\vec{B}^2)\rangle_{\rho_N}$ is
numerically unimportant].
The  dimension-four condensates are expanded to first order in the
nucleon density using Eq.~(\ref{rho_exp}).
In order to implement this expansion, one must
first determine the vacuum values
of these condensates.
For example, the vacuum value of the strange quark condensate
$\langle s^\dagger iD_0 s\rangle_{\rho_N}$ is given by
\begin{equation}
\langle s^\dagger iD_0 s\rangle_{\rm vac}
=u^\prime_\mu u^\prime_\nu
\langle\overline{s}\gamma^\mu iD^\nu s\rangle_{\rm vac}
={m_s\over 4}\langle\overline{s}s\rangle_{\rm vac}\ ,
\end{equation}
where $u^\prime_\mu\equiv(1,\vec{0})$.
We have used the fact that
$\langle\overline{s}\gamma^\mu iD^\nu s\rangle_{\rm vac}$ can only be
proportional to $g^{\mu\nu}$.
The vacuum values of the other condensates are determined by similar
considerations.
Thus the remaining dimension-four condensates are expanded as follows:
\begin{eqnarray}
& &\langle q^\dagger iD_0 q\rangle_{\rho_N}
=\langle q^\dagger iD_0 q\rangle_N\rho_N+\cdots\ ,
\\*
& &\langle s^\dagger iD_0 s\rangle_{\rho_N}
={m_s\over 4}\langle\overline{s}s\rangle_{\rm vac}
+\langle s^\dagger iD_0 s\rangle_N\rho_N+\cdots\ ,
\\*
& &\left<{\alpha_s\over\pi}G^2\right>_{\rho_N}
=\left<{\alpha_s\over\pi}G^2\right>_{\rm vac}
+\left<{\alpha_s\over\pi}G^2\right>_N\rho_N+\cdots\ .
\end{eqnarray}

The QCD trace anomaly is used to estimate
$\langle(\alpha_s/\pi)G^2\rangle_N$.
The details are discussed in Refs.~\cite{COH92,DRU90,DRU91};
therefore, we simply quote the result here:
\begin{equation}
\left<{\alpha_s\over\pi}G^2\right>_N
=-\mbox{$8\over 9$}(M_N-\sigma_N-S_N)\ ,
\label{gluoncon-_nuc}
\end{equation}
where $M_N$ is the nucleon mass,
$\sigma_N\equiv(m_u+m_d)\langle\overline{q}q\rangle_N$
is the nucleon $\sigma$ term,
and we define $S_N\equiv m_s\langle\overline{s}s\rangle_N$.
{}From Eq.~(\ref{wagner}), $S_N$ can be parameterized as
\begin{equation}
S_N=\left(m_s\over m_u+m_d\right)(\sigma_N-\sigma_N^0)\ .
\end{equation}
We take $m_s/(m_u+m_d)\simeq 13$ \cite{GAS91};
thus we have the following estimate for the nucleon matrix element in
Eq.~(\ref{gluoncon-_nuc}):
\begin{equation}
\left<{\alpha_s\over\pi}G^2\right>_N
\simeq -0.650\pm 0.150\,\mbox{GeV}\ .
\end{equation}
At nuclear matter saturation density,
$\langle(\alpha_s/\pi)G^2\rangle_{\rho_N}$
is about 5--10\% smaller than its vacuum value .

The matrix elements $\langle q^\dagger iD_0 q\rangle_N$ and
$\langle s^\dagger iD_0 s\rangle_N$ can be
related to moments of parton distribution
functions measured in deep-inelastic scattering experiments
\cite{HAT92,DRU90,DRU91}:
\begin{eqnarray}
& &\langle q^\dagger iD_0 q\rangle_N
  = \mbox{$3\over 8$}M_N A^q_2(\mu^2) \simeq 0.18\pm 0.01\,\mbox{GeV}\ ,
\\*
& &\langle s^\dagger iD_0 s\rangle_N
  = \mbox{$1\over 4$}S_N
       +\mbox{$3\over 8$}M_N A^s_2(\mu^2)
\simeq 0.06\pm 0.04\,\mbox{GeV}
\ ,
\end{eqnarray}
where the renormalization scale $\mu=1$\,GeV is used.
The moments of the parton distribution functions are defined as
\cite{COL82,EFR80,CUR80}
\begin{eqnarray}
& &A^q_n(\mu^2)=2\int_0^1 dx\,x^{n-1}
[q(x,\mu^2)+(-1)^n\overline{q}(x,\mu^2)]\ ,
\label{enterprise}
\\*
& &A^s_n(\mu^2)=2\int_0^1 dx\,x^{n-1}
[s(x,\mu^2)+(-1)^n\overline{s}(x,\mu^2)]\ ,
\end{eqnarray}
where $q(x,\mu^2)$, $s(x,\mu^2)$, $\overline{q}(x,\mu^2)$,
and $\overline{s}(x,\mu^2)$ are the scale-dependent
distribution
functions for quarks and antiquarks in the nucleon \cite{GLU90,GLU92}.
See Ref.~\cite{JIN93} for more details of these estimates.

We can evaluate dimension-five condensates through a combination
of parton distributions and model calculations.
With the exception of
\begin{equation}
\langle q^\dagger iD_0\,iD_0 q\rangle_N
+\mbox{$1\over 12$}\langle g_s q^\dagger\sigma\cdot{\cal G}q\rangle_N
\simeq 0.031\,\mbox{GeV}^2\ ,
\end{equation}
which is obtained from parton distribution functions,
none of the
dimension-five condensates have been determined accurately; however,
terms proportional to these condensates make only small contributions
to the nucleon sum rules.
Thus the sensitivity of our sum-rule results to the precise values of these
condensates is small.
A numerical analysis of this sensitivity is given in Ref.~\cite{JIN94}.

%

\subsection{Results and Qualitative Conclusions}
\label{result:nuc}

In this section, we analyze the sum rules for nucleons in infinite
nuclear matter with a general interpolating field. In the operator
product expansion (OPE) for the nucleon correlator, we work to leading
order in perturbation theory; leading-logarithmic corrections are
included through anomalous-dimension factors. Contributions
proportional to the up and down current quark masses are neglected as
they give numerically small contributions.
In the numerical results, we include pure gluon
condensates up to dimension four and quark and gluon condensates up to
dimension five. At dimension six, we include only the four quark
condensates, which give numerically important contributions to nucleon
sum rule in vacuum and in nuclear matter. All other dimension-six and
higher dimensional condensates are neglected since their contributions
are expected to be small.

\subsubsection{Borel Sum Rules}
\label{sec:bsr}

QCD sum rules for the nucleon follow by equating the
phenomenological representation to the OPE representation.
More generally,
we can exploit the analytic structure of the correlator by considering
integrals over contours
running above and below the real axis, and then closing in the
upper and lower half planes, respectively (see Ref.~\cite{FUR92}).
By approximating the correlator in the different regions of integration
and applying Cauchy's theorem,
we can derive a general class of  sum rules, which
manifest the  duality between the physical hadronic spectrum and
the spectral function calculated in a QCD expansion:
\begin{equation}
	\int_{-\overline\omega_0}^{\omega_0}d\omega\, W(\omega)
                            \rho^{{\rm phen}}(\omega,|\vec{q}|)
	-
	\int_{-\overline\omega_0}^{\omega_0}d\omega\, W(\omega)
                            \rho^{{\rm OPE}}(\omega,|\vec{q}|)
                                          = 0 \ .
	\label{eq:fesr}
\end{equation}
Here $W(\omega)$ is a smooth (entire) weighting function and the
spectral densities $\rho^{{\rm phen}}$ and $\rho^{{\rm OPE}}$ are proportional
to the
discontinuities of the invariant functions across the real axis.
(These sum rules can also be derived by expanding dispersion relations for
retarded and advanced correlators with external frequency $\omega'$
in the limit $\omega' \rightarrow i\infty$.)
The phenomenological spectral density $\rho^{{\rm phen}}$ models the
low-energy physical
spectrum, while the theoretical spectral density $\rho^{{\rm OPE}}$
follows from the
the operator product expansion (OPE).
The QCD sum-rule approach assumes that,
with suitable choices for $W$ and the effective continuum thresholds
$\omega_0$ and $-\overline\omega_0$, each integral can be reliably
calculated and
meaningful results extracted.

In principle, the effective
thresholds are different
for positive ($\omega_0$) and negative ($\overline\omega_0$)
energies and
for the different sum rules.
The former differences are critical in some sum-rule
formulations \cite{FUR94},
but are not numerically important in our formulation.
Furthermore, the
thresholds are relatively poorly
determined by the sum rules and  effects due to different thresholds
in different sum rules
may be absorbed by slight changes in the other parameters. In the
present discussion, we use a universal effective threshold
$\omega_0$ for simplicity.

If we choose the weighting function
$W(\omega)=\omega \e^{-\omega^2/M^2}$,
then the vacuum Borel sum rule is reproduced in the zero-density limit
up to an overall factor of $\e^{-\qvecsq/M^2}$.
This is not an optimal choice
at finite density, because it weights positive and
negative $\omega$ equally.
In order to suppress the negative-energy contribution, we use the weighting
function
\beq
    W(\omega) = {(\omega - \Eqbar)}
                 \, \e^{-\omega^2/M^2}  \ ,  \label{eq:newweight}
\eeq
where $\Eqbar$ is the energy of the negative-energy pole in our
quasiparticle ansatz [see Eq.~\eqref{eq:Eqbar}].
This choice suppresses a sharp excitation completely but also strongly
suppresses (relative to the positive-energy contribution) a broad
excitation in this vicinity.
Furthermore, this choice  reduces to the usual Borel sum rule in the
vacuum: The vacuum spectral densities are odd in $\omega$;
thus the $\Eqbar$ contribution vanishes.

The spectral densities in Eq.~(\ref{eq:fesr}) can be extracted from the
discontinuities in Eqs.~(\ref{eq:tom})--(\ref{eq:harry}) and
(\ref{PisE_ope})--(\ref{PiuO_ope}).
Alternatively, sum rules for
the nucleon with the weighting function in Eq.~(\ref{eq:newweight})
can be constructed as follows:
\beq
\borel[\PiEi(q_0^2,|\vec{q}|)-\Eqb\PiOi(q_0^2,|\vec{q}|)]_{\rm OPE}=
\borel[\PiEi(q_0^2,|\vec{q}|)-\Eqb\PiOi(q_0^2,|\vec{q}|)]_{\rm phen}\ ,
\label{sum_def}
\eeq
for $i=\{s,q,u\}$, where the left-hand side is obtained from the OPE,
the right-hand side from the phenomenological dispersion relations, and
$\borel$ is the Borel transform  operator defined by
\begin{equation}
{\cal B}[f(q_0^2,|\vec{q}|)]\equiv
\lim_{\stackrel{\scriptstyle -q_0^2,n\rightarrow\infty}
{\scriptstyle -q_0^2/n=M^2}}
{(-q_0^2)^{n+1}\over n!}\left(\partial\over\partial q_0^2\right)^n
f(q_0^2,|\vec{q}|)
\equiv\widehat{f}(M^2,|\vec{q}|)\ .
\label{borel_def}
\end{equation}
The only difference from a Borel transform with respect to $Q^2 = -q^2$
is  a factor of $\e^{-\qvecsq/M^2}$ common to all terms, which
will cancel.
In the zero-density limit,
contributions from the second term in
Eq.~\eqref{eq:newweight} vanish, and we once again recover the usual
vacuum sum rules.

Perturbative corrections $\sim\alphas^n$ can be taken into account in
the leading logarithmic approximation through anomalous-dimension
factors~\cite{SHI79}. After the Borel transform, the effect of these
corrections is to multiply each term on the OPE side by the
factor \cite{SHI79,IOF81,FUR92}
\beq
L^{-2\Gamma_\eta+\Gamma_{O_n}}
\equiv\left[\ln(M/\Lambda_{\rm QCD})
\over\ln(\mu/\Lambda_{\rm QCD})\right]^{-2\Gamma_\eta+\Gamma_{O_n}}\ ,
\eeq
where $\Gamma_{\eta}$ is the anomalous dimension of the interpolating
field $\eta$, $\Gamma_{O_n}$ is the anomalous dimension of the
corresponding local operator (including the current quark masses), $\mu$
is the normalization point of the operator product expansion, and
$\Lambda_{\rm QCD}$ is the QCD scale parameter.
In general, the absolute predictions of QCD sum rules are sensitive to
the precise choices of $\mu$ and $\Lambda_{\rm QCD}$.  However,
when  ratios of finite-density to zero-density
quantities are taken, as we do here,
the predictions are insensitive for $100\,\mbox{MeV}
< \Lambda_{\rm QCD} < 200\,\mbox{MeV}$ \cite{JIN94}.%
\footnote{Values of $\Lambda_{\rm QCD}$ much larger than
$200\,\mbox{MeV}$, as obtained from some experimental analyses,
 are  problematic for the QCD sum rules (at any density).
Shifman has recently
argued that  such values are incompatible with crucial features of QCD
and should be regarded with skepticism \cite{SHI94}.}

Applying Eq.~(\ref{sum_def}) to
the ans\"atze in Eqs.~(\ref{eq:tom})--(\ref{eq:harry}) and the OPE results in
Eqs.~(\ref{PisE_ope})--(\ref{PiuO_ope}), we obtain three
sum rules---one for
each invariant function:
\beqa
\lambda_N^{\ast 2}M_N^\ast\e^{-(E_q^2-\vec{q}^2)/M^2}
&=&-{c_1\over 16\pi^2}M^4 E_1 \rhome{\qbarq}
   -{c_1\over 3}
     \Eqb \rhome{\qbarq}\rhome{\qdaggerq} + \cdots \ ,
   \label{sum_fs}
\\[7.2pt]
\lambda_N^{\ast 2} \e^{-(E_q^2-\vec{q}^2)/M^2}
&=&{c_4\over 256\pi^4}M^6 E_2 L^{-4/9}
+{c_4\over 24\pi^2}\Eqb M^2 E_0\rhome{\qdaggerq}L^{-4/9}
\nonumber\\*[7.2pt]
& &\null
-{c_4\over 72 \pi^2}M^2
\left(5E_0-{8\vec{q}^2\over M^2}\right)\rhome{\qdagger iD_0 q}L^{-4/9}
\nonumber\\*[7.2pt]
& &\null
+{c_4\over 256\pi^2}M^2 E_0 \rhomeadjust{\gluonconA}L^{-4/9}
\nonumber\\*[7.2pt]
& &\null
+{c_1\over 6}\rhome{\qbarq}^2 L^{4/9}+{c_4\over 6}\rhome{\qdaggerq}^2 L^{-4/9}
+ \cdots \ ,
\label{sum_fq}
\\[7.2pt]
\lambda_N^{\ast 2}\Sigma_v \e^{-(E_q^2-\vec{q}^2)/M^2} &=&{c_4\over
12\pi^2}M^4E_1\rhome{\qdaggerq}L^{-4/9}
\nonumber\\*[7.2pt]
& &\null
+ {5c_4\over
18\pi^2}\Eqb M^{2}E_0\rhome{\qdagger iD_0 q}L^{-4/9}
+{c_4\over 3}\Eqb\rhome{\qdaggerq}^2L^{-4/9} + \cdots  \ .
\label{sum_fu}
\eeqa
Here we have defined
\beqa
& &E_0\equiv 1-\e^{-s_0^*/M^2}\ ,
\\*[7.2pt]
\label{con_0}
& &E_1\equiv 1-\e^{-s_0^*/M^2}\left({s_0^{*}\over M^2}+1\right)\ ,
\\*[7.2pt]
\label{con_1}
& &E_2\equiv 1-\e^{-s_0^*/M^2}\left({s_0^{*^2}\over 2M^4}+{s_0^*\over
M^2}+1\right)\ ,
\label{con_2}
\eeqa
where we define $s_0^*\equiv\omega_0^2-\vec{q}^2$.
These factors arise from the continuum model, which approximates
the contributions from higher-energy states by the OPE spectral
density, starting at a sharp energy threshold $\omega_0$.

We see that the above sum rules explicitly involve the Borel mass
$M^2$. If both the QCD expansion and phenomenological description were
exact, predictions for the spectral
parameters would be independent of $M^2$. In practice, both sides
are represented imperfectly. The hope is that there exists a range in
$M^2$ for which the two sides have a good overlap.


\subsubsection{Simplified Sum Rules}

The Ioffe formula [Eq.~(\ref{nucleon_mass_Ioffe})]
manifests the principal physical content of the nucleon
sum rule in vacuum in a highly truncated form, which is
justified {\it a posteriori\/} by examining the full sum rule.
Here we construct  the analogous
simplified finite-density sum rules for the nucleon,
which follow by keeping in each of the three sum rules of
Eqs.~(\ref{sum_fs})--(\ref{sum_fu}) only the
quasinucleon-pole contribution to the phenomenological side
({\it i.e.\/}, no continuum factors)
and only the leading term in the operator product expansion on the OPE
side (without anomalous-dimension corrections):
\begin{eqnarray}
  \lambda_N^2 M_N^\ast \e^{-(E_q^2-\vec{q}^2)/M^2} &=&
 -{1\over{4\pi^2}}M^4\langle\overline{q}q\rangle_{\rho_N}
  \ , \label{eq:simp2} \\[4pt]
  \lambda_N^2 \e^{-(E_q^2-\vec{q}^2)/M^2} &=&
      {1\over{32\pi^4}}M^6  \ , \label{eq:simp1} \\[4pt]
  \lambda_N^2 \Sigma_v \e^{-(E_q^2-\vec{q}^2)/M^2} &=&
 {2\over{3\pi^2}}M^4 \langle q^\dagger q\rangle_{\rho_N}
         \ .  \label{eq:simp3}
\end{eqnarray}
Here Ioffe's current [Eq.~(\ref{int_field})],
{\it i.e.}, Eq.~(\ref{def_ge}) with $t=-1$, has been used.
By considering ratios of these sum
rules evaluated at a value of the Borel mass
in the middle of the range where the full sum rules will be considered,
one hopes to extract the basic physics.
As one can see later,
this truncation is qualitatively reasonable {\it except\/} for the
large contribution of four-quark condensates.

Taking  ratios
of Eqs.~(\ref{eq:simp2})--(\ref{eq:simp3}), one obtains the
simple expressions
\begin{eqnarray}
M_N^\ast &=& -{{8\pi^2}\over{M^2}}\langle\overline{q}q\rangle_{\rho_N}\ ,
\label{eq:one}
\\*
\Sigma_v &=& {{64\pi^2}\over{3M^2}}\langle q^\dagger q\rangle_{\rho_N}\ ,
\label{eq:two}
\end{eqnarray}
which might be expected to apply up to nuclear matter saturation density.
Using Eqs.~(\ref{qdaggerq}) and (\ref{qbar-im}), one can determine the
scalar and vector self-energies in terms of the nucleon density:
\begin{eqnarray}
\Sigma_s&=&-{{4\pi^2}\over{M^2}}{\sigma_N\rho_N \over
        {m_q}}
\ ,\label{eq:simsigs}\\[4pt]
\Sigma_v &=&{{32\pi^2}\over{M^2}}\rho_N \ .\label{eq:simsigv}
\end{eqnarray}
One observes that both $\Sigma_s$ and $\Sigma_v$ are proportional to $\rho_N$.
Taking the ratio of Eqs.~(\ref{eq:simsigs}) and (\ref{eq:simsigv}),
one finds that the explicit dependence of the self-energies on the
Borel mass and the density drops out:
\begin{equation}
{\Sigma_s\over\Sigma_v}=-{{\sigma_N}\over{8m_q}} \ .
      \label{eq:griegel}
\end{equation}
For typical values of $\sigma_N$ and light quark masses,
this ratio is close to $-1$, indicating a substantial cancellation
of $\Sigma_s$ and $\Sigma_v$ in the medium.
Thus the predictions of the simplest sum rules are in qualitative
agreement with several features of relativistic phenomenology:
the self-energies scale with the density, they are weakly dependent on
the nucleon state (three-momentum), and scalar and vector self-energies
cancel.

Alternatively, one can normalize the self-energies
to the nucleon mass determined by taking the zero-density limit of
Eq.~(\ref{eq:one}):
\begin{equation}
M_N=-{8\pi^2\over M^2}\langle\overline{q}q\rangle_{\rm vac}\ .
\label{eq:simvac}
\end{equation}
The hope is that this will reduce the sensitivity to particular details of the
sum rules and to the level of truncation,
provided one works to the same level of approximation
at finite  and zero density.
Adopting the same Borel mass for both finite and zero density and
taking ratios,
one obtains
results independent of $M^2$:
\begin{eqnarray}
     {M_N^\ast\over M_N} &=&  1 + {\Sigma_s\over M_N}  =
  {\langle\overline{q}q\rangle_{\rho_N}\over\langle
      \overline{q}q\rangle_{\rm vac}}
            = 1 - {\sigma_N\rho_N\over m_\pi^2 f_\pi^2}
            \ ,  \label{eq:mstarm}  \\[6pt]
    {\Sigma_v\over M_N} &=&
   -{8\over 3}{\langle q^\dagger q\rangle_{\rho_N}\over\langle
           \overline{q}q\rangle_{\rm vac}}
   = {8m_q\rho_N \over m_\pi^2 f_\pi^2}
    \ . \label{eq:sigmavm}
\end{eqnarray}
The last equalities in Eqs.~(\ref{eq:mstarm}) and (\ref{eq:sigmavm})
follow from Eqs.~(\ref{qdaggerq}), (\ref{qbar-im}),
and (\ref{GMOR}).
[Note: The independence of $M^2$ in the ratios in Eqs.~(\ref{eq:mstarm}) and
(\ref{eq:sigmavm}) should not be interpreted as evidence that
the {\it individual\/} sum rules are weakly dependent on $M^2$.]
For typical values of the relevant condensates and other parameters,
$M_N^\ast/ M_N \sim 0.6\mbox{--}0.7$ and $\Sigma_v/ M_N\sim 0.3\mbox{--}0.4$.
This is in good agreement with the values used in relativistic mean-field
models that provide good fits to
bulk properties of finite nuclei \cite{SER86}.

The key feature that assures qualitative agreement with
relativistic phenomenology is that
Eq.~(\ref{eq:simp1}) is density independent to leading order.
In the simple sum rule, this implies that the pole position and residue
do not vary much with density.
This in turn implies the results of Eqs.~(\ref{eq:one}) and
(\ref{eq:two}), in which the effective mass naturally follows
$\langle\overline{q}q\rangle_{\rho_N}$ and the vector self-energy follows
$\langle q^\dagger q\rangle_{\rho_N}$.
In the more complete sum-rule analysis considered below, these basic
results survive if the correction terms to Eqs.~(\ref{eq:simp2})
and (\ref{eq:simp3}) are not overly large {\it and\/}
if Eq.~(\ref{eq:simp1}) remains weakly density dependent.
The latter condition turns out to be problematic.


\subsubsection{Detailed Sum-Rule Analysis}

In principle, the predictions based on sum rules should be independent of the
auxiliary parameter $M^2$.
In practice, however, one has to truncate the OPE and use a simple
phenomenological ansatz for the spectral density, so one expects at best that
the two descriptions overlap only in a limited range of $M^2$.
As a result, one expects to see a ``plateau'' in the predicted quantities as
functions of $M^2$
(although not necessarily a local extremum).
Nucleon sum rules in vacuum truncated
at dimension-six condensates do
not provide a  plateau
\cite{FUR92,IOF84,LEI90}; nevertheless, it will be  assumed here that the sum
rules  have a region of overlap, although imperfect.\footnote%
{Including direct-instanton effects in nucleon sum rules in vacuum leads to a
more convincing plateau \cite{DOR90,FOR93}.}
We normalize the finite-density predictions for all self-energies
to the zero-density prediction for the mass, with the expectation
that this will compensate for at least some of the limitations of the
truncated sum rules.

To analyze the sum rules and extract the self-energies, one can sample the sum
rules
in the fiducial region, which is the overlap between the region where the sum
rule is dominated by the quasinucleon contribution and the region where the
truncated OPE is reliable.
In choosing the fiducial region, one may introduce a lower bound of the Borel
mass such that the highest-dimensional condensate contributes no more than
$\sim 10\%$ to the total of the terms on the right-hand sides of
Eqs.~(\ref{sum_fs})--(\ref{sum_fu}) and an upper bound such that the continuum
contribution is less than
$\sim 50\%$ of the total phenomenological contribution
({\it i.e.\/}, the sum of the
quasinucleon pole contribution and the continuum contribution).
To quantify the fit of the left- and right-hand sides, one can apply the
logarithmic
measure
\begin{equation}
\delta(M^2)=\ln\left[{\mbox{max}}\{\lambda_N^{\ast 2}
\e^{-(E_q^2-\vec{q}^2)/M^2},
\Pi_s^\prime/M_N^\ast,\Pi_q^\prime,\Pi_u^\prime/\Sigma_v\}
\over{\mbox{min}}\{\lambda_N^{\ast 2}\e^{-(E_q^2-\vec{q}^2)/M^2},
\Pi_s^\prime/M_N^\ast,\Pi_q^\prime,\Pi_u^\prime/\Sigma_v\}\right]\ ,
\end{equation}
which we average over 150 points evenly spaced within the fiducial region of
$M^2$.
Here $\Pi_s^\prime$, $\Pi_q^\prime$, and $\Pi_u^\prime$ denote the right-hand
sides of Eqs.~(\ref{sum_fs})--(\ref{sum_fu}), respectively.
The predictions for $M_N^\ast$, $\Sigma_v$, $s_0^\ast$, and
$\lambda_N^{\ast 2}$
are obtained by minimizing the averaged measure $\delta$.
This approach weights the fits in the region where the
continuum contribution is
minimal and reduces the sensitivity to the endpoints of the optimum
region \cite{LEI90}.
To get a prediction for the nucleon mass in vacuum, one applies the same
procedure
to the sum rules evaluated in the zero-density limit.

In the analysis to follow, the quasinucleon three-momentum is fixed
at $|\vec{q}|=270\,\mbox{MeV}$
({\it i.e.\/}, approximately the Fermi momentum)
and the nucleon $\sigma$ term is taken
to be $\sigma_N=45\,\mbox{MeV}$.
The dimension-five  nucleon
matrix elements not given earlier are taken to be \cite{JIN94}
\begin{eqnarray}
& &\langle\overline{q} iD_0\, iD_0 q\rangle_N
       +\mbox{$1\over 8$}\langle g_s\overline{q}
       \sigma\cdot {\cal G} q\rangle_N=0.3\,
           \mbox{GeV}^2\ ,
\\*
& &\langle g_s\overline{q}\sigma\cdot
            {\cal G} q\rangle_N=3.0\,\mbox{GeV}^2\ ,
\\*
& &\langle g_s q^\dagger\sigma\cdot
            {\cal G} q\rangle_N=-0.33\,\mbox{GeV}^2\ .
\label{matrix:mid}
\end{eqnarray}
The values of vacuum condensates are taken to be
$\langle\overline{q}q\rangle_{\rm vac}=-(245\,\mbox{MeV})^3$,
$\langle(\alpha_s/\pi)G^2\rangle_{\rm vac}=(330\,\mbox{MeV})^4$,
and $\langle g_s\overline{q}\sigma\cdot{\cal G}q\rangle_{\rm vac}=
m_0^2\langle\overline{q}q\rangle_{\rm vac}$
with $m_0^2=0.8\,\mbox{GeV}^2$.
Nuclear matter
saturation density is taken to be $\rho_N=(110\,\mbox{MeV})^3$.

Four-quark condensates are numerically important in both the vacuum and the
finite-density nucleon sum rules, because they contribute in tree diagrams and
do not carry the numerical suppression factors typically associated with loops.
In the sum rules in Eqs.~(\ref{sum_fs})--(\ref{sum_fu}), the
contributions from the four-quark condensates in their in-medium
factorized forms are included; however, the factorization approximation
may not be
justified in nuclear matter.
In the case of  the ``scalar-vector'' and ``vector-vector'' four-quark
condensates,
$\langle\overline{q}q\rangle_{\rho_N}\langle q^\dagger q\rangle_{\rho_N}$
and $\langle q^\dagger q\rangle_{\rho_N}^2$,
such concerns are unimportant, since these condensates give minimal
contributions to the nucleon sum rules \cite{JIN94}.
Thus their factorized forms will be used here for simplicity.
However, the ``scalar-scalar'' four-quark condensate
$\langle\overline{q}q\rangle_{\rho_N}^2$ {\it does\/} give important
contributions to the nucleon sum rules.

In its factorized form, the scalar-scalar four-quark condensate has a very
strong density dependence,
which may not be justified.
Therefore we parameterize the density dependence in terms of a new
parameter $f$.
Specifically,
the scalar-scalar four-quark condensate is parameterized so
that it interpolates between its factorized form in free space and its
factorized form in nuclear matter:
\begin{equation}
\langle\overline{q}q\rangle_{\rho_N}^2\longrightarrow
\langle\widetilde{\overline{q}q}\rangle_{\rho_N}^2
\equiv(1-f)\langle\overline{q}q\rangle_{\rm vac}^2
+f\langle\overline{q}q\rangle_{\rho_N}^2\ .
\label{4quark-p}
\end{equation}
The density dependence of the scalar-scalar four-quark condensate is then
parameterized by $f$ and the density dependence of
$\langle\overline{q}q\rangle_{\rho_N}$ [see Eq.~(\ref{qbar-im})].
The factorized condensate $\langle\overline{q}q\rangle_{\rho_N}^2$ appearing in
Eq.~(\ref{sum_fq}) will be replaced by
$\langle\widetilde{\overline{q}q}\rangle_{\rho_N}^2$ in the calculations to
follow.
Values of $f$ in the range $0\leq f\leq 1$ will be considered;
$f=0$ corresponds to the assumption of no density dependence, and $f=1$
corresponds to the in-medium factorization assumption.

\begin{figure}
\vbox to 3.7in{\vss\hbox to 5.625in{\hss
{\includegraphics{nucl-d1.ps}}\hss}}
\caption{Optimized sum-rule predictions for $M_N^\ast/M_N$, $\Sigma_v/M_N$,
and
$E_q/M_N$ as functions of $f$, with Ioffe's interpolating field ($t=-1$).}
\label{nucl-d1}
\end{figure}

The sum rules are analyzed with the Borel window fixed at
$0.8\leq M^2\leq 1.4\,\mbox{GeV}^2$, which is identified by Ioffe and Smilga
\cite{IOF84} as the fiducial region for the nucleon sum rules in vacuum (with
the contributions from up to dimension-nine condensates included).
Here these boundaries are adopted as the maximal limits of applicability of
the sum
rules at finite density.
We start from Ioffe's interpolating field ({\it i.e.\/}, $t=-1$).
The optimized results for the ratios $M_N^\ast/M_N$, $\Sigma_v/M_N$, and
$E_q/M_N$ as functions of $f$ are plotted in Fig.~\ref{nucl-d1}.
One can see from Fig.~\ref{nucl-d1} that $M_N^\ast/M_N$ and $E_q/M_N$
vary rapidly
with $f$, while $\Sigma_v/M_N$ is relatively insensitive to $f$.
Therefore, the sum-rule prediction for the scalar self-energy depends
{\it strongly\/} on the density dependence of the scalar-scalar four-quark
condensate.
For small values of $f$ ($0\leq f\leq 0.3$), the predictions are
\begin{eqnarray}
M_N^\ast/M_N&\simeq &0.63\mbox{--}0.72\ ,
\\*
\Sigma_v/M_N&\simeq &0.24\mbox{--}0.30\ ,
\end{eqnarray}
which are comparable to typical values from relativistic phenomenology.
On the other hand, for large $f$ ($0.7\leq f\leq 1$), one finds
$\Sigma_v/M_N\simeq 0.34\mbox{--}0.37$, which is still reasonable.
In contrast, the predictions for $M_N^\ast$ and $E_q$ turn out to be
$M_N^\ast/M_N\simeq 0.84\mbox{--}0.94$ and $E_q/M_N\simeq 1.24\mbox{--}1.36$,
which imply $\Sigma_s/M_N\simeq -(0.06\mbox{--}0.16)$ and a significant
shift of
the quasinucleon pole relative to the nucleon pole in vacuum (the net
self-energy is repulsive).
Thus a significant density dependence of the scalar-scalar four-quark
condensate
leads to an essentially vanishing scalar self-energy and a strong vector
self-energy with a magnitude of a few hundred MeV.
The predictions for the ratios $\lambda_N^{\ast 2}/\lambda_N^2$ and
$s_0^\ast/s_0$ also depend on $f$.
For small $f$, the continuum threshold $s_0^\ast$ is close to the vacuum value
while the residue $\lambda_N^{\ast 2}$ drops about 20\% relative to the
corresponding vacuum value.
(Note that these quantities are relatively poorly determined by the sum rules.)
For large $f$, the continuum threshold increases by 20--25\% relative to the
vacuum value and the residue increases by about 20\%, implying a significant
rearrangement of the spectrum.
For intermediate values of $f$, both the continuum threshold and the residue
are
very close to the corresponding vacuum values.

\begin{figure}
\vbox to 3.7in{\vss\hbox to 5.625in{\hss
{\includegraphics{nucl-d2.ps}}\hss}}
\caption{Ratios $M_N^\ast/M_N$ and $\Sigma_v/M_N$ as functions of Borel $M^2$,
with optimized predictions for $E_q$, $\overline{E}_q$ and the continuum
thresholds.
The solid, dashed, and dotted curves corresponds to $f=0$, 0.5, and 1,
respectively.}
\label{nucl-d2}
\end{figure}

{}From the sum rules in Eqs.~(\ref{sum_fs})--(\ref{sum_fu}), it is easy to see
that the ratios $\Pi_s^\prime/\Pi_q^\prime$ and $\Pi_u^\prime/\Pi_q^\prime$
give
$M_N^\ast$ and $\Sigma_v$ as functions of Borel $M^2$, and
$\Pi_s^\prime/\Pi_q^\prime$ in the zero-density limit yields $M_N$ as a
function
of $M^2$.
In Fig.~\ref{nucl-d2}, the ratios $M_N^\ast/M_N$ and $\Sigma_v/M_N$ are
plotted as
functions of $M^2$ for three different values of $f$, with $E_q$,
$\overline{E}_q$, and the continuum threshold fixed at their optimized values.
The curves for $f=0$ and $f=0.5$ (solid and dashed curves respectively) are
quite flat in the optimum region, and thus imply a weak dependence of the
predicted ratios on $M^2$ (even though the individual sum-rule predictions
before taking ratios are not flat).
For $f=1$ (dotted curves), the ratio $\Sigma_v/M_N$ is flat, indicating
again a
weak dependence on $M^2$; in contrast, $M_N^\ast/M_N$ changes significantly
in the region of interest.

\begin{figure}
\vbox to 3.7in{\vss\hbox to 5.625in{\hss
{\includegraphics{nucl-d3-a.ps}}\hss}}
\vbox to 3.7in{\vss\hbox to 5.625in{\hss
{\includegraphics{nucl-d3-b.ps}}\hss}}
\caption{(a) The left- and right-hand sides of the finite-density sum rules as
functions of Borel $M^2$ for $t=-1$ and $f=0$, with the optimized values for
$M_N^\ast$, $\Sigma_v$, $s_0^\ast$, and $\lambda_N^{\ast 2}$.
The four curves correspond to $\Pi_s^\prime/M_N^\ast$ (solid), $\Pi_q^\prime$
(dashed), $\Pi_u^\prime/\Sigma_v$ (dot-dashed), and
$\lambda_N^{\ast 2}\e^{-(E_q^2-\vec{q}^2)/M^2}$ (dotted).
(b) The left- and right-hand sides of the corresponding vacuum sum rules,
with the optimized values for $M_N$, $s_0$, and $\lambda_N^2$.
The three curves correspond to $\Pi_s^\prime/M_N$ (solid), $\Pi_q^\prime$
(dashed), and $\lambda_N^2 \e^{-M_N^2/M^2}$ (dot-dashed) at the zero-density
limit.}
\label{nucl-d3}
\end{figure}

In Fig.~\ref{nucl-d3}(a),
$\lambda_N^{\ast 2}\e^{-(E_q^2-\vec{q}^2)/M^2}$, $\Pi_s^\prime/M_N^\ast$,
$\Pi_q^\prime$, and $\Pi_u^\prime/\Sigma_v$ are plotted as functions of $M^2$
for $f=0$, with the predicted values for $M_N^\ast$, $\Sigma_v$,
$s_0^\ast$, and $\lambda_N^{\ast 2}$.
If the sum rules work well, one should expect the four curves to
coincide with each other.
It is found that their $M^2$ dependence in the region of interest turns
out to
be equal up to 15\%.
The overlap of the corresponding vacuum sum rules
({\it i.e.\/}, the zero-density
limit)
is illustrated in Fig.~\ref{nucl-d3}(b).
One can see that the quality of the overlap for the finite-density sum rules
is
similar to that of the corresponding sum rules in vacuum.
As $f$ increases, the overlap of the sum rules gets better; however, this does
not necessarily imply that the results with large $f$ are more trustworthy,
because other corrections (such as the contributions from higher-dimensional
condensates, the corrections from the higher-order density dependence of
condensates, {\it etc.}) will change the behavior of the overlap.

\begin{figure}
\vbox to 3.7in{\vss\hbox to 5.625in{\hss
{\includegraphics{nucl-d4.ps}}\hss}}
\caption{Optimized sum-rule predictions for $M_N^\ast/M_N$ and $\Sigma_v/M_N$
as
functions of $t$.
The three curves correspond to $f=0.2$ (solid), 0.5 (dashed), and 1.0
(dotted).}
\label{nucl-d4}
\end{figure}

All of the results above use Ioffe's interpolating field ($t=-1$); now
the results for the general interpolating field [Eq.~(\ref{def_ge})] are
 presented.
In Fig.~\ref{nucl-d4}, the predicted ratios $M_N^\ast/M_N$ and
$\Sigma_v/M_N$ as functions of $t$ for three different values of $f$
have been displayed.
The ratio $\Sigma_v/M_N$ increases as $t$ increases; the rate of increase is
essentially the same for all values of $f$.
For $f=1$, the ratio $M_N^\ast/M_N$ decreases slowly as $t$ increases; for
$f=0.5$, $M_N^\ast/M_N$ is nearly independent of $t$; for $f=0.2$,
$M_N^\ast/M_N$ increases slowly as $t$ increases over the range of $t$ that is
of interest.
It is found that the continuum contributions increase and the residue decreases
as $t$ increases.
On the other hand, the overlap of the sum rules gets better as $t$ increases.
The predictions for the continuum thresholds depend only weakly on $t$.
It is also found that for $f<0.2$ and $-1.15\leq t\leq -1.05$, the numerical
optimizing procedure converges slowly and the predicted continuum threshold
and residue are much smaller than those for $f\geq 0.2$.
In this case, the continuum contributions dominate the sum rules, making the
predictions for $M_N^\ast$ and $\Sigma_v$ unreliable.

%

The sensitivity of the sum-rule results to other factors,
such as variations in the fiducial interval and changes in various
condensates and parameters, have been examined in Refs.~\cite{JIN94,JIN93b}.
In general, one finds that the sum-rule predictions are largely insensitive
to details, with the important exception
of the assumed density dependence of the scalar-scalar four-quark condensate.

\subsubsection{Qualitative Conclusions}

The most concrete conclusion we can draw is that
QCD sum rules predict a positive vector self-energy with a magnitude of a few
hundred MeV for a quasinucleon in nuclear matter.
This qualitative feature is {\it robust\/} and, for the most part,
independent of the details of the calculation.
For Ioffe's interpolating field and typical values of the relevant condensates
and other input parameters, one obtains $\Sigma_v/M_N\simeq 0.24$--0.37,
which
is a range very similar to that found for vector self-energies in relativistic
nuclear physics phenomenology.
On the other hand, the prediction for the scalar self-energy depends strongly
on
the value of the in-medium scalar-scalar four-quark condensate, which is not
well established, and on the value of the nucleon $\sigma$ term.
This means that the conclusions about the quasinucleon scalar self-energy must
still be somewhat indefinite.
Nevertheless, it should be emphasized that predictions with different
values of the four-quark
condensate give different physical features that are not equally compatible
with
known nuclear phenomenology.

\begin{figure}
\vbox to 4.0in{\vss\hbox to 5.625in{\hss
{\includegraphics{opt_fig1g.ps}}\hss}}
\caption{Optimized sum-rule predictions for $M_N^\ast/M_N$ (diamonds)
and $\Sigma_v/M_N$ (squares) as functions of density, with $f=0$
(parameters are specified in Ref.~\protect\cite{FUR92}).
The predictions from a nonlinear
relativistic mean-field model \protect\cite{SER92}
are shown for comparison (dashed and dot-dashed curves).}
\label{fig:dens}
\end{figure}

If the scalar-scalar four-quark condensate depends only weakly on the nucleon
density ({\it i.e.\/}, if $f$ is small),
the prediction for $M_N^\ast/M_N$ is insensitive to the Borel mass.
The predicted scalar self-energy
is large and negative, which is consistent with relativistic
phenomenology.
In this case, there is a significant degree of cancellation between the scalar
and vector self-energies, which
leads to a quasinucleon energy close to the free-space nucleon
mass.
This result is compatible with the empirical observation that the quasinucleon
energy is shifted only slightly in nuclear matter relative to the free-space
mass.
The density dependence of the self-energies for this case is shown in
Fig.~\ref{fig:dens}.
The prediction for the continuum threshold is close to the vacuum value and the
residue at the quasinucleon pole drops slightly relative to the corresponding
vacuum value.
This is also compatible with experiment; there is no evidence for a strong
rearrangement of the spectrum at nuclear matter saturation density, merely a
spreading of strength over energy scales too small to be resolved by the sum
rules.

In contrast, if the four-quark condensate has a strong density
dependence ({\it i.e.\/}, if $f$ is large), the predicted ratio
$M_N^\ast/M_N$ varies strongly with the Borel mass, and the magnitude of
$M_N^\ast/M_N$ is close to unity, implying that the scalar self-energy is
essentially zero.
The predicted vector self-energy, on the other hand, is larger than it is with
small $f$.
Thus the resulting quasinucleon energy is significantly larger than the nucleon
mass.
This result is unrealistic and is totally different from known nuclear
phenomenology.
Moreover, both the continuum threshold and the residue at the quasinucleon pole
are well above their values in vacuum; this suggests a significant
rearrangement of the spectrum in nuclear matter, which is inconsistent with
experiment.

For intermediate values of $f$, the predicted scalar self-energy is negative
with a sizable magnitude.
The vector self-energy is still strong.
The magnitudes of the self-energies and the degree of cancellation between them
depend on the choice of interpolating field and the values of the
condensates and input parameters.
The quasinucleon energy, the residue at the quasinucleon pole, and the
continuum
threshold are close to their corresponding vacuum values.

The qualitative features discussed above can also be identified through the
dominant behavior of the OPE sides of the sum rules.
{}From Eq.~(\ref{sum_fq}), one notes that $\Pi_q^\prime$ is mainly determined
by
the density-independent leading-order perturbative term and the scalar-scalar
four-quark condensate.
For small $f$, $\Pi_q^\prime$ is close to its zero-density value; this implies
that the quasinucleon energy, residue, and continuum are essentially unchanged
from their values in vacuum.
Since $\Pi_s^\prime$ is dominated by the leading-order term proportional to
the in-medium quark condensate, the significant reduction of
$\langle\overline{q}q\rangle_{\rho_N}$ from its vacuum value
$\langle\overline{q}q\rangle_{\rm vac}$ implies a significant reduction of
$M_N^\ast$ from $M_N$.
The vector self-energy tends to follow the nucleon density, as the
leading-order
term proportional to $\langle q^\dagger q\rangle_{\rho_N}$  gives the largest
contribution to $\Pi_u^\prime$.
For large values of $f$, $\Pi_q^\prime$ is significantly reduced from its
vacuum
value; this leads to a shift in the quasinucleon energy and a significant
rearrangement of the spectrum.
As $\Pi_s^\prime$ and $\Pi_u^\prime$ are independent on $f$, one expects
$M_N^\ast$ and $\Sigma_v$ to increase due to the reduction of
$\Pi_q^\prime$.
Clearly, further study of the in-medium four-quark condensates is very
important.

The sum-rule predictions are fairly sensitive to the choice of interpolating
field, reflecting the dependence of the truncated OPE on this choice.
In the region $-1.15\leq t\leq -1$, the scalar
and vector self-energies (recall $\Sigma_s=M_N^\ast-M_N$) each have a
magnitude of a few hundred MeV with opposite signs for $f<0.5$, which is in
qualitative agreement with relativistic phenomenology.
As $t$ gets larger, smaller $f$ values are needed to produce large and
canceling
scalar and vector self-energies.
It is worth emphasizing that, in the interval of $t$ considered here, the
contributions of higher-order terms in the OPE become more important for larger
magnitudes of $t$, while the
continuum contributions become larger and the coupling of
the interpolating field to the quasinucleon states becomes weaker for smaller
magnitudes  of $t$.
Since there is less information about the higher-dimensional condensates, and
only the quasinucleon state is of interest here, one should not use a $t$ with
a magnitude that is too large or too small.
Within the range of $t$ considered in this section, the vector self-energy
is always large; the scalar self-energy is mainly controlled by the value of
$f$.

The nucleon $\sigma$ term $\sigma_N$ is a crucial phenomenological input in the
finite-density nucleon sum rules; its value determines the degree of chiral
restoration in the nuclear medium (to first order in the nucleon density).
The scalar self-energy strongly depends on $\sigma_N$ through
$\langle\overline{q}q\rangle_{\rho_N}$ and the four-quark
condensates \cite{JIN94,JIN93b}.
One observes that the large and canceling self-energies found with small and
moderate $f$ values mainly depend on the ratio $\sigma_N/m_q$.
An understanding of the cancellation between scalar and vector components will
be essential in making connections between QCD and nuclear physics.

Even if one assumes that the scalar-scalar four-quark condensate has weak
or moderate density dependence (so that the sum-rule predictions are
consistent with known relativistic phenomenology), there are still
important open questions to confront.
One must test sum rules for other baryons as well as for other nucleon
properties.
In Sec.~\ref{sec:hyp},
the finite-density sum-rule approach will be extended
to study the self-energies of hyperons in nuclear matter; there are
experimental data and phenomenological models to confront with
QCD sum-rule predictions.
Since the $\Delta$ sum rule is
especially sensitive to the scalar-scalar
four-quark condensate \cite{IOF81,LEI90}, one may obtain  some additional
phenomenological constraints on its density
dependence \cite{JIN94d}.

\subsection{Other Approaches}

We have focused our study of nucleons at finite density
on sum rules that use the formalism outlined in Ref.~\cite{FUR92}
to calculate {\it individually\/} the scalar and vector
self-energies of a nucleon in infinite nuclear matter.
(This approach was also discussed in Ref.~\cite{HAT91}.)
However, there are several other approaches to analyzing nucleons  at
finite density.
All of these approaches
start with the same correlator we have been studying, but
calculate different quantities.
Furthermore, in some cases a different OPE and different
dispersion relations are used.

The work by Drukarev and Levin pioneered many of the common features of
finite-density sum rules, such as the linear treatment of the quark and
gluon condensates \cite{DRU90}.
Their approach is presented in detail in a review article \cite{DRU91}
and several journal
articles \cite{DRU90,DRU91a}
and most recently by Drukarev and Ruskin \cite{DRU94}.
Here we simply comment on differences between their approach and ours.

One difference is in the choice of external kinematics,
which leads  to a different OPE.
Rather than exploiting the similarities to the vacuum sum rules, as we
have done by choosing unphysical kinematics that imply a
short-distance expansion, Drukarev and Levin
advocate an analogy to deep-inelastic scattering.
In this approach, the spacelike limit $q^2 \rightarrow -\infty$ is
taken with $q^2/\qdotu$ fixed, which is similar to the Bjorken limit.
This ensures that only the light cone is probed, but does not imply a
short-distance expansion.
Thus the appropriate OPE is a light-cone expansion, and one must
deal with all operators of a given twist.

These  authors also consider
finite-density dispersion relations
in $q^2$ (rather than $q_0$), with $s=(p+q)^2$ fixed (instead
of fixing $\vec q$).
Here $p^\mu$ is the four-momentum of a nucleon in the medium \cite{DRU91}.
These dispersion relations are assumed, rather than constructed
through a Lehmann representation.
One difficulty with this approach is that the identification of
quasinucleon intermediate states becomes obscured.

Drukarev and Levin use their finite-density sum rules to describe nuclear
matter saturation properties.
(They have also calculated $g_A$ in the medium \cite{DRU91}.)
Thus, rather than focusing on individual scalar and vector self-energies,
they study the shift of the quasinucleon pole.
Cancellations between scalar and vector contributions are still present
in the sum rules and were pointed out in Ref.~\cite{DRU91}.
Since the empirical pole shift is quite small on hadronic scales,
however, its
determination from sum rules is likely to be very uncertain.
For example, one would require detailed knowledge of the density
dependence of the condensates; it is clear that our present
understanding of this density dependence is insufficient.
We argue that one can establish whether QCD predicts large and canceling
scalar and vector self-energies far
more reliably than one can quantitatively
predict the net single-particle energy.

The work by Henley and Pasupathy \cite{HEN93}
is based heavily on the development of
Drukarev and Levin, but with a focus on different observables.
Instead of nuclear-matter saturation properties, Henley and Pasupathy
calculate nucleon-nucleus scattering.
In particular, they expand the correlator
to linear order in the density,
which allows them to isolate the forward nucleon-nucleus
scattering amplitude from the residue of the double pole
at the nucleon mass.
The advantage to this calculation
is that they make a more direct connection to actual physical observables,
the scattering amplitudes.
However, as noted in the Introduction, QCD sum rules are not well suited
for calculating such quantities because of delicate cancellations.

In a recent Letter \cite{KON93},
Kondo and Morimatsu advocate the
calculation of nucleon-nucleon scattering lengths using QCD sum-rule methods.
At the same time, they argue that the calculations we have described
in this review are ill-founded.
We disagree strongly with both aspects of their discussion.
The essential criticism is given in Refs.~\cite{FUR94,FUR94b}; here
we just summarize the main points.

The first problem is that the QCD sum-rule approach is not appropriate
for the calculation of $NN$ scattering lengths.
The large magnitudes of the empirical $NN$ scattering lengths reflect the
slightly and nearly bound states in the two scattering channels.
It is well known from conventional nuclear physics that
making very small changes in the strength of an assumed
$NN$ potential can make calculated
scattering lengths arbitrarily large or change their signs.  Sum rules
cannot hope to address the fine details of this physics, which
is determined at relatively large distances or times.

Despite this,
the sum-rule calculations of Kondo and Morimatsu appear to
predict scattering lengths qualitatively similar to the experimental
data.
However, their sum rules suffer from a subtle flaw that has been
alluded to earlier in this review.
Their predictions are acutely sensitive to asymmetries in the
spectral density between positive- and negative-energy states.
They neglected to account for this asymmetry in the effective
continuum thresholds in their analysis, and this led to spurious
results (see Ref.~\cite{FUR94} for details).
If one accounts properly for these asymmetries in their approach,
one finds results consistent with ours \cite{FUR94}.

\section{Hyperons in Nuclear Medium}
\label{sec:hyp}
\subsection{Overview}

Large and canceling
Lorentz scalar and vector self-energies for the in-medium nucleon,
which are predicted by relativistic phenomenology,
are made plausible by the nucleon sum-rule results.
Yet, as we have stressed, these results are also inconclusive at present;
thus it is important to test the sum-rule framework against
finite-density phenomenology and experiment in other contexts.
In this section, we discuss finite-density QCD sum rules
for the self-energies of $\Lambda$ and $\Sigma$ hyperons in nuclear matter.

In hypernuclei, hyperons
can occupy the lowest (innermost) shell-model orbits in a nucleus,
which should allow reasonable comparisons to theoretical
predictions for hyperons in uniform nuclear matter.
Furthermore, various
investigators have extended relativistic phenomenology to the
study of hypernuclear physics
\cite{BRO77,NOB80,JCO87,JEN90,GAT91,JCO91,MAR93,MAR94,GLE93,RUF90,SCH92}.
In these relativistic models,
the hyperon propagating in the nuclear medium is described by a Dirac
equation with scalar and vector potentials.
The potential depths result from
the coupling of the hyperons to the same scalar and vector fields
as the nucleon, but with different coupling strengths;
however, these coupling strengths are not well determined.
By solving the Dirac equation, one
can obtain the binding energies and spin-orbit splittings for
different states in hypernuclei, which can be confronted with
experimental data.

One of the compelling successes of relativistic models in describing
nucleon-nucleus interactions is the naturally large spin-orbit force
for nucleons in a finite nucleus.
This force depends on the
derivatives of the scalar and vector optical potentials, which add
constructively.
An analogous prediction for the $\Lambda$ hyperon
would seem to be problematic, if one adopts the {\it naive\/} quark-model
prediction that each coupling for the $\Lambda$ should be
$2/3$ the coupling for the
nucleon \cite{PIR79,DOV84,JEN90,GAT91,JCO91,MAR93,MAR94}.
[That is, if one assumes that the scalar
($\sigma$) and vector ($\omega$) mesons couple exclusively to the up
and down (constituent) quarks and not to the strange quark.]
In the $\Lambda$-nucleus system, recent experiments indicate that the
spin-orbit force is small, and even consistent with
zero \cite{BOU77,MIL88}.
This has raised questions about the
validity of relativistic nuclear phenomenology for hyperons.

An early work \cite{BRO77} achieved a small spin-orbit force by
taking the potentials ({\it i.e.\/}, the couplings) for the $\Lambda$ to be a
factor of three smaller than for the nucleon.
More recently, however, it has been
suggested that scalar and vector coupling strengths
consistent with the naive quark-model predictions can be used if a tensor
coupling between hyperons
and the vector meson ($\omega$) \cite{JEN90,GAT91,JCO91,MAR93,MAR94},
motivated from a quark-model picture \cite{JEN90,JCO91,MAR93}, is introduced.
In this picture, the tensor couplings of the hyperons
($\Lambda$, $\Sigma$, $\Xi$) to the vector meson differ in
their magnitudes and signs, and hence their contributions to the
spin-orbit force are different.
(This picture yields a negligible tensor coupling for the nucleon.)
For the $\Lambda$, this picture leads to a tensor coupling with strength
equal in magnitude to the corresponding vector coupling, but with the
opposite sign.
The net result, in combination with the scalar contribution,
is a small spin-orbit force.
A recent global
fit to hypernuclear data favors smaller couplings (from 0.3 to 0.5
times the nucleon-meson couplings, depending on details \cite{RUF90}),
but does not include this new tensor coupling.

While experimental evidence of $\Sigma$ hypernuclei is
lacking at present, a number of authors have extended the
relativistic phenomenology to $\Sigma$ hypernuclei and presented
theoretical predictions \cite{JCO91,MAR93,MAR94,GLE93}.
In Refs.~\cite{JCO91,MAR93,MAR94}, the
naive quark-model prediction for the coupling strengths of the $\Sigma$ to the
scalar and vector mesons was adopted
(the vector coupling for the $\Sigma$ is $2/ 3$ the coupling for the nucleon,
and the scalar coupling for the $\Sigma$ is slightly smaller than $2/ 3$
the coupling for the nucleon);
in addition, the tensor coupling between the $\Sigma$ and the
vector meson  was included.
In the quark-model picture of
Ref.~\cite{JCO91}, the tensor coupling of the
$\Sigma$ to $\omega$ has the {\it same\/} sign as the corresponding
vector coupling, in contrast to the $\Lambda$ case.
With the quark-model values for the tensor coupling, the spin-orbit
force for the $\Sigma$ was found to be comparable with the nucleon
spin-orbit force.
In Ref.~\cite{GLE93}, the tensor coupling was
omitted and universal couplings were assumed for {\it all\/} hyperons;
the ratio of the $\Sigma$ to $\Lambda$ spin-orbit force obtained
is about $0.9$.

QCD sum-rule predictions for the scalar and vector self-energies
of hyperons may offer
independent information on the scalar and vector couplings (or potential
depths)
for hyperons in the nuclear medium.
In addition, they may offer new insight into possible
SU(3) symmetry-breaking effects in the baryon self-energies
and into the spin-orbit forces for the hyperons.
We note, however, that the
prediction of tensor couplings is not tested in the sum-rule
results described below.

The finite-density sum rules for hyperons can be obtained following
the same methods as discussed in the previous sections of this review.
Here we will omit the details of the derivations and the actual sum
rules, and focus on the predictions for hyperons.
The reader is referred to Refs.~\cite{JIN94a,JIN94c} for further details.

\subsection{$\Lambda$ Hyperons}

The in-medium correlator for the $\Lambda$ is defined by
\begin{equation}
\Pi_\Lambda(q)\equiv i\int d^4 x\, \e^{iq\cdot x}
\langle\Psi_0|T\eta_{\Lambda}(x)\overline{\eta}_{\Lambda}(0)|\Psi_0\rangle\ ,
\label{corrl_lambda_def}
\end{equation}
where $\eta_{\Lambda}$ is a color-singlet interpolating field with the
spin, isospin, and strangeness of the $\Lambda$.
As in the nucleon case, we consider
interpolating fields that contain no derivatives and couple to
spin-${1\over 2}$ only.
Here we choose the interpolating field
\begin{equation}
\eta_{\Lambda}=\sqrt{2\over 3}\epsilon_{abc}
[(u_a^T C\gamma_\mu s_b)\gamma_5\gamma^\mu d_c
-(d_a^T C\gamma_\mu s_b)\gamma_5\gamma^\mu u_c]\ ,
\label{intfield_lambda}
\end{equation}
which has been used in the studies of vacuum sum rules for
the $\Lambda$ \cite{REI83,REI85,CHI85}.
In the SU(3) limit, this interpolating field leads to
the same results as Ioffe's current for the nucleon.

Since the $\Lambda$ hyperon decays only weakly, its width in free space
can be neglected on hadronic scales.
Experimentally, bound $\Lambda$ single-particle states with
well-defined energies have been observed in $\Lambda$ hypernuclei,
which indicates that a quasiparticle description of the $\Lambda$ is
reasonable; thus, one can adopt a pole approximation for the
quasilambda.
As in the nucleon sum-rule analysis, we choose a weighting function
to suppress the contribution
from negative-energy excitations (corresponding to the antiparticle).

The operator product expansion for the $\Lambda$ correlator is a direct
generalization of the nucleon OPE and is given in detail in
Ref.~\cite{JIN94a}.
The new features are the contributions of condensates involving the strange
quark (such as $\langle\overline{s}s\rangle_{\rho_N}$,
see Sec.~\ref{condensates})
and the strange-quark mass $m_s$.
There are two dimension-six scalar-scalar four-quark condensates in
the $\Lambda$ sum rules: $\langle\overline{q}q\rangle_{\rho_N}^2$ and
$\langle\overline{q}q\rangle_{\rho_N}\langle\overline{s}s\rangle_{\rho_N}$
\cite{JIN94a}.
As stressed in the nucleon case, the in-medium factorization
approximation may not be justified in nuclear matter.
Thus, we follow
the parametrization in Eq.~(\ref{4quark-p}) for the density dependence of
$\langle\overline{q}q\rangle_{\rho_N}^2$ and parametrize the density
dependence of
$\langle\overline{q}q\rangle_{\rho_N}\langle\overline{s}s\rangle_{\rho_N}$ as
\begin{equation}
\langle\widetilde{\overline{q}q}\rangle_{\rho_N}
\langle\widetilde{\overline{s}s}\rangle_{\rho_N}
=(1-f^\prime)
\langle\overline{q}q\rangle_{\rm vac}\langle\overline{s}s\rangle_{\rm vac}
+f^\prime
\langle\overline{q}q\rangle_{\rho_N}\langle\overline{s}s\rangle_{\rho_N}\ ,
\label{mix_4q_prap}
\end{equation}
where $f^\prime$ is a real parameter.
The $f^\prime$ values are
taken to be in the range $0\leq f^\prime\leq 1$ such that the condensate
interpolates between its factorized form in free space and
its factorized form in the nuclear medium.

In Fig.~\ref{lambda-res}, the optimized results for the normalized
self-energies $M_\Lambda^\ast/M_\Lambda$ and $\Sigma_v/M_\Lambda$ are
displayed as functions of $f^\prime$ for $m_s=150\,\mbox{MeV}$,
$y=0.1$,
$|\vec{q}|=270\,\mbox{MeV}$, and three different values of $f$.
One notes that $\Sigma_v/M_\Lambda$ is insensitive to both $f$ and $f^\prime$.
$M_\Lambda^\ast/M_\Lambda$, however, varies rapidly with $f$ and $f^\prime$;
therefore, the sum-rule prediction for the scalar self-energy is
{\it strongly\/} dependent on the density dependence of the four-quark
condensates.
For $f=0.25$ and values of $f^\prime$ in the range
$0.6\leq f^\prime\leq 1$, the predictions are
$M_\Lambda^\ast/M_\Lambda\simeq\mbox{0.85--0.94}$ and
$\Sigma_v/M_\Lambda\simeq\mbox{0.09}$.
On the other hand, for $f=0.25$ and small values of $f^\prime$
($0\leq f^\prime\leq 0.3$), one finds $\Sigma_v/M_\Lambda\simeq\mbox{0.08}$ and
$M_\Lambda^\ast/M_\Lambda\simeq\mbox{0.68--0.76}$.
As $f$ increases, $M_\Lambda^\ast/M_\Lambda$ decreases, which
implies an even larger magnitude for the scalar self-energy.
(In the nucleon case, $M_N^\ast/M_N$ {\it increases\/} as $f$ increases.)
Thus, a weak density dependence of {\it both\/}
scalar-scalar four-quark condensates leads to a moderate vector
self-energy and a very strong scalar self-energy (the net self-energy
is strong and attractive).
(A strong density dependence for $\langle\overline{q}q\rangle_{\rho_N}^2$
also yields a strong scalar self-energy.)
The sum-rule predictions for the self-energies turn out to be
insensitive to the values of $m_s$, $y$, and $|\vec{q}|$ \cite{JIN94a}.

\begin{figure}
\vbox to 3.0in{\vss\hbox to 5.625in{\hss
{\includegraphics{lambda.ps}}\hss}}
\caption{Optimized sum-rule predictions for $M^{\ast}_\Lambda/M_\Lambda$
and $\Sigma_v/M_\Lambda$ as functions of $f^\prime$.
The three curves correspond
to $f=0$ (solid), $f=0.25$ (dashed), and $f=0.5$ (dotted).}
\label{lambda-res}
\end{figure}

We observe that
the prediction for the normalized vector self-energy
$\Sigma_v/M_\Lambda$ is insensitive to the details of the calculations.
For typical values of the relevant condensates and other input
parameters, $\Sigma_v/M_\Lambda\simeq\mbox{0.08--0.09}$.
The finite-density {\it nucleon\/} sum rules predict
$\Sigma_v/M_N\simeq\mbox{0.25--0.30}$.
Thus, one finds
$(\Sigma_v)_\Lambda/(\Sigma_v)_N\simeq\mbox{0.3--0.4}$.
This result, if interpreted in terms of a relativistic hadronic model,
would imply that the coupling of the $\Lambda$ to the Lorentz vector
field is weaker than the corresponding nucleon coupling by the same ratio.
This compares to the naive quark-model prediction of
$2/3$, which is obtained by assuming that the mesons couple
directly to constituent quarks.

The predictions for the scalar self-energy are quite sensitive
to the undetermined density dependence of certain four-quark condensates.
If one assumes that the four-quark condensate
$\langle\overline{q}q\rangle_{\rho_N}^2$ depends only weakly on the
nucleon density ({\it i.e.\/}, if $f$ is small) and the four-quark condensate
$\langle\overline{q}q\rangle_{\rho_N}\langle\overline{s}s\rangle_{\rho_N}$
has a strong density dependence ({\it i.e.\/}, if $f^\prime$ is large),
one finds $M_\Lambda^\ast/M_\Lambda\simeq\mbox{0.85--0.94}$,
which implies $\Sigma_s/M_\Lambda\simeq -\mbox{(0.06--0.15)}$.
With the nucleon sum-rule prediction,
$M_N^\ast/M_N\simeq\mbox{0.65--0.70}$, one obtains
$(\Sigma_s)_\Lambda/(\Sigma_s)_N\simeq\mbox{0.2--0.4}$.
In a hadronic model,
this implies again a much smaller coupling of the $\Lambda$ to the
Lorentz scalar field than for the nucleon.
In this case, there is a significant degree of cancellation between the
scalar and vector self-energies.
This result is compatible with the
empirical observation that the $\Lambda$ single-particle energy (the
quasiparticle pole position) is shifted only slightly in nuclear matter
relative to its free space mass.
That is, the $\Lambda$ is very weakly bound.

On the other hand, if {\it both\/} $\langle\overline{q}q\rangle_{\rho_N}^2$ and
$\langle\overline{q}q\rangle_{\rho_N}\langle\overline{s}s\rangle_{\rho_N}$
depend either weakly or moderately on the nucleon density,
the predicted ratio $M_\Lambda^\ast/M_\Lambda$ is
significantly smaller than unity, implying that the scalar self-energy
is large and negative.
The predicted vector self-energy, on the other hand, is still moderate.
Thus, in this case the sum rules predict incomplete cancellation and
the resulting quasilambda energy is significantly smaller than the
mass.
This result is inconsistent with experiment.

\subsection{$\Sigma$ Hyperons}

The sum rules for $\Sigma$ hyperons are obtained by studying
the correlator
\begin{equation}
\Pi_\Sigma(q)\equiv i\int d^4 x\, \e^{iq\cdot x}
\langle\Psi_0|T\eta_\Sigma(x)\overline{\eta}_\Sigma(0)|\Psi_0\rangle\ ,
\label{corr_def_sigma}
\end{equation}
where $\eta_\Sigma$ is an interpolating field
with the quantum numbers of a $\Sigma$.
The interpolating field for the $\Sigma^+$ can be
obtained directly by an SU(3) transformation of Ioffe's current
for the nucleon \cite{REI83,REI85}:
\begin{equation}
\eta_{\Sigma^+}=\epsilon_{abc}
(u_a^T C\gamma_\mu u_b)\gamma_5\gamma^\mu s_c\ .
\label{intfield_sigma}
\end{equation}
The analogous interpolating field for $\Sigma^0$ follows
by changing one of the up-quark fields into a down-quark field
and then symmetrizing the up and down quarks, which can also
be written as:
\begin{equation}
\eta_{\Sigma^0}=\sqrt{2}\epsilon_{abc}
[(u_a^T C\gamma_\mu s_b)\gamma_5\gamma^\mu d_c
+(d_a^T C\gamma_\mu s_b)\gamma_5\gamma^\mu u_c]\ .
\end{equation}
(Note that $\eta_{\Sigma^0}$ and $\eta_{\Lambda}$ in
Eq.~(\ref{intfield_lambda}) have an identical structure, except for the
isospin couplings of the up- and down-quark fields.)
The results obtained by using $\eta_{\Sigma^+}$ and
$\eta_{\Sigma^0}$ are the same due to isospin symmetry.

The width of the $\Sigma$ in free space is small and can be
ignored on hadronic scales.
At finite density, the width of the $\Sigma$ will be broadened due to
strong conversions; here we assume that the broadened width is
relatively small on hadronic scales and that a quasiparticle
description of the $\Sigma$ is reasonable.
In the
context of relativistic phenomenology, the $\Sigma$ is assumed to
couple to the same scalar and vector fields as a nucleon in
nuclear matter and is treated as a quasiparticle with real Lorentz
scalar and vector self-energies.
Thus a pole ansatz is adopted
in our study, and Eq.~(\ref{sum_def}) can be utilized to minimize the
sensitivity to the negative-energy excitations.

The  expressions
for $\Pi_\Sigma$ and the three finite-density sum rules
are given in Ref.~\cite{JIN94c}.
Note that only the scalar-scalar four-quark condensate
$\langle\overline{q}q\rangle_{\rho_N}^2$
appears in the $\Sigma$ sum rules.
We use the parametrization of Eq.~(\ref{4quark-p})
for its density dependence.

The optimized sum-rule predictions for the ratios
$M_\Sigma^\ast/M_\Sigma$ and $\Sigma_v/M_\Sigma$
are plotted as functions of $y$ in Fig.~\ref{sigma-res} for
$m_s=150\,\mbox{MeV}$,
$|\vec{q}|=270\,\mbox{MeV}$, and three different values of $f$.
Again, the ratio $\Sigma_v/M_\Sigma$ is
insensitive to both $y$ and $f$.
The ratio $M_\Sigma^\ast/M_\Sigma$,
on the other hand, changes rapidly with $y$ and $f$, which implies
that the sum-rule prediction for the scalar self-energy is {\it strongly\/}
dependent on the strangeness content of the nucleon and on the density
dependence of the four-quark condensate.
For $f=0$ and values of $y$ in the range $0.4\leq y\leq 0.6$,
the predictions are
$M_\Sigma^\ast/M_\Sigma\simeq\mbox{0.78--0.85}$ and
$\Sigma_v/M_\Sigma\simeq\mbox{0.18--0.19}$.
On the other hand, for $f=0$ and small values of $y$ ($0\leq y\leq 0.2$),
one gets $\Sigma_v/M_\Sigma\simeq 0.18$ and
$M_\Sigma^\ast/M_\Sigma\simeq\mbox{0.92--0.98}$.
As $f$ increases, $M_\Sigma^\ast/M_\Sigma$ increases, which implies an even
smaller magnitude of the scalar self-energy.
The sum-rule predictions are
insensitive to the values of $m_s$ and $|\vec{q}| $\cite{JIN94c}.

\begin{figure}
\vbox to 3.0in{\vss\hbox to 5.625in{\hss
{\includegraphics{sigma.ps}}\hss}}
\caption{Optimized sum-rule predictions for
$M^{\ast}_\Sigma/M_\Sigma$ and $\Sigma_v/M_\Sigma$ as functions of
$y$. The three curves correspond to $f=0$ (solid), $f=0.25$
(dashed), and $f=0.5$ (dotted).}
\label{sigma-res}
\end{figure}

The sum-rule prediction for the normalized vector self-energy
$\Sigma_v/M_\Sigma$ is apparently insensitive to the details of the
calculations.
For typical values of the relevant condensates and other
input parameters, one obtains $\Sigma_v/M_\Sigma\simeq 0.18$--0.21.
The finite-density {\it nucleon\/} sum rules predict
$\Sigma_v/M_N\simeq\mbox{0.25--0.30}$; thus, we find
$(\Sigma_v)_\Sigma/(\Sigma_v)_N\simeq\mbox{0.8--1.1}$.
This result, if interpreted in terms of a relativistic hadronic model,
would imply that the coupling of the $\Sigma$ to the Lorentz vector
field is very similar to the corresponding nucleon coupling.
This compares to the naive quark-model prediction of
$2/ 3$, which is obtained by assuming that the mesons couple
directly to constituent
quarks \cite{PIR79,DOV84,JEN90,GAT91,JCO91,MAR93,MAR94}.

The sum-rule prediction for the scalar self-energy is sensitive
to the strangeness content of the nucleon and the density
dependence of the four-quark condensate
$\langle\overline{q}q\rangle_{\rho_N}^2$.
If we assume that the nucleon has
a large strangeness content ({\it i.e.\/}, $y\simeq 0.4$--0.6)
and the four-quark
condensate $\langle\overline{q}q\rangle_{\rho_N}^2$ depends only weakly
on the nucleon density ({\it i.e\/}, if $f\simeq 0$), we find
$M_\Sigma^\ast/M_\Sigma\simeq\mbox{0.77--0.84}$, which implies
$\Sigma_s/M_\Sigma\simeq -(\mbox{0.16--0.23})$.
With the nucleon sum-rule
prediction $M_N^\ast/M_N\simeq 0.65$ (for $f\simeq 0$), we obtain
$(\Sigma_s)_\Sigma/(\Sigma_s)_N\simeq\mbox{0.6--0.85}$.
In a hadronic model,
this implies again a coupling of the $\Sigma$ to the Lorentz scalar
field close to that for the nucleon.
In this case, there is a
significant degree of cancellation between the scalar and vector
self-energies, which is compatible with that implemented in the
relativistic phenomenological models.
[We note, however, that Eq.~(\ref{wagner}) suggests values of the strangeness
content in the range $y\simeq 0$--0.45.]

In contrast, if the strangeness content of the nucleon is small
({\it i.e.\/}, $y\leq 0.2$) or if $\langle\overline{q}q\rangle_{\rho_N}^2$ has
a significant dependence on the nucleon density, the predicted ratio
$M_\Sigma^\ast/M_\Sigma$ is close to unity, implying that the scalar
self-energy is very small.
The predicted vector self-energy, on the
other hand, is still essentially the same as the nucleon vector self-energy.
Thus, in this case, the sum rules predict incomplete
cancellation and, hence, a sizable repulsive net self-energy for the
$\Sigma$.
This result is at odds with the predictions of relativistic models.

\subsection{Summary}


The QCD sum rules discussed here predict that the $\Lambda$ vector self-energy
is significantly smaller than the nucleon vector self-energy
while the $\Sigma$
vector self-energy is close to that for the nucleon.
However, just as for the nucleon, the sum-rule predictions
for the  hyperon scalar self-energies are obscured
by the strong dependence on the values of four-quark
condensates and on the strangeness content of the nucleon
 (in the $\Sigma$ case).

Despite the uncertainties, one finds in general that
sum-rule predictions for the scalar and vector self-energies imply
a much weaker spin-orbit force for the $\Lambda$ in a nucleus than that felt
by a nucleon.
The predictions for the $\Sigma$ scalar and vector
self-energies seem to imply a spin-orbit force for the
$\Sigma$ comparable to that for a nucleon, but much stronger than
that for a $\Lambda$.
These results are compatible with  experiment and with predictions of
 relativistic phenomenological models with an extra tensor coupling
between the hyperons and the vector meson
\cite{JEN90,GAT91,JCO91,MAR93,MAR94}.
Note that the magnitude of the tensor contribution cannot be estimated
from QCD sum rules  in uniform nuclear matter.

It is worth emphasizing that in
Refs.~\cite{JEN90,GAT91,JCO91,MAR93,MAR94}, scalar and vector couplings
consistent with the naive quark-model predictions have been adopted, and
it is the extra tensor coupling of the hyperons to the vector meson
that reduces the $\Lambda$ spin-orbit force and enhances the $\Sigma$
spin-orbit force.
The sum-rule predictions, on the other hand, suggest that it is the
weak scalar and vector couplings for the $\Lambda$ and strong scalar
and vector couplings for the $\Sigma$, deviating from the naive
quark-model prediction, that lead to a small $\Lambda$ spin-orbit force
and a large $\Sigma$ spin-orbit force, respectively.
We also note that the sum-rule predictions do not agree with the
universal coupling assumption ({\it i.e.\/}, all hyperons couple to the
scalar and vector fields with the same strength) suggested in
Ref.~\cite{GLE93}.

\section{Vector Mesons in Medium}

\subsection{Introduction}

This review has primarily stressed the problem of baryons---nucleons
and hyperons---propagating in nuclear matter.
In this section, we will briefly consider a different problem, namely
the propagation of vector mesons in the nuclear medium.
Much of the sum-rule machinery carries over directly from our previous
discussion; further details can be found in the cited literature.

The possibility that the properties of vector mesons might change
significantly with increasing density is of considerable current
theoretical interest.
For example, Brown and Rho \cite{BRO91} have proposed a scenario in
which the vector-meson masses in the medium decrease.
Experimentally, the question of how vector mesons behave at finite
density is an open question.
If the masses decrease, then one effect should be an increased range of
propagation, so that the effective size of the nucleons as ``seen'' in
hadron reactions mediated by vector mesons will increase.
Unfortunately, it is difficult to isolate the meson-mass effect of
``swollen'' nucleons due to decreased vector-meson masses from other
physics arising from nuclear many-body effects or the substructure of
the hadrons.
Given this generic difficulty, attempts to extract the meson-mass
effect are fraught with ambiguity.

Various pieces of experimental evidence have been proposed in support
of the picture of decreasing meson masses.
One example is the quenching of the longitudinal response (relative to
the transverse response) in quasielastic electron scattering
\cite{ALT80} and $(e,e'p)$ \cite{REF88} reactions, which might reflect
``swollen'' nucleons \cite{NOB81,CEL85} in the medium.
As noted in Refs.~\cite{BRO89,SOY93}, this swollen-nucleon picture
emerges naturally if the electromagnetic current couples to the
nucleons through the vector meson (at least partly) and if the
vector-meson mass in medium drops.
A second example is the discrepancy between the total cross section in
$K^+$-nucleus scattering on $^{12}$C and that predicted from an impulse
approximation calculation using $K^+$-nucleon scattering amplitudes
(extracted from $K^+$-D elastic scattering).
The discrepancy is removed \cite{BRO88} if there is an effective
increase in the nucleon's in-medium cross section due to a decrease in
the mass of the $\rho$ meson mediating the interaction.

More direct investigations of vector-meson masses in the nuclear medium
have also been proposed.
One proposal is to study dileptons as a probe of vector mesons in the
dense and hot matter formed during heavy-ion collisions \cite{DIL}.
The dilepton mass spectra should allow one to reconstruct the mass of
vector mesons decaying electromagnetically.
A potentially cleaner probe (less leptonic background) would be via
virtual compton scattering---the ($\gamma,e^-e^+$) reaction \cite{MOS94}.

There are a number of potential difficulties with these probes.
One concerns the lifetimes of the vector mesons.
The $\omega$ and $\phi$ mesons are rather long-lived; thus, there is a
large probability that, regardless of where they are created, they will decay
{\it outside\/} the region in which the matter is dense.
This raises the question of whether these decays will provide useful
information about the dense matter region.
A second difficulty is that, in the nuclear medium, the position of the
pole (assuming pole dominance) is not a function of the four-momentum
$q^2$ only; it can also depend explicitly on the three-momentum $\vec{q}$.
Thus, studies of the invariant mass of the dileptons may not be
sufficient to pick out cleanly the effects of mass shifts.
Finally, there is no reason to suppose that the longitudinal and
transverse components behave in the same way---indeed one expects them
to be different, and their difference may complicate attempts to extract
the masses from data.


\subsection{Formulation of the Sum Rule}

The study of vector mesons in the nuclear medium via QCD sum rules
offers a theoretical complement to these experimental efforts.
The pioneering studies at finite density were by Hatsuda and Lee
\cite{HAT92}, and there have been subsequent calculations by Asakawa and
Ko \cite{ASA93}.
The treatment is quite similar to the treatment of the baryons
discussed in previous sections of this review.
Here we will only sketch the analysis, with an emphasis on the
differences between the baryon and vector-meson cases.
Our notation will not follow the notation of Refs.~\cite{HAT92,ASA93}
precisely; instead we introduce a more general form for the correlator
that allows for calculations away from the $\vec{q}=0$ point.

The interpolating fields used for these studies and for the studies of
vector mesons in the vacuum \cite{SHI79} are the conserved vector
currents of QCD:
\begin{equation}
J_\mu^\rho=\mbox{$1\over 2$}
(\overline{u}\gamma_\mu u-\overline{d}\gamma_\mu d)\ ,
\qquad
J_\mu^\omega=\mbox{$1\over 2$}
(\overline{u}\gamma_\mu u+\overline{d}\gamma_\mu d)\ ,
\qquad
J_\mu^\phi=\overline{s}\gamma_\mu s \ .
\end{equation}
The correlator is defined as
\begin{equation}
\Pi_{\mu\nu}(q)\equiv\int d^4 x\, \e^{iq\cdot x}
\langle\Psi_0|TJ_\mu(x)J_\nu(0)|\Psi_0\rangle\ ,
\end{equation}
where $|\Psi_0\rangle$ is the nuclear matter ground state, $T$ is the
covariant time-ordering operator \cite{ITZ80},
and $J_\mu$ represents any of the three vector currents.

To proceed, it is necessary to identify the various tensor structures
that can appear in this problem.
A simplification is that each of these currents is exactly conserved:
$\partial_\mu J^\mu=0$.
The covariant time-ordered product is defined such that \cite{ITZ80}
\begin{equation}
q^\mu\Pi_{\mu\nu}(q)=q^\nu\Pi_{\mu\nu}(q)=0\ .
\label{cons}
\end{equation}
This constraint plus the fact that there are two four-vectors, $q_\mu$
and $u_\mu$, from which to construct tensor structures allows us to
deduce that there are two independent tensor structures:
\begin{equation}
\Pi_{\mu\nu}(q) = (q_\mu q_\nu-q^2 g_{\mu\nu})\Pi_T(q^2,(q\cdot u)^2)
+q^2\left(n_\mu n_\nu\over n^2\right)
[\Pi_T(q^2,q\cdot u)-\Pi_L(q^2,(q\cdot u)^2)]\ ,
\label{vecform}
\end{equation}
where we define $n_\mu\equiv u_\mu-(q\cdot u/q^2)q_\mu$.
It is easy to see that each of these structures satisfies
Eq.~(\ref{cons}).
The first term in Eq.~(\ref{vecform}) is the familiar structure seen in
the vacuum correlator; clearly, the second term must vanish as the
density goes to zero.
We choose to normalize the longitudinal and transverse correlators,
$\Pi_L$ and $\Pi_T$, such that, with $q^\mu=(\omega,0,0,|\vec{q}|)$,
$\Pi_{00}=\vec{q}^2\Pi_L$, $\Pi_{11}=\Pi_{22}=q^2\Pi_T$, and
$\Pi_{33}=\omega^2\Pi_L$.

In principle, one can derive sum rules for $\Pi_L$ and $\Pi_T$, or any
linear combination of the two.
In practice, all calculations done to date have been with $\vec{q}=0$
(in the rest frame of the matter); as $\vec{q}\rightarrow 0$ the
longitudinal and transverse polarizations become degenerate, since
there is no way to distinguish longitudinal from transverse.
The issue of whether sum rules for certain combinations of invariant
functions are more effective than others will arise, however, as soon
as calculations away from $\vec{q}=0$ are attempted.

In the baryon case, we noted that the correlation function, considered
in the rest frame as a function of $q_0$, has both even and odd parts.
The reason is that, at finite baryon density, a baryon in medium
propagates differently than an antibaryon, yielding a correlation
function that is asymmetric in the energy variable.
In contrast, for the correlators in the vector channels, both time
orderings correspond to the creation of the vector meson, which is its
own antiparticle.
Accordingly, the spectral function is necessarily an even function of
energy, and one can write the dispersion relation as an integral over
the energy squared.%
\footnote{This is true for {\it neutral\/} vector mesons in any type
of baryonic matter.
For charged $\rho$ mesons, however, $\rho^+$ is the antiparticle of
$\rho^-$.
In an isoscalar medium, they are degenerate and the spectral function
is even.
If, however, one considers a medium with a net isospin density (such as
neutron matter), they cease to be degenerate and the spectral function
is no longer even.}
Since we work here at $\vec{q}=0$ in the rest frame of the medium, we
can write the  correlator as $\Pi_L(\omega^2)$ with
$\omega\equiv q_0$.

Having decided to fix $\vec{q}=0$, the calculation goes forward in the
standard way.
One starts with a dispersion relation
\begin{equation}
\Pi_L(\omega^2)=\frac{1}{\pi}\int_0^\infty ds\,
\frac{{\rm Im}\,\Pi_L(s)}
{s-\omega^2-i\epsilon}\ .
\end{equation}
In general we would need to make a subtraction, but as usual we convert
to a more effective weighting function in the integral by applying a
Borel transform.
The result is:
\begin{equation}
\widehat{\Pi}_L(M^2)=\frac{1}{\pi}\int_0^\infty ds\,
W(s)\ {\rm Im}\,\Pi_L(s)\ , \ \ \
W(s)=\e^{-s/M^2}
\end{equation}
In the nucleon case, we had to isolate the nucleon contribution from
the antinucleon contribution, which led us to an asymmetric weighting
function designed to minimize this contamination (see Sec.~\ref{sec:bsr}).
In the present case, this issue does not arise and we can use the
Borel weighting identical to the vacuum case.

To proceed, one follows the same steps described for baryon sum rules:
\begin{enumerate}
  \item  compute the Wilson coefficients;
  \item  estimate the corresponding condensates;
  \item  devise a phenomenological model for the spectral density that
  contains a few free parameters;
  \item  determine the parameters by finding
  the best match of the phenomenology to the OPE.
\end{enumerate}
We focus here on some issues raised in these sum rules and refer the
reader to the literature \cite{HAT92,ASA93}
for specific details of the actual calculations and results (including
results for the $\phi$ meson).

The calculation of the Wilson coefficients involves no new subtleties;
the coefficients can be calculated using the techniques outlined in
Sec.~\ref{wilson:coe}.
Since finite-density condensates are properties of the medium and not
the probe, the same condensates estimated earlier for baryon sum rules
enter here.
The reader should be reminded that these estimates are rather crude.
Typically we only estimate these condensates by folding nuclear matrix
elements with the nuclear density.
While this may be adequate for normal nuclear matter
density, it is likely to be problematic at high densities
({\it e.g.}, 2--3 times normal nuclear density).
This should be kept in mind in the context of heavy-ion collisions,
which produce such high densities.
The appropriate nuclear matrix elements come from a variety of sources,
and, in some cases, modeling is necessary.

In the case of the nucleon, we found that the sum-rule predictions were
extremely sensitive to the value of certain dimension-six four-quark
condensates.
The hypothesis of ground-state saturation led to results in strong
contradiction with experiment; furthermore, the theoretical prejudices
in favor of the hypothesis are not strongly founded.
This issue is also extremely important in the context of the
vector-meson sum rules.
In the vacuum sum rule for the vector mesons, four-quark condensates
provide the dominant correction to the perturbative term in the OPE.
(Note, however, that these are not the same four-quark condensates as
in the nucleon case.)
In the finite-density sum rules, the density dependences of these
condensates largely determine the predicted density dependence of the
vector-meson mass.
Since all in-medium vector-meson calculations to date assume
factorization, we must interpret the results with caution.

To implement the sum rule, one needs a model for the spectral
function.
In Ref.~\cite{HAT92}, the following ansatz is adopted:
\begin{equation}
R\,{\rm Im}\Pi_L(\omega^2)=\rho_{\rm sc}\delta(\omega^2)
+F\delta(\omega^2-m^{\ast 2})
+\left(1+\frac{\alpha_s}{\pi}\right)\theta(\omega^2-s_0)\ ,
\label{rhoansatz}
\end{equation}
with $R=8\pi$ for the $\rho$ and $\omega$ mesons, and $R=4\pi$ for the
$\phi$ meson.
The effective mass $m^\ast$ corresponds to the position of the propagator
pole.
The second and third terms of Eq.~(\ref{rhoansatz}) represent
a pole-plus-continuum ansatz, exactly as in the vacuum sum rules.

The first term is a scattering term, which corresponds to Landau
damping.
The form of this scattering term requires some explanation.
For a nonzero three-momentum $\vec{q}$, the imaginary part of the
correlator has support in the {\it spacelike\/} region
($\omega^2<\vec{q}^2$).
Indeed, in electron scattering these spacelike excitations are all that
one can probe.
For the propagation of modes, these spacelike excitations act as
damping terms.
In general, the analytic structure of the correlator in the $\omega$
plane corresponding to these spacelike excitations is a cut.
However, as $\vec{q}\rightarrow 0$, this cut gets squeezed to the
$q^2=0$ point and reduces to a delta function in the imaginary
part.
A detailed description of the scattering term can be found in
Ref.~\cite{FUR90}.
In practice, the contribution from this term is quite small for the
in-medium vector meson sum rules \cite{HAT92}.

It is not yet clear to what extent the ansatz in Eq.~(\ref{rhoansatz}) is
justified at finite density.
A central question is whether there are regions of large spectral
density away from the meson pole.
One possible source for the $\rho$-meson channel is
from $\pi$-$\Delta$-hole states.
In model calculations \cite{HER92}, these give rise to significant
spectral strength well isolated from the $\rho$ pole.
If there is a large low-$\omega^2$ contribution from the $\pi$-$\Delta$-hole
physics that is not included in the model for the spectral density, the
fitted mass for the $\rho$ pole will be significantly lower than the
actual mass.
The calculation of Asakawa and Ko \cite{ASA93} attempts to include the
$\pi$-$\Delta$-hole contributions.
The results of their sum-rule calculations are quite insensitive to the
effects of the  $\pi$-$\Delta$-hole contributions.
On the other hand, the spectral densities extracted by Asakawa and Ko
are quite different from those of the model calculations \cite{HER92}.

In the end, Hatsuda and Lee find a substantial reduction in the $\rho$
and $\omega$ masses in medium.
For low densities, they find \cite{HAT92}
\begin{equation}
\frac{m^\ast_{\rho,\omega}}{m_{\rho,\omega}}
\simeq 1-(0.18 \pm 0.05)\frac{\rho_N}{\rho_N^{\rm sat}}\ .
\end{equation}
The decreasing mass arises primarily due to the density dependence of
the four-quark condensates and $\langle\overline{q}iD_0 q\rangle_{\rho_N}$.
Asakawa and Ko find a somewhat stronger density dependence for the
$\rho$ meson, with its mass reduced to about 530 MeV at nuclear matter
density.
These results are consistent with the hypothesis of decreasing vector
meson masses in the nuclear medium, although the reader should keep in
mind the many caveats we have cited.
In particular, we are concerned about the applicability of a
quasiparticle-pole
ansatz and the use of factorized four-quark condensates.
Future work should consider the density dependence of the coupling constant
and the momentum dependence of the longitudinal and transverse effective
masses.

\section{Discussion}

\subsection{Critical  Assessment of the QCD Sum-Rule Approach}

The QCD sum-rule approach to hadronic properties is now rather mature.
The seminal work of Shifman, Vainshtein, and Zakharov
is more than 15
years old and has
spawned an entire industry---the original  paper \cite{SHI79} has more
than 1400 citations.
Sum-rule methods have produced
phenomenological  successes for a wide range of problems,
some far beyond the
original application of meson masses and couplings.
The
current state of affairs is  summarized in the
review/reprint volume edited by Shifman \cite{SHI92}.

Despite the popularity and apparent  success of QCD sum rules, there
remain concerns about the validity of various aspects of
the approach.   Many of these issues are addressed in
Ref.~\cite{SHI92}. Here we briefly
review some of the questions that carry over in particular
to the problem of hadrons in medium.

One central issue is
the  predictive power of the approach for
low-energy hadronic observables.
While the OPE may predict with reasonable accuracy certain weighted
integrals over physical spectral densities,%
\footnote{Assuming that the relevant condensates and Wilson
coefficients can be determined with sufficient accuracy.}
this is insufficient to make solid
predictions for properties of stable particles or resonances.
QCD sum rules require nontrivial knowledge
of the spectral density, which is used to formulate a spectral ansatz.
That is,
in order for the sum rules to be predictive, one must already know
something about the spectral functions---one cannot make pure {\it a
priori\/} predictions.  This point  should be kept in mind when
discussing sum rules for hadrons in medium;  one's knowledge about the
form of spectral densities for this problem is certainly less than for
the vacuum case.    If assumptions about the finite-density spectral
density are qualitatively wrong, then the
sum rule may produce  highly misleading results.  It is accordingly
very important to have the best possible model for the phenomenological
side.

A second issue at the core
of the approach is the validity of the OPE {\em as actually
implemented in QCD sum-rule calculations}. It has been argued \cite{NOV85}
that there is {\em always\/} an OPE that separates long-distance physics
(summarized in condensates) from short-distance physics in Wilson
coefficients.  The question that must be addressed is the extent to
which the Wilson coefficients may be calculated via low-order
perturbation theory.
Shifman argues \cite{SHI92} that the
success of the QCD sum-rule approach phenomenologically justifies the
``put all nonperturbative contributions into the condensates and calculate the
Wilson coefficients perturbatively'' approach for most cases.
There
are cases in which this approach seems to fail, such as for glueball channels.
Others have called it into question for the nucleon channel.  For
example, in Refs.~\cite{DOR90,FOR93}, it is claimed that direct-instanton
effects in the Wilson coefficients  may
play an important role in stabilizing the nucleon sum rule.   These
nonperturbative direct-instanton effects can occur, since instantons
may be smaller than the
separation scale used in the OPE, and  will still play a nonperturbative role.
In
general the role of instantons in the OPE remains an open question.
Our assumption has been
that  perturbation theory is sufficient
to calculate the Wilson coefficients with acceptable accuracy.

A third issue is the estimation of
condensates.  To the extent that the sum rules are only
sensitive to a few low-dimensional condensates, it is probably
reasonable to assume that the important  condensates can be determined
with sufficient accuracy by fitting to a number of different channels.
So long as there are fewer important condensates then there are
observables, one will  be able to deduce their values from some
set of observables and still have predictive power for other observables.

A central question, however, is  the validity of the vacuum-saturation or
factorization hypothesis for the four-quark condensates,
which play an essential
role in sum rules for light-quark hadrons.
The
importance of these condensates despite their relatively high dimension
(six) is simply understood---the Wilson coefficients associated with
these condensates are anomalously large, because they
do not have the small numerical factors
associated with loop integrals \cite{SHI79}.

In the original work of SVZ \cite{SHI79},
a duality argument was given in
favor of the vacuum-saturation hypothesis.  Subsequently a large-$N_c$
argument was also advanced \cite{NOV84}.  In our view,
neither argument is especially compelling.
The large-$N_c$ argument can be questioned since higher-order terms in
the $1/N_c$ expansion are not necessarily negligible given the fact that
$N_c$ is only three.
The duality argument was
based on saturating $\langle \overline{q}q \overline{q} q \rangle$ with
both two-pion states and the vacuum:
\begin{equation}
 \langle \overline{q}q \overline{q} q \rangle = | \, \langle 0
 |  \overline{q}q | 0 \rangle \, |^2  \, + \, \sum_{2 \pi\  {\rm
 states}} | \, \langle 0 |  \overline{q}q | \rm{2 \pi} \rangle
 \, |^2   \ .
\end{equation}
If one assumes that the two-pion intermediate
states are {\em free\/} noninteracting pions, then one can show using
PCAC-type arguments that the contribution from the pions is much
smaller than the vacuum term unless one sums up to very large
relative pion energies $\sim 2$\,GeV.  If the sum were cut off by other
physics at a more moderate scale ({\it e.g.\/}, the $\rho$-meson mass),
then the
two-pion contribution is  small.
However, the two-pion states need not dominate the correction to the
vacuum-saturation contribution:  multiple pion states can contribute.
Moreover, the assumption that the two-pion states behave as free pions is
only valid right at threshold.

Apart from these two arguments, the only real justification for the
factorization hypothesis is {\it a posteriori}.   Namely, it seems to
work phenomenologically in the context of QCD sum rules.  For example,
Shifman \cite{SHI92}  writes that for most channels the phenomenology is
consistent with deviations from factorization at only the 10\% level
or less.

As we have noted, the values of the dimension-six four-quark
condensates are the most significant uncertainties in QCD sum-rule
treatments of hadrons in nuclear  matter.  The most naive
generalization of vacuum saturation  to this problem is ground-state
saturation.    We can argue based on phenomenological concerns that it
is {\em not\/} a good approximation at nuclear matter densities.  In
this context it is worth noting that none of the three arguments in
favor of vacuum saturation---duality, large $N_c$, and
phenomenology---give much support to the ground-state saturation
hypothesis at finite density.
Results given earlier imply that
phenomenology is  inconsistent with
ground-state saturation, at least so far as the nucleon sum rules are
concerned.

While neither the duality nor large-$N_c$
arguments are compelling for the vacuum case,
for the case of finite
density matter they are even weaker.  The duality argument was based on
the contribution of two-pion states.  At finite densities, however, in
addition to two-pion states there are very many low-lying particle-hole
type states in the system.  Thus, it is by no means clear that the ground
state
will dominate over contributions from these new states.  With
regard to the large-$N_c$ argument, it should be noted that the $1/N_c$
expansion alone is known to be unreliable
for estimating the relative sizes of nuclear matter observables.
For example,
a large-$N_c$ analysis gives the kinetic energy of nucleons in nuclear
matter as being $O(N_c^{-1})$, while the potential energy is
$O(N_c)$.  In realistic nuclei at nuclear matter densities, the
two quantities are, in fact, of similar magnitudes \cite{COH89}.

A final problem to consider is that our results may to some degree be
artifacts of the interpolating fields chosen for the analysis.
There is always this danger when predictions are not independent of the
interpolating fields, even though there are compelling arguments why one
choice of field is superior to another.
This sort of question is best answered by numerical simulations.
Unlike the case of the vacuum, the prospects for useful numerical simulations
of nuclear observables is quite unclear.
One can hope, however, for some future
lattice calculations that
will test the basic ideas of scalar and vector physics that
we have explored with sum rules and
provide some
definite insight into the density dependence of the four-quark condensates.

\subsection{Simple Spectral Ansatz}
\label{sec:ansatz}

As emphasized above,
in order for sum rules to be useful in practice, one must be able to
make a simple parameterization of the spectral function.  Typically,
for vacuum sum rules,
one uses a pole plus continuum ansatz, where the continuum is modeled
as the perturbative spectral density (plus OPE contributions) starting at
a sharp threshold.  Clearly, such  a
parameterization is a major simplification of a complicated spectrum.
The important question is whether
this simplification leads to  significant uncertainties in the application
of QCD sum rules.

While there have not been direct tests of the continuum model
based on the
Borel-transformed correlator,  there have been tests of a lattice
analog---the correlator as a function of imaginary time.
At three-momentum $\vec{q}=0$ (achieved by summing uniformly over spatial
directions), the correlator at Euclidean time $\tau$
can be written as%
\footnote{Note that if $\tau$ is small, only small $x^2$ is
probed, so a short-distance OPE is appropriate in this regime.\/}
\begin{equation}
  \Pi (\tau,\vec q=0)=\int_0^{\infty}
          d\omega\, \e^{-\omega\tau}\rho(\omega)\ ,
          \label{eq:latt}
\end{equation}
where $\rho$ is the spectral density.
This lattice correlator
simply corresponds to a weighted integral of the spectral density with a
different weighting function than in the Borel transform.
Like the Borel transform, this
weighting  suppresses the high-energy states and improves
convergence of the OPE \cite{SHU84}.
(Note, however, that the suppression is greater with Borel weighting.)
We also observe natural association of short times with a broad smearing of
the spectral function.

Recently, Leinweber has tested the simple spectral
ansatz in fits to lattice data of the nucleon correlator \cite{LEI94}.
He found
that the pole-plus-continuum model
did a good job of  describing the
correlator over a surprisingly large  range of  $\tau$, including most of
the intermediate range between the perturbative regime and the large-$\tau$
regime.
Only the latter regime is dominated by the nucleon pole.
Moreover, fits to the
nucleon mass based on the ansatz in the
intermediate-$\tau$ range gave remarkable
agreement with the nucleon mass extracted from the correlator evaluated
at large times.  This test supports the pole-plus-continuum ansatz
commonly used in sum-rule applications.  Some caveats should be made
concerning this test \cite{LEI94}:   The lattice calculations were done in the
quenched approximation, required a sizable   extrapolation in
quark mass $m_q$, and were confined to distances somewhat larger
than those used in QCD sum-rule fits.
We hope similar calculations without these limitations will be made
eventually.

While lattice calculations suggest that the simple
pole-plus-continuum model
works well in (or at least near) the regime of interest,
there are concerns on the horizon.  Recently it was pointed
out \cite{GRI94} that one can use the known behavior of various quantities
with changes in
the quark mass (in the small quark-mass limit) to test the
consistency of the pole-plus-continuum ansatz.
When studying the nucleon correlator with the OPE, one finds terms
proportional to $\langle\overline{q}q\rangle_{\rm vac}$, which is known
from chiral perturbation theory \cite{NOV81} to behave as follows near
the chiral limit:
\begin{equation}
\langle\overline{q}q\rangle_{\rm vac}
=\langle\overline{q}q\rangle_{\rm vac}^\chi
\left(1 - \frac{3}{32\pi^2 f_\pi^2}
m_\pi^2\ln{m_\pi^2\over M_0^2}+O(m_\pi^4)\right)\ ,
\end{equation}
where $\langle\overline{q}q\rangle_{\rm vac}^\chi$ is the value of the
quark condensate in the chiral limit and $M_0$ is some mass parameter.
Thus, the OPE side
of the sum rule has a term that behaves like $m_\pi^2\ln m_\pi^2$.
On the other hand, the leading nonanalytic term (in $m_q$ or
$m_\pi^2$) in the nucleon mass is known to be
$-(3g_A^2/32\pi f_\pi^2)m_\pi^3$ \cite{GAS88}; there is no
$m_\pi^2\ln m_\pi^2$ term.
To make the phenomenology consistent with the OPE,  the continuum
must produce a term that goes as $m_\pi^2\ln m_\pi^2$ to
exactly cancel the term in the OPE.
Such a contribution to the continuum is not possible in a simple
pole-plus-continuum model.
On the other hand, it is
possible to use soft-pion theorems to show that, if the continuum model
includes pions, then the pion contributions automatically reproduce the
chiral log seen in the OPE \cite{CHO94}.
A rough estimate of the
uncertainty in the nucleon mass due to the neglect of pions in the
continuum model is 100 MeV \cite{GRI94}, the size of the contribution of the
$m_\pi^2\ln m_\pi^2$ term with $M_0$ chosen to have a
``reasonable'' value of $\sim 1$\,GeV.

\subsection{Does Dirac Phenomenology Make Sense?}
\label{sec:diracsense}

One of the principal motivations for using QCD sum rules to describe
nucleons propagating in nuclear matter is to test the essential
qualitative physics underlying Dirac phenomenology.   In light of this
goal, some comments on the controversy
about the validity of the Dirac
approach \cite{BRO84,THE85,COO86,BLE87,ACH88,JAR91,COH92,WAL94}
are appropriate. This
controversy has been long running and  is not considered as
settled.  However,
we argue that the main criticism of the Dirac phenomenology
is not germane  to the question of interest here: the magnitudes and
signs of the scalar and vector parts of the optical potential.

To those who doubt the entire Dirac approach, the question  ultimately
comes down to whether it makes sense to describe a composite particle
such as the nucleon by a Dirac equation \cite{BRO84,BLE87,ACH88,JAR91}.
Often this is expressed in
terms of the role of
Z graphs, {\it i.e.\/}, scattering from a positive-energy state into a
negative-energy state and back.  A fundamental difference between a
Dirac description and a Schroedinger description is the
possibility of  these Z graphs.  The essence of the criticism of
Dirac phenomenology is that the composite nature of the nucleon
suppresses Z graphs and, in doing so, suppresses the relativistic
effects \cite{BRO84,BLE87,JAR91}.
While one cannot yet directly address this
question in QCD, it is clear, in the context of toy models
\cite{BLE87,JAR91,COH92,WAL94} and large-$N_c$ QCD \cite{COH92}, that
the composite nature of the nucleon does suppress the coupling to
nucleon-antinucleon pairs (relative to what one expects with pointlike
Dirac nucleons).
What these simple models show is the unsurprising fact that a
composite nucleon will not behave the same way as a Dirac nucleon.  On
the basis of this result one may be tempted to throw out Dirac
phenomenology, which is, after all, based on the Dirac equation.

However, this observation is completely beside the point.
The key question is the nature of nucleon propagation: What are the on-shell
self-energies or
optical potentials?
Suppose we wish to describe
proton-nucleus scattering with an optical potential.  To do so, one must
suppress explicit reference to both the excitations of the nucleus and
the internal excitations of the nucleon (which is a necessary
consequence of the nucleon's composite structure).  The cost of
suppressing  explicit references to these degrees of freedom is
  that we get complex and energy-dependent potentials.  Now
suppose that one wishes to describe the nucleon scattering problem
covariantly as well.
The result will be a Dirac optical potential.  While we do
not know how to calculate the optical potential from first principles, we can
use general symmetry considerations to deduce its Lorentz structure.
We learn that
such an optical potential has both scalar and vector pieces.  It is
ultimately an empirical question whether the scalar and vector
potentials are separately large or small.

How can one reconcile  large scalar optical potentials (and hence large
Z-graph contributions) in the optical-model description with the fact
that  compositeness suppresses the amplitude to create $\overline{N}N$
pairs?  The simple answer is that there is nothing to
reconcile; they are completely different things.  The Z graphs in the
optical potential need not be associated with virtual $ \overline{N}N$
pairs.  The intuitive connection between Z graphs and virtual
$\overline{N}N$ pairs is the hole picture of Dirac.  While this is
certainly a useful picture for free or weakly interacting point
fermions, for strongly interacting or  highly composite fermions, there
is no necessary connection.

Recently, Wallace, Gross, and Tjon \cite{WAL94} constructed a solvable toy
model of a composite comprised of a fermion-boson bound state.  In this
model, one sees  explicitly how, in a complete treatment of the
composite in an external field, virtual composite-anticomposite
pairs are suppressed (relative to what would be expected from a point
Dirac fermion).   At the same time, they show how a
description of  the same physics without explicit reference to
compositeness requires large Z-graph contributions at the level of the
effective theory.  The point here is that the effects of Z graphs
in the effective theory (without explicit reference to the internal
structure)  does {\em not\/} correspond to contributions from virtual
pairs of composite particles and antiparticles.

To reiterate our perspective on this issue:  The essential physics
underlying the success of Dirac phenomenology is the large and
canceling scalar and vector parts of the optical potential (self-energy).
Questions concerning the role of virtual $\overline{N}N$ states or the
validity of the Dirac equation for composite particles, while
interesting in their own right, are not relevant in this
context.

\section{Summary and Outlook}

The extension of QCD sum-rule techniques to finite density can provide a
bridge between nuclear phenomenology and the underlying theory of quantum
chromodynamics.
However, when making connections to QCD, one must recognize that most familiar
nuclear observables are associated with energy scales much smaller than
those of hadronic observables.
Furthermore,
many or most nuclear {\it experimental\/} observables are the
result of fine cancellations.

QCD sum rules, which extrapolate from short times with limited resolution,
cannot be used to resolve small energy differences or to
predict physics that is
determined at long times or distances.
Therefore one should not be too ambitious in applying sum-rule
techniques at finite density.
Detailed experimental predictions of conventional nuclear observables
are more appropriately studied in the context of effective models of QCD.
In this class we would include attempts to understand nuclear matter
saturation or nucleon-nucleon scattering lengths quantitatively
(although {\em qualitative\/} insight might be possible).
The alternative that we have proposed is to focus on phenomenology that
is associated with energy scales more conducive to a sum-rule treatment.

The most solid result to date
connected with finite-density QCD sum rules is the
behavior of the chiral condensate with density.
The model-independent result [Eq.~(\ref{qbar-im})],
which predicts the linear density dependence
of the condensate, appears to be robust, at least up to nuclear matter
saturation density.
Although there are still sizable uncertainties in the value of the nucleon
$\sigma$
term, it is known well enough to imply a very substantial change
(30--40\% reduction) in the
chiral condensate inside of nuclei.
Furthermore, the change with density has been convincingly related to
two-pion exchange physics \cite{BIR94},
which provides a natural connection to the
scalar meson of meson-exchange phenomenology and establishes that the
range is sufficiently large that short-range correlations should not cause
drastic changes.
The latter point is very important; if correlations dominated the
physics, then the value of the {\it average\/} condensate would not be
very relevant.

Given a large change in $\langle\overline q q\rangle_{\rho_N}$,
one might anticipate  compensating
changes in the hadronic spectrum.
QCD sum rules offer a direct way to make a
connection between such QCD ground-state properties and properties of
observed states.
Here we have used the sum rules to associate
$\langle\overline q q\rangle_{\rho_N}$ with the change in the scalar
self-energy of a quasinucleon.
Similarly, changes in the vector self-energy (of opposite
sign) are primary associated with the vector density
$\langle q^\dagger q\rangle_{\rho_N}$.
Assuming the density dependence of the four-quark condensates to be weak,
the resulting self-energies are in qualitative agreement with
relativistic phenomenology.

A disclaimer that must be made is that in another context,
sum rules for nucleons
at finite temperature,
changes in the quark condensate have been {\em mistakenly\/} identified with a
shift in the nucleon mass.
The critical
point is that one must adequately model the phenomenological spectral
density; if not, changes that show up as new features in the spectral
density, could be attributed instead to a shift in the mass
by the fitting procedure.%
\footnote{One should note that even when this happens,
it is not so much a failure
of the sum-rule approach as of the application of the methods.
Thus, with further insight or checks from alternative approaches such as
chiral perturbation theory, one can refine and correct
the sum-rule calculation.}
We know of no analogous problems with our analysis of baryons in
nuclear matter.
While the spectral ansatz is
quite simple and could, in principle, be missing important features,
it is consistent with successful nuclear phenomenology.

Many QCD sum-rule results we have cited are rather indefinite
(at least at present).
However,
the nature of the sum-rule approach should be taken into account when
accessing the value of its predictions.
As in the zero-density case, it is not the quantitative predictions for hadron
masses and other properties that are most important.
Indeed, lattice calculations will ultimately provide far more precise
determinations of the hadronic spectrum.
It is the quantitative relations
and qualitative insight that are most
valuable, and which will persist to challenge any numerical or model
predictions.

It has been said (about vacuum sum rules)
that QCD sum rules only work well when making ``postdictions''
rather than predictions.
This is true to some extent, although there are also notable exceptions.
More precisely, one can say
that the reliability of sum-rule applications, particularly
when applied to new observables and in
new domains, is not guaranteed without some feedback from experiment.
The problem is that subtleties can be missed!
One should not, at the same time,
underestimate the usefulness of postdictions.  After all,
the data {\it is\/} measured experimentally.  What we seek is {\it
understanding\/} of how it fits into a larger picture.  This is what the
sum rules can provide.

For example,
one might ask whether the decomposition into scalar and vector pieces
made in relativistic phenomenology is merely an artificial construction.
The sum-rule results tell us there is a reason to pay attention
to these pieces separately:  The physics is different!
Specifically, the scalar self-energy is predominantly
associated with changes in the scalar condensate, which is in turn
related to chiral symmetry restoration.
On the other hand, the vector self-energy is largely determined by the
vector condensate, or quark density, which is not dynamical.

The ultimate stumbling block to drawing solid quantitative conclusions from
the finite-density sum rules is the density dependence of the
four-quark condensates.
We have repeatedly emphasized the importance of matrix elements of
four-quark field operators in QCD sum rules for light-quark hadrons.
This is true at zero density, finite density, and finite temperature.
In the vacuum, the factorization hypothesis plays an important role in
reducing the inputs to the sum rules to a small number of
phenomenological parameters.
The validity of factorization at zero density
seems to be dependent on the particular four-quark
condensate and seems quite reasonable in some cases.
However, one must remember that a factor of two uncertainty
in a dimension-six condensate
only propagates as a sixth root of two uncertainty in the predicted
mass of a hadron.
Thus one only needs factorization to determine the mass scale rather crudely.

In contrast, our concern is with the density dependence of hadronic
properties, which requires the density dependence of condensates.
Factorization implies a very rapid density dependence for four-quark
condensates, which  has drastic and adverse
effects on the nucleon sum-rule predictions.
Nevertheless, there are too many uncertainties associated with the
finite-density sum rules to insist that our results imply that the
four-quark condensates vary slowly with density.

There are two obvious ways to proceed to clarify the situation,
which are being pursued at present.
The first is to better model the four-quark condensates \cite{SHA94}.
The second way is to concentrate on sum rules that do not rely on the
four-quark condensates.
The key here is to find new, independent sum rules for the nucleon,
taking advantage of interpolating fields with both spin-1/2 and spin-3/2
parts.
By considering a mixed correlator with Ioffe's current and
the spin-3/2--spin-1/2 current,
spin-1/2 intermediate states (including the nucleon)
are still projected, but one generates
additional sum rules for $M_N^*$ that
are independent of the problematic four-quark condensates
\cite{LEI90,FUR95}.
This sum-rule
analysis should provide a clean test of the density dependence of
$M_N^\ast$.

Finally, we comment on sum-rule predictions
of changes in hadron
properties under the extreme conditions of density that can be reached
experimentally in relativistic heavy-ion collisions.
The untangling of experimental signatures for such changes is a formidable
challenge.
The most promising observables appear to be vector-meson masses,
which might be measured by monitoring dilepton production.
There are several difficulties with the quantitative sum-rule predictions
of these masses at present:
\begin{itemize}
 \item[$\bullet$]
  The principal determining factor in the density dependence of masses are
  certain
  four-quark condensates.  Predictions to date of large changes rely,
  once again, on a
  factorization assumption.
 \item[$\bullet$]
  Even accepting factorization, one also needs to extrapolate the density
  dependence of the quark condensate far beyond ordinary nuclear
  densities. This extrapolation is very uncertain at this point in time
  since predictions from model calculations begin to
  diverge from each other  around nuclear matter
  saturation density.
 \item[$\bullet$]
  An essential ingredient of the sum-rule approach is a simple but
  complete model of the phenomenological spectral density.
  There is the definite possibility of missed physics (such as new
  excitations and widths) in the vector-meson channel
  that get translated into spurious predictions
  for mass shifts.
\end{itemize}

One can hope that the sum-rule picture and approach
will be refined as experimental data is accumulated.
At present, we must conclude that extrapolations
to high density based on QCD sum rules are not quantitatively reliable.


\begin{thebibliography}{199}


\bibitem{PRE75} M.~A. Preston and R.~K. Bhaduri,
    {\it Structure of the Nucleus} (Addison-Wesley, Reading, 1975).

\bibitem{WEI90}S. Weinberg, Phys.\ Lett.\ B {\bf 251} (1990) 288;
     Nucl.\ Phys.\ {\bf B363} (1991) 3;
     Phys.\ Lett.\ B {\bf 295} (1992) 114.

\bibitem{LAT94}T.  Draper, S.  Gottlieb, A.  Soni,  D.  Toussaint, eds.
  {\it Lattice '93.    Proceedings, International  Symposium On Lattice
  Field Theory}  (Nucl.\  Phys.\  B, Proc.\  Suppl.\  {\bf 34} (1994)).

\bibitem{HAS83} P. Hasenfratz and F. Karsch,
   Phys.\ Lett.\ {\bf 125B} (1983) 308.

\bibitem{SHI79}M.~A. Shifman, A.~I. Vainshtein, and V.~I. Zakharov,
   Nucl.\ Phys.\ {\bf B147} (1979) 385;
   {\bf B147} (1979) 448;
   {\bf B147} (1979) 519.

\bibitem{IOF81}B.~L. Ioffe, Nucl.\ Phys.\ {\bf B188} (1981) 317;
           {\bf B191} (1981) 591(E).

\bibitem{CHU81}Y.~Chung, H.~G.\ Dosch, M.~Kremer, and D.~Schall,
      Phys.\ Lett.\ {\bf 102B} (1981) 175;
      Nucl.\ Phys.\ {\bf B197} (1982) 55.

\bibitem{REI85}L.~J. Reinders, H.~R. Rubinstein, and S. Yazaki,
     Phys.\ Rep.\ {\bf 127} (1985) 1.

\bibitem{NAR89} S. Narison, {\it QCD Spectral Sum Rules} (World
  Scientific, Singapore, 1989).

\bibitem{SHI92}M.~A. Shifman, ed., {\it Vacuum Structure and QCD Sum
   Rules} (North Holland, Amsterdam, 1992).

\bibitem{DOS94}H.~G. Dosch, Prog.\ Part.\ Nucl.\
               Phys.\ {\bf 33} (1994) 121.

\bibitem{DRU90}E.~G. Drukarev and E.~M. Levin,
     Pis'ma Zh.\ Eksp.\ Teor.\ Fiz.\ {\bf 48} (1988) 307
     [JETP Lett.\ {\bf 48} (1988) 338];
     Zh.\ Eksp.\ Teor.\ Fiz.\ {\bf 95} (1989) 1178
     [Sov.\ Phys.\ JETP {\bf 68} (1989) 680];
     Nucl.\ Phys.\ {\bf A511} (1990) 679;
     {\bf A516} (1990) 715(E).

\bibitem{HAT91}T. Hatsuda, H. H{\o}gaasen, and M. Prakash,
          Phys.\ Rev.\ C {\bf 42} (1990) 2212;
          Phys.\ Rev.\ Lett.\ {\bf 66} (1991) 2851.

\bibitem{COH91}T.~D. Cohen, R.~J. Furnstahl, and D.~K. Griegel,
                Phys.\ Rev.\ Lett.\ {\bf 67} (1991) 961.

\bibitem{ADA91}C. Adami and G.~E. Brown, Z. Phys.\ A {\bf 340} (1991) 93.

\bibitem{DRU91a} E.~G. Drukarev and E.~M. Levin, Nucl.\ Phys.\ {\bf
    A532} (1991) 695.

\bibitem{DRU91} E.~G. Drukarev and E.~M.  Levin,
   Prog.\ Part.\ Nucl.\ Phys.\ {\bf 27} (1991) 77.

\bibitem{FUR92} R.~J. Furnstahl, D.~K. Griegel, and T.~D. Cohen, Phys.\
   Rev.\ C {\bf 46} (1992) 1507.

\bibitem{HAT92} T. Hatsuda and S.~H. Lee, Phys.\ Rev.\ C {\bf 46}  (1992) R34.

\bibitem{JIN93}X. Jin, T.~D. Cohen, R.~J. Furnstahl, and D.~K.
    Griegel, Phys.\  Rev.\ C {\bf 47} (1993) 2882.

\bibitem{SCH93} T. Sch\"afer, V. Koch, and G.~E. Brown,  Nucl.\  Phys.\
   {\bf A562} (1993) 644.

\bibitem{JIN94a}X. Jin and  R.~J. Furnstahl,  Phys.\  Rev.\
    C {\bf 49} (1994) 1190.

\bibitem{HEN93} E.~M. Henley and J. Pasupathy, Nucl.\ Phys.\
         {\bf A556} (1993) 467.

\bibitem{ASA93}M. Asakawa and C.~M. Ko, Nucl.\ Phys.\ {\bf A560} (1993)
    399;  Phys.\ Rev.\ C {\bf 48} (1993) R526.

\bibitem{JIN94}X. Jin, M. Nielsen, T.~D. Cohen, R.~J. Furnstahl,
   and D.~K. Griegel,
   Phys.\ Rev.\ C {\bf 49} (1994) 464.

\bibitem{DRU94} E.~G. Drukarev and M.~G. Ryskin,
    Nucl.\ Phys.\ {\bf A472} (1994) 560.

\bibitem{JIN94c}X. Jin and M. Nielsen,
         Phys.\ Rev.\ C {\bf 51} (1995) 347.

\bibitem{LEH54} H. Lehmann, Nuovo Cimento {\bf 11} (1954) 342; the
  Lehmann representation is discussed in most good texts on quantum field
  theory.  See, for example, Ref.~\cite{ITZ80}.

\bibitem{WIL64}K. G. Wilson, {\it On Products of Operators at Short Distance}
   (Cornell Report 1964); Phys.\ Rev.\ {\bf 179}  (1969) 1499.

\bibitem{COL84} J.~C. Collins,  {\it Renormalization\/} (Cambridge
       University Press, New York, 1984).



\bibitem{CLA83a}B.~C. Clark, S. Hama, and R.~L. Mercer,
    in {\it The Interaction Between Medium Energy Nucleons in Nuclei},
  edited by H.~O. Meyer (American Institute of Physics, New York, 1983).

\bibitem{CLA83b} B.~C. Clark, R.~L. Mercer, and P. Schwandt, Phys.\ Lett.\
    {\bf 122B} (1983) 211.

\bibitem{CLA83c} B.~C. Clark, S. Hama, R.~L. Mercer, L. Ray, and B.~D.
     Serot,  Phys.\ Rev.\ Lett.\ {\bf 50} (1983) 1644.

\bibitem{MCN83} J.~A. McNeil, J.~R. Shepard, and
    S.~J. Wallace, Phys.\ Rev.\ Lett.\ {\bf 50} (1983) 1439.

\bibitem{SHE83}J.~R. Shepard, J.~A. McNeil, and
   S.~J. Wallace, Phys.\ Rev.\ Lett.\ {\bf 50} (1983) 1443.

\bibitem{TJO87} J.~A. Tjon and S.~J. Wallace, Phys.\
   Rev.\ C {\bf 36}  (1987) 1085.

\bibitem{WAL87}S.~J. Wallace, Annu.\ Rev.\ Nucl.\ Part.\ Sci.\
        {\bf 37} (1987) 267.

\bibitem{HAM90}S. Hama, B.~C. Clark, E.~D. Cooper, H.~S. Sherif, and
     R.~L. Mercer, Phys.\ Rev.\ C {\bf 41} (1990) 2737.

\bibitem{SER86}B.~D. Serot and J.~D. Walecka,
                 Adv.\ Nucl.\ Phys.\ {\bf 16} (1986) 1.

\bibitem{SER92} B.~D. Serot, Rep.\ Prog.\ Phys.\  {\bf 55} (1992) 1855.

\bibitem{BRO84} S.~J. Brodsky, Comm.\ Nucl.\ Part.\ Phys.\ {\bf 12} (1984) 213.

\bibitem{THE85} M. Thies, Phys.\  Lett.\  {\bf 162B} (1985) 255;
         {\bf 166B} (1986) 23.

\bibitem{COO86} E.~D. Cooper and B.~K. Jennings,
       Nucl.\ Phys.\ {\bf A458} (1986) 717.

\bibitem{BLE87} E. Bleszynski, M. Bleszynski, and T. Jaroszewicz,
   Phys.\ Rev.\ Lett.\ {\bf 59} (1987) 423.

\bibitem{ACH88} J. Achtzehnter and L. Wilets, Phys.\ Rev.\ C {\bf 38} (1988) 5.

\bibitem{JAR91} T. Jaroszewicz and S.~J. Brodsky, Phys.\ Rev.\ C {\bf
    43} (1991) 1946.

\bibitem{COH91a}T.~D. Cohen, Phys.\ Rev.\ C {\bf 45} (1992) 833.

\bibitem{WAL94} S.~J. Wallace, F. Gross, and
   J.~A. Tjon,  University of Maryland Report No.\ 95--020 (1994).

\bibitem{HOO80} G. `t Hooft, in {\it Recent Developments in Gauge
   Theories}, edited by G. `t Hooft {\it et al.\/} (Plenum, New York, 1980).

\bibitem{BIR94} M.~C. Birse, J.\ Phys.\ G {\bf 20} (1994) 1537.

\bibitem{COH92}T.~D. Cohen, R.~J. Furnstahl, and D.~K. Griegel,
  Phys.\ Rev.\ C {\bf 45} (1992) 1881.

\bibitem{BRO89} G.~E. Brown and M. Rho, Phys.\ Lett.\ {\bf B222} (1989) 324.

\bibitem{SOY93} M. Soyeur, G.~E. Brown, and M. Rho, Nucl.\ Phys.\ {\bf
   A556} (1993) 355.

\bibitem{BRO88} G.~E. Brown, C.~B. Dover, P.~B. Siegel, and W. Wiese,
    Phys.\ Rev.\ Lett.\ {\bf 60} (1988) 2723.

\bibitem{DIL} The Phoenix collaboration  at RHIC and the
       Hades collaboration at GSI will look for this effect.

\bibitem{MOS94} U. Mosel at {\it INT/CEBAF Workshop on Real
   and Virtual Compton Scattering}, Crystal Mountain,
   Washington, Sept.\ 24--29, 1994.

\bibitem{RIN80}P. Ring and P. Schuck, {\em The Nuclear Many-Body
      Problem\/} (Springer, Berlin, 1980).

\bibitem{SHI92b}M.~A.\ Shifman, QCD Sum Rules: The Second Decade, in
   {\it QCD: 20 Years Later}, edited by P.~M. Zerwas
   and H.~A. Kastrup (World Scientific, Singapore, 1993) 775-794.

\bibitem{NOV81}  V.~A.  Novikov, M.~A.  Shifman,
   A.~I.  Vainshtein, and V.I.  Zakharov, Nucl.\  Phys.\ {\bf B191} (1981) 301.

\bibitem{MAN84}A. Manohar and H. Georgi, Nucl.\ Phys.\ {\bf B234}
        (1984) 189.

\bibitem{ITZ80}C. Itzykson and J.-B. Zuber,  {\it Quantum Field Theory}
       (McGraw-Hill, New York, 1980).

\bibitem{IOF83}B.~L. Ioffe,
          Z.\ Phys.\ C {\bf 18} (1983) 67.

\bibitem{BJO65}J.~D. Bjorken and S.~D. Drell,
       {\it Relativistic Quantum Fields}
     (McGraw-Hill, New York, 1965).

\bibitem{VAI85}A.~I. Vainshtein, V.~I. Zakharov, V.~A. Novikov,
    and M.~A. Shifman, Yad.\ Fiz.\ {\bf 41} (1985) 1063
    [Sov.\ J. Nucl.\ Phys.\ {\bf 41} (1985) 683], and references
    therein.

\bibitem{NOV85}  V.~A.  Novikov, M.~A.  Shifman,
    A.~I. Vainshtein, and V.~I. Zakharov, Nucl.\ Phys.\ {\bf B249} (1985) 445.

\bibitem{DAV86}F. David, Nucl.\ Phys.\ {\bf B263} (1986) 637,
     and references therein.

\bibitem{PAS84}P. Pascual and R. Tarrach, {\it QCD: Renormalization
   for the Practitioner\/} (Springer-Verlag, New York, 1984).

\bibitem{IOF84}B.~L. Ioffe and A.~V. Smilga, Nucl.\ Phys.\ {\bf B232}
      (1984) 109.

\bibitem{REI84}L. J. Reinders,
   Acta Physica Polonica {\bf B15} (1984) 329.

\bibitem{OVC88}A.~A. Ovchinnikov and A.~A. Pivovarov,
  Yad.\ Fiz.\ {\bf 48} (1988) 1135
  [Sov.\ J.\ Nucl.\ Phys.\ {\bf 48} (1988) 721].

\bibitem{LEI94} D.~B.  Leinweber, Ohio State University Report No.\
     94--0332 (1994).

\bibitem{DOR90}A.~E. Dorokhov and N.~I. Kochelev,
             Z. Phys.\ C {\bf 46} (1990) 281.

\bibitem{FOR93} H. Forkel and M.~K. Banerjee, Phys.\ Rev.\ Lett.\ {\bf
         71} (1993) 484.

\bibitem{DEJ91}F. de Jong and R. Malfliet, Phys.\ Rev.\ C
        {\bf 44} (1991) 998.

\bibitem{AMO92}A. Amorim and J. A. Tjon, Phys.\ Rev.\ Lett.\
               {\bf 68} (1992) 772.

\bibitem{LI93} G. Q. Li and R. Machleidt, University
     of Idaho Reports Nos.\ UI-NTH-9305, UI-NTH-9307, and UI-NTH-9310 (1993).

\bibitem{RUS94}J. J. Rusnak and R. J. Furnstahl, Ohio State University
   Report No.\ 94--220 (1994) unpublished.

\bibitem{MAH85}C. Mahaux, P.~F. Bortignon, R.~A. Broglia, and
     C.~H. Dasso, Phys.\ Rep.\ {\bf 120} (1985) 1.

\bibitem{JAM89}M. Jaminon and C. Mahaux,
    Phys.\ Rev.\ C {\bf 40} (1989) 354.

\bibitem{FUR94}R.~J.\ Furnstahl, Phys.\ Rev.\ C {\bf 50} (1994) 1735.

\bibitem{GEL68}M. Gell-Mann, R.~J. Oakes, and B. Renner Phys.\ Rev.\ {\bf
   175} (1968) 2195.

\bibitem{GAS91}
       J. Gasser, H. Leutwyler, and M.~E. Sainio, Phys.\ Lett.\
        B {\bf 253} (1991) 252, and references therein.

\bibitem{BIR93} M.~C. Birse and J.~A. McGovern,
   Phys.\  Lett.\  B {\bf 309} (1993) 231.

\bibitem{CEL93} L.~S. Celenza, C.~M. Shakin, W.-D. Sun, and X. Zhu,
    Phys.\ Rev.\ C {\bf 48} (1993) 159.

\bibitem{CHA93} G. Chanfray  and M. Ericson, Nucl.\ Phys.\
         {\bf A556} (1993) 427.

\bibitem{ERI93} M.  Ericson, Phys.\  Lett.\  B {\bf 301} (1993) 11.

\bibitem{NYF93} A. Nyffeler, Z.\ Phys.\ C {\bf 60} (1993) 159.

\bibitem{ERI94} T.~E.~O. Ericson, Phys.\  Lett.\  B {\bf 321} (1994) 312.

\bibitem{MER70} E. Merzbacher, {\it Quantum Mechanics} (Wiley, New
    York, 1970), 2nd ed.

\bibitem{COH94} T.~D. Cohen and W. Broniowski,
Phys.\ Lett.\ B {\bf 342} (1995) 25.

\bibitem{BHA88}R.~K. Bhaduri,
  {\it Models of the Nucleon: From Quarks to Soliton}
  (Addison-Wesley, Redwood City, 1988).

\bibitem{FET71}A.~L. Fetter and J.~D. Walecka,
   {\it Quantum Theory of Many-Particle Systems}
   (McGraw-Hill, New York, 1971).

\bibitem{FOC37}V. Fock,
  Phys.\ Z.  Sowjetunion {\bf 12} (1937) 404.

\bibitem{SCH70}J. Schwinger,
    {\it Particles, Sources, and Fields}
    (Addison-Wesley, Reading, MA, 1970), Vol.~I.

\bibitem{CRO80}C. Cronstr\"om,
  Phys.\ Lett.\ {\bf 90B} (1980) 267.

\bibitem{SMI82}A.~V. Smilga,
   Yad.\ Fiz.\ {\bf 35} (1982) 473
   [Sov.\ J.\ Nucl.\ Phys.\ {\bf 35} (1982) 271].

\bibitem{HUB82}W. Hubschmid and S. Mallik,
  Nucl.\ Phys.\ {\bf B207} (1982) 29.

\bibitem{NOV84b}V.~A. Novikov, M.~A. Shifman, A.~I. Vainshtein,
  and V.~I. Zakharov,
   Fortschr.\ Phys.\ {\bf 32} (1984) 585.

\bibitem{SHI80}M.~A. Shifman,
   Nucl.\ Phys.\ {\bf B173} (1980) 13.

\bibitem{JIN93b}X. Jin, Ph.D. Thesis, University of Maryland
    at College Park (1993).

\bibitem{TAR82}R. Tarrach,
    Nucl.\ Phys.\ {\bf B196} (1982) 45.

\bibitem{GAS82}J. Gasser and H. Leutwyler,
   Phys.\ Rep.\ {\bf 87} (1982) 77.

\bibitem{BEL82}V.~M. Belyaev and B.~L. Ioffe,
  Zh.\ Eksp.\ Teor.\ Fiz.\ {\bf 83} (1982) 876
  [Sov.\ Phys.\ JETP {\bf 56} (1982) 493];
  {\bf 84} (1983) 1236
  [{\bf 57} 716 (1983)].

\bibitem{LEI90}D.~B. Leinweber,
    Annals of Physics {\bf 198} (1990) 203.

\bibitem{VAI78}A.~I. Vainshtein, V.~I. Zakharov,
    and M.~A. Shifman,
    Pis'ma Zh.\ Eksp.\ Teor.\ Fiz.\ {\bf 27} (1978) 60
    [JETP Lett.\ {\bf 27} (1978) 55].

\bibitem{SHU88}E.~V. Shuryak,
  {\it The QCD Vacuum, Hadrons, and the Superdense Matter}
   (World Scientific, Singapore, 1988), and references therein.

\bibitem{KRE87}M. Kremer and G. Schierholz,
  Phys.\ Lett.\ B {\bf 194} (1987) 283.

\bibitem{BAK91}A.~P. Bakulev and A.~V. Radyushkin,
  Phys.\ Lett.\ B {\bf 271} (1991) 223.

\bibitem{MIK92}S.~V. Mikhailov and A.~V. Radyushkin,
        Phys.\ Rev.\ D {\bf 45} (1992) 1754.

\bibitem{SHU89}E.~V. Shuryak,
  Nucl.\ Phys.\ {\bf B328} (1989) 85.

\bibitem{NOV84}  V.~A.  Novikov, M.~A.  Shifman,
   A.~I.  Vainshtein, M.~B.  Voloshin, and V.~I.  Zakharov, Nucl.\ Phys.\
   {\bf B237} (1984) 525.

\bibitem{ZHI85}A.~R. Zhitnitski\u{\i},
    Yad.\ Fiz.\ {\bf 41} (1985) 805
    [Sov.\ J.\ Nucl.\ Phys.\ {\bf 41} (1985) 513];
    {\bf 41} (1985) 1035 [{\bf 41} (1985) 664];
    {\bf 41} (1985) 1331 [{\bf 41} (1985) 846]

\bibitem{GOV87}J. Govaerts, L.~J. Reinders, F. de Viron, and J. Weyers,
  Nucl.\ Phys.\ {\bf B283} (1987) 706.

\bibitem{COL82}J.~C. Collins and D.~E. Soper,
  Nucl.\ Phys.\ {\bf B194} (1982) 445.

\bibitem{EFR80}A.~V. Efremov and A.~V. Radyushkin,
   Riv.\ Nuovo Cimento {\bf 3}(2) (1980) 1, and references
  therein.

\bibitem{CUR80}G. Curci, W. Furmanski, and R. Petronzio,
  Nucl.\ Phys.\ {\bf B175} (1980) 27.

\bibitem{GLU90}M. Gl\"uck, E. Reya, and A. Vogt,
 Z.\ Phys.\ C {\bf 48} (1990) 471.

\bibitem{GLU92}M. Gl\"uck, E. Reya, and A. Vogt,
  Z.\ Phys.\ C {\bf 53} (1992) 127.

\bibitem{SHI94}M.~Shifman, University of Minnesota preprint,
      HEP-PH-9501222, 1995.

\bibitem{JIN94d}X. Jin, Phys.\ Rev.\ C {\bf 51} (1995), in press.

\bibitem{KON93}Y. Kondo and O. Morimatsu, Phys.\ Rev.\ Lett.\
       {\bf 71} (1993) 2855.

\bibitem{FUR94b}R.~J.\ Furnstahl and T. Hatsuda, Phys.\
                 Rev.\ Lett.\ {\bf 72} (1994) 3128.

\bibitem{BRO77}R. Brockmann and W. Weise, Phys.\ Lett.\ {\bf 69B}
  (1977) 167.

\bibitem{NOB80}J.~V. Noble, Phys.\ Lett. {\bf 89B} (1980) 325.

\bibitem{JCO87}J. Cohen and R.~J. Furnstahl,
   Phys.\ Rev.\ C {\bf 35} (1987) 2231;
   J. Cohen, Phys.\ Rev.\ C {\bf 48} (1993) 1346.

\bibitem{JEN90}B.~K. Jennings, Phys.\ Lett.\ B {\bf 246} (1990) 325.

\bibitem{GAT91}A.~O. Gattone, M. Chiapparini, and E.~D. Izquierdo,
  Phys.\ Rev.\ C {\bf 44} (1991) 548;
  M. Chiapparini, A.~O. Gattone, and B.~K. Jennings,
   Nucl.\ Phys.\ {\bf A529} (1991) 589.

\bibitem{JCO91}J. Cohen and H.~J.\ Weber, Phys.\ Rev.\ C {\bf 44}
            (1991) 1181.

\bibitem{MAR93}J. Mare\v{s}, Particles and Nuclei XIII International
Conference,
   Book of Abstracts Vol. 2, (1993) 661.

\bibitem{MAR94}J. Mare\v{s} and B.~K. Jennings, Phys.\ Rev.\ C
        {\bf 49} (1994) 2472.

\bibitem{GLE93}N. K. Glendenning {\it et al.}, Phys.\ Rev.\ C {\bf 48}
  (1993) 889.

\bibitem{RUF90}M. Rufa, J. Schaffner, J. Maruhn, H. St\"ocker,
    W. Greiner, and P.-G. Reinhard, Phys.\ Rev.\ C {\bf 42}
       (1990) 2469.

\bibitem{SCH92}J. Schaffner, C. Greiner, and H. St\"ocker, Phys.\ Rev.\ C
    {\bf 46} (1992) 322.

\bibitem{PIR79}H.~J. Pirner, Phys.\ Lett.\ {\bf 85B} (1979) 190.

\bibitem{DOV84}C. B. Dover and A. Gal, Prog.\ Part.\ Nucl.\ Phys. {\bf 12}
         (1984) 171.

\bibitem{BOU77}A. Bouyssy, Nucl.\ Phys.\ {\bf A290} (1977) 324.

\bibitem{MIL88}D.~J. Millener, C.~B. Dover, and A. Gal, Phys.\ Rev.\ C
   {\bf 38} (1988) 2700; D.~J.\ Millener, A. Gal, C.~B.\ Dover,
    and R.~H. Dalitz,
    Phys.\ Rev.\ C {\bf 31} (1985) 499.

\bibitem{REI83}L.~J. Reinders, H.~R. Rubinstein, and S. Yazaki, Phys.\ Lett.
  {\bf 120B} (1983) 209.

\bibitem{CHI85}C.~B. Chiu, J. Pasupathy, and S.~L. Wilson, Phys.\ Rev.\ D
  {\bf 32} (1985) 1786.

\bibitem{BRO91} G.~E. Brown and M. Rho,   Phys.\  Rev.\  Lett.\
    {\bf 66} (1991) 2720.

\bibitem{ALT80} R. Altemus {\it et al.\/}, Phys.\ Rev.\ Lett.\ {\bf 44}
   (1980) 965;  P. Barreau  {\it et al.\/},  Nucl.\ Phys.\ {\bf A402} (1983)
   515;  Z.~E.  Meziani {\it et al.\/},
    Phys.\  Rev.\  Lett.\ {\bf 52} (1984) 2130; Z.~E.  Meziani {\it et al.\/},
   Phys.\  Rev.\  Lett.\ {\bf 54} (1985) 1233; A. Zghiche {\it et al.\/},
   Nucl.\ Phys.\ {\bf A572} (1994) 513.

\bibitem{REF88} D. Reffay-Pikeroen {\it et al.}, Phys.\  Rev.\  Lett.\
  {\bf 60} (1988) 776; A. Magnon  {\it et al.}, Phys.\   Lett.\ {\bf B222}
  (1989) 352; J.~E. Ducret {\it et al.},  Nucl.\ Phys.\ {\bf A556} (1993) 373.

\bibitem{NOB81} J.~V. Noble, Phys.\ Rev.\ Lett.\ {\bf 46} (1981) 412.

\bibitem{CEL85} L.~S. Celenza, A. Rosenthal, and C.~M. Shakin,  Phys.\
   Rev.\ C {\bf 31} (1985) 232.

\bibitem{FUR90} R.~J.  Furnstahl,  T.  Hatsuda, and S.~H.  Lee, Phys.\
    Rev.\ D {\bf 42} (1990) 1744.

\bibitem{HER92}  M.  Herrmann,
  B.~L.  Friman, and W.  N\"orenberg,  Nucl.\  Phys.\  {\bf A545} (1992)
  267c;  {\bf A560} (1993) 411.

\bibitem{COH89} T.~D. Cohen, Nucl.\ Phys.\  {\bf A495} (1989) 545.

\bibitem{SHU84} E.~V. Shuryak, Phys.\ Rep.\ {\bf 115} (1984) 151.

\bibitem{GRI94} D.~K.  Griegel and T.~D.  Cohen,  Phys.\ Lett.\  B
      {\bf 333} (1994) 27.

\bibitem{GAS88} J. Gasser, M.~E. Sainio, and A. \v{S}varc, Nucl.\ Phys.\
     {\bf B307} (1988) 779.

\bibitem{CHO94} S.~H.\ Lee, S. Choe, T.~D. Cohen, D.~K. Griegel,
         Phys.\ Lett.\ B (to be published).

\bibitem{SHA94}L. S. Celenza, C. M. Shakin, W.-D. Sun,
          J. Szweda, Brooklyn College Report No.\ BCCNT-94-081-239-R, 1994.

\bibitem{FUR95}R.~J.\ Furnstahl, X.~Jin, and D.~B.\ Leinweber,
          in preparation.



\end{thebibliography}
\end{document}